\documentclass[a4paper, 10 pt, journal, twoside]{IEEEtran}

\ifCLASSINFOpdf
\else
\fi

\usepackage{gensymb}

\makeatletter
\newcommand\notsotiny{\@setfontsize\notsotiny\@vipt\@viipt}
\makeatother

\usepackage{amsfonts}
\usepackage{amsmath}
\usepackage{amssymb}
\usepackage{balance}
\usepackage{blindtext}
\usepackage{centernot}
\usepackage{array}
\usepackage{graphicx}
\usepackage{adjustbox}
\usepackage{hyperref}
\usepackage[latin1]{inputenc}
\usepackage{lineno}
\usepackage[ruled,vlined,linesnumbered]{algorithm2e}
\usepackage{algpseudocode}

\usepackage{mathrsfs}
\usepackage{mathtools}
\usepackage{todonotes}
\usepackage{verbatim}
\usepackage{float}
\usepackage{upgreek}
\usepackage{mathdots}
\usepackage{listings}

\usepackage{times}
\usepackage{titlesec}

\usepackage{dirtytalk}
\usepackage{wrapfig}

\usepackage[backend=biber,style=numeric-comp,sorting=none,maxbibnames=99]{biblatex}

\newcommand{\cost}{\text{cost}}

\newcommand{\diag}{\text{diag}}
\newcommand{\df}{d_f}
\newcommand{\dx}{d_x}
\newcommand{\dz}{d_z}
\newcommand{\E}{\mathbb{E}}
\newcommand{\Exp}{\text{Exp}}
\newcommand{\F}{\mathcal{F}}
\newcommand{\IMU}{\text{IMU}}

\newcommand{\Log}{\text{Log}}

\newcommand{\nf}{p}
\newcommand{\nx}{n}
\newcommand{\N}{\mathbb{N}}

\newcommand{\ra}{\rightarrow}
\newcommand{\R}{\mathbb{R}}

\newcommand{\X}{\mathcal{X}}

\newcommand{\Z}{\mathcal{Z}}

\newtheorem{theorem}{Theorem}[section]

\newtheorem{definition}[theorem]{Definition}

\newtheorem{remark}{Remark}[section]

\bibliography{references}

\begin{document}


\title{Simultaneous Localization and Mapping: Through the Lens of Nonlinear Optimization}
  \author{Amay Saxena$^\star$,
      Chih-Yuan Chiu$^\star$ $^\dagger$,
      Ritika Shrivastava,
      Joseph Menke,
      and~Shankar~Sastry
      
  \thanks{Manuscript received: February 24th, 2022; Revised May 20th, 2022; Accepted May 28th, 2022.}
  
  \thanks{This paper was recommended for publication by Editor Sven Behnke upon evaluation of the Associate Editor and Reviewers' comments.}

  \thanks{Research supported by NSF Grant DMS 2013985 THEORINet: Transferable, Hierarchical, Expressive, Optimal, Robust and Interpretable Networks and U.S. Office of Naval Research MURI grant N00014-16-1-2710.}
  
  \thanks{$^\star$Equal contribution.}
  
  \thanks{$^\dagger$Corresponding author.}
  
  \thanks{All authors are with the Department of EECS at the University of California, Berkeley, CA 94720 USA (emails: \texttt{\{amaysaxena, chihyuan\_chiu, joemenke, ritishri, sastry\} at berkeley dot edu.})
  }
  
  \thanks{Digital Object Identifier (DOI): see top of this page.}
  
  }

\maketitle

\thispagestyle{plain}
\pagestyle{plain}

\allowdisplaybreaks

\begin{abstract}

Simultaneous Localization and Mapping (SLAM) algorithms perform visual-inertial estimation via filtering or batch optimization methods. Empirical evidence suggests that filtering algorithms are computationally faster, while optimization methods are more accurate. This work presents an optimization-based framework that unifies these approaches, and allows users to flexibly implement different design choices, e.g., the number and types of variables maintained in the algorithm at each time. We prove that filtering methods correspond to specific design choices in our generalized framework. We then reformulate the Multi-State Constrained Kalman Filter (MSCKF) and contrast its performance with that of sliding-window based filters. Our approach modularizes state-of-the-art SLAM algorithms to allow for adaptation to various scenarios. Experiments on the EuRoC MAV dataset verify that our implementations of these algorithms are competitive with the performance of off-the-shelf implementations in the literature. Using these results, we explain the relative performance characteristics of filtering and batch-optimization based algorithms in the context of our framework. We illustrate that under different design choices, our empirical performance interpolates between those of state-of-the-art approaches.

\end{abstract}




\section{Introduction}
\label{sec: Introduction}

In Simultaneous Localization and Mapping (SLAM), a robotic agent maps its uncharted environment while locating itself in the constructed map \cite{LeonardDurrantWhyte1991SimultaneousMapBuilding}. Applications include military map construction, search-and-rescue missions, augmented and virtual reality, and 3D scene capture \cite{Cadena2016PastPresentandFuture,
davison2018futuremapping1, davison2019futuremapping2}.


Typical modern SLAM algorithms consist of \textit{front} and \textit{back ends}. The front end performs feature extraction, data association, and outlier rejection on raw sensor data. 
The back end then uses dynamics and measurement models for inference over the processed data, and produces compatible state estimates.
Back end algorithms are often considered one of two classes---\textit{Gaussian filtering} or \textit{batch optimization} based. Filtering methods iteratively refine the distribution of recent states under a Gaussian prior \cite{Sola2014SLAMWithEKF, Mourikis2007MultiStateConstraintKalmanFilter, LiMourikis2012ImprovingTA}, while optimization methods iteratively estimate states as solutions to an optimization problem, with objective constructed from inertial measurement unit (IMU) and image reprojection error terms.
In particular, factor graph-based approaches efficiently solve optimization problems over past variables via factorization schemes that maintain the sparsity of the underlying least squares problem \cite{dellaert2012gtsam, KaessDellaert2009ISAM, DellaertKaess2006SquareRootSAM, KaessDellaert2012iSAM2}. Keyframe-based methods are optimization-based approaches that retain only a small subset of maximally informative frames (\say{keyframes}) spaced arbitrarily far apart in time in the optimization window, while dropping all other poses \cite{Leutenegger2015KeyframebasedVO}. 
Empirically, both classes of algorithms attain state-of-the-art performance, though the latter often attain higher accuracy at the cost of longer compute times \cite{Cadena2016PastPresentandFuture, Leutenegger2015KeyframebasedVO, Eckenhoff2019ClosedformPM}. 

Prior literature contrasted theoretical and empirical properties of filtering and batch optimization algorithms. Scaramuzza and Fraundorfer compared filtering and bundle adjustment-based methods for visual odometry \cite{ScaramuzzaFraundorfer2011VisualOdometryTutorialPartI, FraundorferScaramuzza2012VisualOdometryTutorialPartII}. Frese et al. surveyed the use of grid-based and pose graph-based SLAM algorithms from a practitioner's perspective 
\cite{Frese2010SLAMOverviewFromUserPerspective}. Huang and Dissanyake conducted a theoretical study of the consistency, accuracy, and computational speed of filtering, optimization-based, and pose-graph SLAM \cite{HuangDissanayake2016ACritiqueOfCurrentDevelopmentsInSLAM}. Khosoussi et al. exploited sparsity in SLAM problems by  conditioning on estimates of robot orientations \cite{Khosoussi2016ASparseSeparableSLAMBackEnd}.  
Strasdat et al. conducted Monte Carlo experiments on visual SLAM algorithms \cite{
StrasdatDavison2012VisualSLAMWhyFilter}, revealed that including more features in the back end increased accuracy more (compared to including more frames), and concluded that bundle adjustment (BA) outperforms filtering, since its computation time increases less drastically with the number of features. 

In this work, we build upon prior literature 
by formulating a unified optimization-based framework for the SLAM back end that encompasses a large class of
existing, state-of-the-art SLAM algorithms. We use this unified framework to recast the Extended Kalman Filter (EKF), Multi-State Constrained Kalman Filter (MSCKF), and Open Keyframe Visual-Inertial SLAM algorithm (OKVIS) as optimization-based back-end algorithms, and compare the empirical performance of the reformulated MSCKF with that of sliding window optimization-based back-end algorithms, including the keyframe-based approach of Open Keyframe Visual-Inertial SLAM \cite{Leutenegger2015KeyframebasedVO}. Somewhat surprisingly, the MSCKF outperforms sliding window filters (SWF) of comparable sizes on several datasets, despite not performing multiple Gauss-Newton updates. We use our generalized framework to analyze these empirical findings.

\section{SLAM: Formulation on Euclidean Spaces}
\label{sec: Problem Formulation on Euclidean Spaces}

\subsection{SLAM on Euclidean Spaces}
\label{subsec: SLAM on Euclidean Spaces: Dynamics and Measurement Maps}


SLAM estimates two types of variables: \textit{states} and \textit{features}. 
    The state at each time $t$, denoted $x_t \in \R^{\dx}$, encodes information describing the robot, e.g., camera positions and orientations (poses). Feature positions available at time $t$ in a global frame, denoted $\{f_k| j = 1, \cdots, p\} \subset \R^{\df}$, can be obtained by analyzing information from image measurements $\{z_{t,k}| j = 1, \cdots, p\} \subset \R^{\dz}$ and state estimates; these describe the relative position of the robot in its environment. 
    States and features are described by an infinitely differentiable (i.e., $C^\infty$) dynamics map $g: \R^{\dx} \ra \R^{\dx}$ and a $C^\infty$ measurement map $h: \R^{\dx} \times \R^{\df} \ra \R^{\dz}$, via additive noise models:
    \begin{align} \label{Eqn: Problem Formulation, Dynamics}
        x_{t+1} &= g(x_t) + w_t, \hspace{5mm} w_t \sim \mathcal{N}(0, \Sigma_w), \\
        \label{Eqn: Problem Formulation, Measurement}
        z_{t,k} &= h(x_t, f_k) + v_{t,k}, \hspace{5mm} v_{t,k} \sim \mathcal{N}(0, \Sigma_v),
    \end{align}
    where $\Sigma_w \in \R^{\dx \times \dx}, \Sigma_w \succeq 0$ and $\Sigma_v \in \R^{\dz \times \dz}, \Sigma_v \succeq 0$.
    
    
    
    
    

For localization and mapping, SLAM algorithms maintain a \textit{full state (vector)} $\overline{x_t} \in \R^d$, in which a number of past states and feature positions are concatenated. The exact number and time stamps of these states and features vary with the design choice of each SLAM algorithm. For example, sliding window filters (SWFs) may define the full state $\overline{x_t} := (x_{t-n+1}, \cdots, x_t, f_{\nf-q+1}, \cdots, f_{\nf}) \in \R^d$, with $d := \dx n + \df q$, to be a sliding window of the most recent $n$ states, consisting of one pose each, and the most recent estimates, at time $t$, of a collection of $q$ features \cite{Sola2014SLAMWithEKF, Mourikis2007MultiStateConstraintKalmanFilter}. Batch optimization methods, on the other hand, maintain all states and features encountered in the problem up to the current time \cite{ DellaertKaess2006SquareRootSAM, dellaert2012gtsam, KaessDellaert2009ISAM, KaessDellaert2012iSAM2}.

Equations \eqref{Eqn: Problem Formulation, Dynamics} and \eqref{Eqn: Problem Formulation, Measurement} do not involve overparameterized state variables, e.g., quaternion representations for poses, which are discussed in Section \ref{sec: Problem Formulation on General Manifolds}.

\subsection{SLAM as an Optimization Problem on Euclidean Spaces}
\label{subsec: SLAM on Euclidean Spaces: the Optimization Problem}


SLAM estimates state and feature positions that best enforce constraints posed by given dynamics and measurement models, as well as noisy state and feature measurements collected over time. This is formulated as the minimization of the sum of weighted residual terms representing these constraints. For example, weighted residuals associated with the prior distribution over  $\overline{x_t} \in \R^d$, the dynamics constraints between states $x_i, x_{i+1} \in \R^{\dx}$, and the reprojection error of feature $f_j \in \R^{\df}$ corresponding to the state $x_i \in \R^{\dx}$ and image measurement $z_{t,k} \in \R^{\dz}$, may be given by $\Sigma_0^{-1/2} (\overline{x_t} - \mu_0) \in \R^d$, $\Sigma_w^{-1/2} \big(x_{i+1}-g(x_i) \big) \in \R^{\dx}$, and $\Sigma_v^{-1/2} \big(z_{i, j}-h(x_i,f_k) \big) \in \R^{\dz}$, respectively (here, $1 \leq i \leq n-1$, $1 \leq j \leq q$).
We define the running cost, $c: \R^{\dx n + \df q} \ra \R$, as the sum of weighted norm squares of these residuals. For example, for a SWF algorithm for SLAM:
\begin{align} \label{Eqn: Cost, General}
    c(\overline{x_t}) 
    &:= \Vert \overline{x_t}-\mu_0 \Vert_{\Sigma_0^{-1}}^2 + \sum_{i=t-n+1}^{t-1} \Vert x_{i+1}-g(x_{i}) \Vert_{\Sigma_w^{-1}}^2 \\ \nonumber
    &\hspace{1cm} + \sum_{j= \nf-q+1}^{\nf} \sum_{i=t-n+1}^{t}  \Vert z_{i,j}-h(x_i,f_k) \Vert_{\Sigma_v^{-1}}^2,
\end{align}
where $\Vert v \Vert_A^2 := v^\top Av$ for any real vector $v$ and real matrix $A$ of compatible dimension.

To formulate SLAM as a nonlinear least-squares problem, we stack all residual terms
into one residual vector $C(\overline{x_t})$. For example, for the SWF given above:
\begin{align*}
    &C(\overline{x_t}) := \Big[ \big( \Sigma_0^{-1/2}(\overline{x_t} - \mu_0) \big)^\top \\
    &\hspace{3mm} \big( \Sigma_w^{-1/2} (x_{t-n+1} - g(x_{t-n})) \big)^\top \cdots \big( \Sigma_w^{-1/2} (x_t - g(x_{t-1})) \big)^\top \\
    &\hspace{3mm} \big( \Sigma_v^{-1/2} ( z_{t-n+1, \nf-q+1} - h(x_{t-n+1}, f_{\nf-q+1})) \big)^\top \cdots \\
    &\hspace{3mm} \big( \Sigma_v^{-1/2} ( z_{t-n+1, p} - h(x_{t-n+1}, f_{\nf})) \big)^\top \cdots \\
    &\hspace{3mm} \big( \Sigma_v^{-1/2} ( z_{t, \nf-q+1} - h(x_t, f_{\nf-q+1})) \big)^\top \cdots \\
    &\hspace{3mm} \big( \Sigma_v^{-1/2} (z_{t, p} - h(x_t, f_{\nf})) \big)^\top \Big]^\top 
    \in \R^{(2n-1)\dx + nq \dz}.
\end{align*}
Thus, $c(\overline{x_t}) = C(\overline{x_t})^\top C(\overline{x_t})$, and the SLAM problem is now reduced to the nonlinear least squares problem below:
\begin{align}
    \label{Eqn: Problem Formulation, Inner Product Cost}
    \min_{\overline{x_t}}. c(\overline{x_t}) = \min_{\overline{x_t}}. C(\overline{x_t})^\top C(\overline{x_t})
\end{align}
Section \ref{sec: Main Algorithm} introduces the main algorithmic submodules used to find an approximate solution to \eqref{Eqn: Problem Formulation, Inner Product Cost}.






\section{SLAM: Formulation on Manifolds}
\label{sec: Problem Formulation on General Manifolds}




Here, we generalize the SLAM formulation in Section \ref{sec: Problem Formulation on Euclidean Spaces} to the case where dynamical states are defined on smooth manifolds rather than Euclidean spaces. SLAM often involves estimating the orientations of rigid bodies, which evolve on a smooth manifold embedded in an ambient space, e.g., rotation matrices expressed as unit quaternions. In such situations, we use \textit{boxplus} ($\boxplus$) and \textit{boxminus} ($\boxminus$) operators, defined below, to perform composition and difference operations in the iterative algorithm presented in Section \ref{sec: Main Algorithm}, while enforcing constraints imposed by the manifold's geometric structure. 


Suppose the full state $x$ evolves on a smooth manifold $\mathcal M$, with dim($\mathcal M$) $=n$. For each $x \in \mathcal M$, let $\pi_x: U_x \rightarrow V_x$ be a diffeomorphic chart from an open neighborhood $U_x \subset \mathcal M$ of $x \in \mathcal M$ to an open neighborhood $V_x \subset \R^n$ of $0 \in \mathbb R^n$. Without loss of generality, suppose $\pi_x(x) = 0$. 
The operators $\boxplus: U_x \times V_x \rightarrow U_x$ and $\boxminus: U_x \times U_x \rightarrow V_x$ are defined by:
\begin{align} \label{Eqn: boxplus def}
    x \boxplus \delta &= \pi_x^{-1}(\delta) \\ \label{Eqn: boxminus def}
    y \boxminus x &= \pi_x(y)
\end{align}
In essence, $\boxplus$ adds a perturbation $\delta \in \R^n$, in local coordinates, to a state $x \in \mathcal M$, while $\boxminus$ extracts the difference $\delta \in \R^n$, in local coordinates, between states $x, x' \in \mathcal M$ covered by the same chart. Below, \say{$\delta$} often describes an error or increment to a nominal state on the manifold.

\subsection{Manifold Examples}
\label{subsec: Manifold Examples}

This subsection gives examples of the $\boxplus$, $\boxminus$ and $\pi$ operators for manifolds that occur widely in SLAM: the set of unit quaternions, $\mathbb H_u$, and the set of rotation matrices, $SO(3)$.

Each $q \in \mathbb H_u$ is expressed as $q = (q_u, \vec q_v)$ where $q_u \in \mathbb R$ and $\vec q_v \in \mathbb R^3$ denote the scalar and vector (imaginary) parts, respectively, with $\Vert q \Vert = \sqrt{q_u^2 + \Vert \vec q_v \Vert_2^2} = 1$ (JPL convention). Here, the coordinate map $\pi: \mathbb H_u \rightarrow \mathbb R^3$ is defined as the Log map on $\mathbb H_u$; its inverse $\pi^{-1}$ is the Exp map. Specifically, we write each $q \in \mathbb H_u$ as $q = \big(\cos(\frac{\theta}{2}), \ \sin(\frac{\theta}{2}) \vec\omega \big)$ for some $\theta \in [0, \pi]$, $\vec\omega \in \mathbb R^3$ with $\Vert\vec\omega\Vert = 1$, i.e., the quaternion $q$ implements a rotation about the axis $\vec\omega$ by $\theta$ radians counterclockwise.  Then, $\pi: \mathbb H_u \rightarrow \mathbb R^3$ and $\pi^{-1}: B_\pi(0) \ra \mathbb H_u$ are defined by: ($B_\pi(0) := \{x \in \R^3: \Vert x \Vert_2 < \pi \}$ denotes the image of $\pi$)
\begin{align} \nonumber
        \pi(q) &= \operatorname{Log}(q) = \theta \vec \omega, \\ \nonumber
    \pi^{-1}(\theta\vec\omega) &= \operatorname{Exp}(\theta\vec\omega) = (\cos(\theta/2), \ \sin(\theta / 2) \vec\omega).
\end{align}
The $\boxplus$ and $\boxminus$ maps are then implemented via the standard quaternion product $\star: \mathbb H_u \times \mathbb H_u \rightarrow \mathbb H_u$:
\begin{align} \nonumber
    q_a\boxplus \vec\omega &= q_a \star \operatorname{Exp}(\vec\omega) \\ \nonumber
    q_a\boxminus q_b &= \operatorname{Log}(q_b^{-1} \star q_a)
\end{align}

For $SO$(3), we define $\boxplus$ and $\boxminus$ similarly, i.e.,
\begin{align} \nonumber
    R_a\boxplus \vec\omega &= R_a \hspace{0.5mm} \operatorname{Exp}(\vec\omega) \\ \nonumber
    R_a\boxminus R_b &= \operatorname{Log}(R_b^T  R_a)
\end{align}

Often, the full state in a SLAM problem exists in the Cartesian product of a finite collection of manifolds, since it contains poses and features on their own manifolds. For a product manifold $\mathcal M_1 \times \mathcal M_2$, with projection, increment, and difference maps already defined on $\mathcal M_1$ and $\mathcal M_2$, we define $\boxplus$ and $\boxminus$ on $\mathcal M_1 \times \mathcal M_2$ by:
\begin{align} \nonumber
    (g_1, g_2) \boxplus (\xi_1, \xi_2) &= (g_1 \boxplus \xi_1, g_2 \boxplus \xi_2) \\ \nonumber
    (g_1, g_2) \boxminus (h_1, h_2) &= (g_1 \boxminus h_1, g_2 \boxminus h_2)
\end{align}

\subsection{SLAM as an Optimization Problem on Manifolds}
\label{subsec: SLAM as an Optimization Problem on Manifolds}

The SLAM problem can be formulated on manifolds using modified cost functions, where plus and minus operations are replaced with $\boxplus$ and $\boxminus$ when necessary. (Appendix \ref{subsubsec: App, SLAM on Manifolds: Dynamics and Measurement Maps}).


\section{Main Algorithm}
\label{sec: Main Algorithm}

\subsection{Algorithm Overview}
\label{subsec: Algorithm Overview}

This section details submodules for a general SLAM algorithm, using state variables and cost terms defined in Sections \ref{sec: Problem Formulation on Euclidean Spaces} and \ref{sec: Problem Formulation on General Manifolds}. We first introduce a formulation on Euclidean spaces. Below, denote the state and concatenated cost vector by $\overline{x_t} \in \R^d$ and $C: \R^d \ra \R^{d_C}$, respectively. (e.g., the SWF in Section \ref{sec: Problem Formulation on Euclidean Spaces} would correspond to $d = \dx n + \df q$ and $d_C = (2n-1) \dx + nq \dz$).
The SLAM problem is then equivalent to solving the nonlinear least-squares problem \eqref{Eqn: Problem Formulation, Inner Product Cost}, reproduced below:
\begin{align*}
    \min_{\overline{x_t}}. c(\overline{x_t}) = \min_{\overline{x_t}}. 
    \Vert C(\overline{x_t}) \Vert_2^2.
\end{align*}





\subsection{Gauss-Newton Descent}
\label{subsec: Gauss-Newton Descent}

Gauss-Newton descent minimizes $c(\overline{x_t})$ via Gauss-Newton steps (Alg. \ref{Alg: gauss-newton}), by iteratively approximating $c(\overline{x_t})$ about a given linearization point $\overline{x_t}^\star$ as a linear least-squares cost term, i.e.,
\begin{align} 
\label{Eqn: Algorithm, Linear Least-Squares}
    \min_{\overline{x_t}}. c(\overline{x_t}) 
    &= \min_{\overline{x_t}}. \Vert \overline{x_t}-\mu_t \Vert_{\Sigma_t^{-1}}^2 + o(\overline{x_t} - \overline{x_t}^\star)
\end{align}
for some $\mu_t \in \R^d$ and $\Sigma_t \in \R^{d \times d}$. The theorem below describes the linearization procedure required to obtain $\mu_t \in \R^d$ and $\Sigma_t \in \R^{d \times d}$, and the approximation involved.

\begin{theorem}(\textbf{Gauss-Newton Step}) \label{Thm: Gauss-Newton Eqns} Let $\overline{x_t}^\star \in \R^d$ be a given linearization point, and suppose $J := \frac{\partial C}{\partial \overline{x_t}} \in \R^{d_C \times d}$ has full column rank. Applying a Gauss-Newton step (Alg. \ref{Alg: gauss-newton}) to the cost $c(\overline{x_t})$, about $\overline{x_t}^\star \in \R^d$ yields the new cost:
\begin{align*}
    c(\overline{x_t}) = \Vert \overline{x_t}-\mu_t \Vert_{\Sigma_t^{-1}}^2 + o(\overline{x_t} - \overline{x_t}^\star),
\end{align*}
where $\mu_t \in \R^{d}$ and $\Sigma_t \in \R^{d \times d}$ are given by:
\begin{align*}
    \Sigma_t &\gets (J^\top J)^{-1}, \\
    \mu_t &\gets \overline{x_t}^{\star}  - (J^\top J)^{-1} J^\top  C(\overline{x_t}^{\star}).
\end{align*} 
\end{theorem}

\begin{proof}
See Appendix (Section \ref{subsubsec: App, Gauss-Newton Steps}). 
\end{proof}

\subsection{Marginalization of States} 
\label{subsec: Marginalization of States}

The marginalization step (Alg. \ref{Alg: marginalization}) reduces the size of the SLAM problem by removing states that are no longer relevant, thus improving computational efficiency. First, we partition the overall state $\overline{x_t} \in \R^{d} = \R^{d_M + d_K}$ into a \textit{marginalized component} $\overline{x_{t,M}} \in \R^{d_M}$, to be discarded from $\overline{x_t}$, and a \textit{non-marginalized component} $\overline{x_{t,K}} \in \R^{d_K}$, to be kept. Then, we partition $c(\overline{x_t})$ into two cost terms: $c_1(\overline{x_{t,K}})$, which depends only on non-marginalized state components, and $c_2(\overline{x_{t,K}}, \overline{x_{t,M}}) $ which depends on both marginalized and non-marginalized state components:
\begin{align*}
    c(\overline{x_t}) &= c(\overline{x_K}, \overline{x_M}) = c_1(\overline{x_K}) + c_2(\overline{x_K}, \overline{x_M}) \\
    &= \Vert C_1(\overline{x_K}) \Vert_2^2 + \Vert C_2(\overline{x_K}, \overline{x_M}) \Vert_2^2.
\end{align*}
Here, $C_1(\overline{x_K}) \in \R^{d_{C,1}}$ and $C_2(\overline{x_K}, \overline{x_M}) \in \R^{d_{C,2}}$ denote the concatenation of residuals associated with
$c_1(\overline{x_K})$ and $c_2(\overline{x_K}, \overline{x_M})$ (with $d_C = d_{C,1} + d_{C,2}$).
To remove $\overline{x_{t,M}} \in \R^{d_M}$ from the optimization problem, observe that:
\begin{align*}
    \min_{\overline{x_t}} c(\overline{x_t}) &= \min_{\overline{x_{t,K}}, \overline{x_{t,M}}} \Big( c_1(\overline{x_{t,K}}) + c_2(\overline{x_{t,K}}, \overline{x_{t,M}}) \Big) \\
    &= \min_{\overline{x_{t,K}}} \Big( \Vert C_1(\overline{x_{t,K}}) \Vert_2^2 + \min_{\overline{x_{t,M}}} \Vert C_2(\overline{x_{t,K}}, \overline{x_{t,M}}) \Vert_2^2 \Big).
\end{align*}
To remove $\overline{x_{t,M}}$, we approximate the solution to the inner minimization problem by a linear least-squares cost, i.e.:
\begin{align*}
    \min_{\overline{x_{t,M}}} \Vert C_2(\overline{x_{t,K}}, \overline{x_{t,M}}) \Vert_2^2 \approx \Vert \overline{x_{t,K}} - \overline{\mu}_{t,K} \Vert_{\overline{\Sigma}_{t,K}^{-1}}^2
\end{align*}
for some $\overline{\mu}_{t,K} \in \R^{d_K}$ and $\overline{\Sigma}_{t,K} \in \R^{d_K \times d_K}$. Since $\Vert C_2(\overline{x_{t,K}}, \overline{x_{t,M}}) \Vert_2^2$ is in general non-convex, we obtain $\overline{\mu}_{t,K}$ and $\overline{\Sigma}_{t,K}$ by minimizing the first-order Taylor expansion of $\Vert C_2(\overline{x_{t,K}}, \overline{x_{t,M}}) \Vert_2^2$ about some linearization point. Below, Theorem \ref{Thm: Marginalization Eqns} details the derivation of $\overline{\mu}_{t,K}$ and $\overline{\Sigma}_{t,K}$. (For the proof, see Appendix \ref{subsubsec: App, Marginalization of States}).

\begin{theorem}[\textbf{\emph{Marginalization Step}}] \label{Thm: Marginalization Eqns}
Let $\overline{x_t}^\star \in \R^d$ be a given linearization point, and suppose $J := \frac{\partial C}{\partial \overline{x_t}} \in \R^{d_C \times d}$ has full column rank. Define $J_K := \frac{\partial C}{\partial \overline{x_{t,K}}} \in \R^{d_C \times d_K}$, $J_M := \frac{\partial C}{\partial \overline{x_{t,M}}} \in \R^{d_C \times d_M}$. If $C(\overline{x_{t,M}}, \overline{x_{t,K}})$ were a linear function of $\overline{x_t} = (\overline{x_{t,M}}, \overline{x_{t, K}})$, then applying a Marginalization step (Alg. \ref{Alg: marginalization}) to the cost $c(\overline{x_t})$, about the linearization point $\overline{x_t}^\star = (\overline{x_{t,K}^\star}, \overline{x_{t,M}^\star}) \in \R^d$, yields:
\begin{align} \label{Eqn: Marginalized Linear}
    \min_{\overline{x_t}} c(\overline{x_{t,K}}, \overline{x_{t,M}}) 
    &= \min_{\overline{x_{t, K}}}. \Big(c_1(\overline{x_{t, K}}) + \Vert \overline{x_{t,K}} - \overline \mu_{t,K} \Vert_{\overline \Sigma_{t,k}^{-1}}^2 \Big), 
\end{align}
where $\Sigma_{t,K} \in \R^{d_K \times d_K}$ and $\mu_{t,K} \in \R^{d_K}$ are given by:
\begin{align} \label{Eqn: Main Alg, Marginalization, Sigma K}
    \Sigma_{t,K} &:= \big( J_K^\top \big[ I - J_M(J_M^\top J_M)^{-1} J_M^\top \big] J_K \big)^{-1} 
    , \\ \label{Eqn: Main Alg, Marginalization, mu K}
    \mu_{t,K} &:= \overline{x_{t,K}^\star} - \Sigma_{t,K} J_K^\top \big[ I - J_M(J_M^\top J_M)^{-1} J_M^\top \big] C_2(\overline{x_t^\star}).
\end{align}
\end{theorem}


\subsection{Main Algorithm on Manifolds}

The Euclidean-space framework above can be directly extended to a formulation on manifolds, by using concepts in Section \ref{sec: Problem Formulation on General Manifolds} to modify the dynamics and measurement maps in Section \ref{sec: Problem Formulation on Euclidean Spaces}, as well as the cost functions, Gauss-Newton steps, and marginalization steps in Sections \ref{subsec: Gauss-Newton Descent}, \ref{subsec: Marginalization of States}. When appropriate, plus and minus operations must be replaced with $\boxplus$ and $\boxminus$ (Appendix \ref{subsubsec: App, SLAM on Manifolds: the Optimization Problem}).

\section{Equivalence of Filtering and Optimization Approaches}
\label{sec: Equivalence of Filtering and Optimization-Based Approaches}

Here, we demonstrate the equivalence of filtering and batch optimization-based SLAM algorithms, using the Extended Kalman Filter (EKF, in Section \ref{subsec: EKF}) and Multi-State Constrained Kalman Filter (MSCKF, in Section \ref{subsec: MSCKF}), as examples. 
Although similar results exist in the optimization literature \cite{Bell1994TheIK}, they do not analyze algorithmic submodules unique to SLAM, e.g., feature incorporation, processing, and discarding. For an introduction to the classical formulations of EKF and MSCKF SLAM, please see Appendices \ref{subsubsec: App, EKF, Setup} and \ref{subsubsec: App, MSCKF, Setup}. 

\subsection{Extended Kalman Filter (EKF), on Euclidean Spaces}
\label{subsec: EKF}



At each time $t$, the EKF SLAM algorithm on Euclidean spaces maintains the full state vector $\tilde{x}_t := (x_t, f_1, \cdots, f_{\nf}) \in \R^{\dx + p \df}$, consisting of the most recent state $x_t \in \R^{\dx}$ and feature position estimates $f_1, \cdots, f_{\nf} \in \R^{\df}$.
At initialization ($t = 0$), no feature has been detected ($p = 0$), and the EKF full state is simply the initial state $\tilde{x}_0 = x_0 \in \R^{\dx}$, with mean $\mu_0 \in \R^{\dx}$ and covariance $\Sigma_0 \in \R^{\dx \times \dx}$. 
Suppose, at the current time $t$, the running cost $c_{EKF,t,0}: \R^{\dx + p\df} \ra \R^{\dx + p\df}$ is:
\begin{align*}
    c_{EKF,t,0} = \Vert \tilde{x}_t - \mu_t \Vert_{\Sigma_t^{-1}}^2,
\end{align*}
where $\tilde{x}_t := (x_t, f_1, \cdots, f_{\nf}) \in \R^{\dx + p \df}$ denotes the EKF full state at time $t$, with mean $\mu_t \in \R^{\dx + p\df}$ and covariance $\Sigma_t \in \R^{(\dx + p\df) \times (\dx + p\df)}$. 
    First, the \textit{feature augmentation step} appends position estimates of new features $f_{\nf+1}, \cdots, f_{\nf+\nf'} \in \R^{\df}$ to the EKF full state $\tilde{x}_t$, and updates its mean and covariance. In particular, feature measurements $z_{t,\nf+1}, \cdots, z_{t,\nf+\nf'} \in \R^{\dz}$ are assimilated by adding measurement residual terms, creating a new cost $c_{EKF,t,1}: \R^{\dx + (\nf+\nf')\df} \ra \R$:
    \begin{align*}
        &c_{EKF,t,1}(\tilde{x}_t, f_{\nf+1}, \cdots, f_{\nf+p'}) \\
        := \hspace{0.5mm} &\Vert \tilde{x}_t-\mu_t \Vert_{\Sigma_t^{-1}}^2 + \sum_{k= \nf+1}^{p+p'} \Vert z_{t,k} -h(x_t, f_k) \Vert_{\Sigma_v^{-1}}^2.
    \end{align*}
    In effect, $c_{EKF,t,1}$ appends positions of new features to $\tilde{x}_t$, and constrains it using feature measurements residuals. We then replace $\nf$ with $\nf + \nf'$.
    
    
    Next, the \textit{feature update} step uses measurements of features originally contained in $\tilde{x}_t$ to update the mean and covariance of $\tilde{x}_t$. More precisely, feature measurements $z_{t,1}, \cdots, z_{t,p} \in \R^{\dz}$, of the $p$ features $f_1, \cdots, f_{\nf}$ included in $\tilde{x}_t$, are introduced by incorporating the corresponding measurement residuals 
    to create a new cost $c_{EKF,t,2}: \R^{\dx + p \df} \ra \R$:
    \begin{align*}
        c_{EKF,t,2}(\tilde{x}_t) 
        &:= \Vert \tilde{x}_t-\mu_t \Vert_{\Sigma_t^{-1}}^2 + \sum_{k=1}^p \Vert z_{t,k} -h(x_t, f_k) \Vert_{\Sigma_v^{-1}}^2.
    \end{align*}
    A Gauss-Newton step then constructs an updated mean $\overline{\mu_t} \in \R^{\dx + p \df}$ and covariance $\overline{\Sigma}_t \in \R^{(\dx + p \df) \times (\dx + p \df)}$ for $\tilde{x}_t$, creating a new cost $c_{EKF,t,3}: \R^{\dx + p \df} \ra \R$:
    \begin{align*}
        c_{EKF,t,3}(\tilde{x}_t) 
        &:= \Vert \tilde{x}_t- \overline{\mu_t} \Vert_{\overline{\Sigma}_t^{-1}}^2,
    \end{align*}
    which returns the running cost to the form of $c_{EKF,t,0}$.
    
    
    The \textit{state propagation} step propagates the EKF full state forward by one time step, via the EKF state propagation map $g: \R^{\dx + p \df} \ra \R^{\dx + p \df}$.
    To propagate $\tilde{x}_t$ forward in time, we add the dynamics residual, creating a new cost $c_{EKF,t,4}: \R^{2\dx
    + p\df} \ra \R$:
    \begin{align*}
        c_{EKF,t,4}(\tilde{x_t}, x_{t+1}) := \Vert \tilde{x}_t - \overline{\mu_t} \Vert_{\overline{\Sigma}_t^{-1}}^2 + \Vert x_{t+1} - g(x_t) \Vert_{\Sigma_w^{-1}}^2.
    \end{align*}
    In effect, $c_{EKF,t,4}$ appends the new state $x_{t+1} \in \R^{\dx}$ to $\tilde{x}_t$, while adding a new constraint posed by the dynamics residuals. A marginalization step, with $\tilde{x}_{t,K} := (x_{t+1}, f_1, \cdots, f_{\nf}) \in \R^{\dx + p\df}$ and $\tilde{x}_{t,M} := x_t \in \R^{\dx}$, then removes the previous state $x_t \in \R^{\dx}$ from the running cost. This step produces a mean $\mu_{t+1} \in \R^{\dx + p\df}$ and a covariance $\Sigma_{t+1} \in \R^{(\dx + p\df) \times (\dx + p\df)}$ for the new EKF full state, $\tilde{x}_{t+1} := \tilde{x}_{t,K}$. The running cost 
    is updated to $c_{EKF,t+1,0}: \R^{\dx + p\df} \ra \R$:
    \begin{align*}
        c_{EKF,t+1,0}(\tilde{x}_{t+1}) 
        &:= \Vert \tilde{x}_{t+1} - \mu_{t+1} \Vert_{\Sigma_{t+1}^{-1}}^2,
    \end{align*}
    which returns the running cost to  the form of $c_{EKF,t,0}$.


The theorems below establish that the 
feature augmentation, 
feature update, and state propagation steps of the EKF, presented above in our optimization framework, correspond precisely to those presented in the standard EKF SLAM algorithm (Alg. \ref{Alg: EKF}) \cite{Thrun2005ProbabilisticRobotics, Sola2014SLAMWithEKF}. (For proofs, see Appendix \ref{subsubsec: App, EKF}).

\begin{theorem} \label{Thm: EKF, Feature Augmentation}
The feature augmentation step of standard EKF SLAM (Alg. \ref{Alg: EKF, Feature Augmentation}) is equivalent to applying a Gauss-Newton step to $c_{EKF,t,1}: \R^{\dx + (p + p') \df} \ra \R$, with:
\begin{align*}
    &c_{EKF,t,1}(\tilde{x}_t, f_{\nf+1}, \cdots, f_{\nf+\nf'}) \\
    = \hspace{0.5mm} &\Vert \tilde{x}_t - \mu_t \Vert_{\Sigma_t^{-1}}^2 + \sum_{k= \nf+1}^{p+p'} \Vert  z_{t,k} - h(x_t, f_k) \Vert_{\tilde{\Sigma}_v^{-1}}^2.
\end{align*}
\end{theorem}


\begin{theorem} \label{Thm: EKF Update}
    The feature update step of standard EKF SLAM (Alg. \ref{Alg: EKF, Feature Update}) is equivalent to applying a Gauss-Newton step on $c_{EKF,t,2}: \R^{\dx + p \df} \ra \R$, with:
    \begin{align*}
        c_{EKF,t,2}(\tilde{x}_t)
        := \hspace{0.5mm} &\Vert \tilde{x}_t-\mu_t \Vert_{\Sigma_t^{-1}}^2 + \sum_{k=1}^p \Vert z_{t,k}-h(x_t, f_k) \Vert_{\Sigma_v^{-1}}^2.
    \end{align*}
\end{theorem}


\begin{theorem} \label{Thm: EKF Propagation}
    The state propagation step of standard EKF SLAM (Alg. \ref{Alg: EKF, State Propagation}) is equivalent to applying a Marginalization step to $c_{EKF,t,4}: \R^{2 \dx + p \df} \ra \R$, with:
    \begin{align*}
        c_{EKF,t,4}(\tilde{x}_t, x_{t+1}) := \hspace{0.5mm} &\Vert \tilde{x}_t - \overline{\mu}_t \Vert_{\overline \Sigma_t^{-1}}^2 + \Vert x_{t+1}-g(x_t) \Vert_{\Sigma_w^{-1}}^2.
    \end{align*}
    where $\tilde{x}_{t,K} := (x_{t+1}, f_1, \cdots, f_{\nf}) \in \R^{\dx + p \df}$ and $\tilde{x}_{t,M} = x_t \in \R^{\dx}$.
\end{theorem}




\begin{algorithm}

{
\small
\SetAlgoLined


\KwData{Prior $\mathcal N(\mu_0, \Sigma_0)$ on $x_0 \in \R^{\dx}$, noise covariances $\Sigma_w$, $\Sigma_v$, dynamics map $g$, measurement map $h$, inverse measurement map $\ell$, time horizon $T$.}
\KwResult{Estimates $\hat x_t \in \R^{\dx}, \hspace{0.5mm} \forall \hspace{0.5mm} t \in \{1, \cdots, T\}$.}

\vspace{2mm}
 $\cost_0(x) \gets \|x_0 - \mu_0\|^2_{\Sigma_0^{-1}} $
 
 
 $\nf \gets 0$.
 
 \vspace{2mm}
 \For{$t = 0, 1, \cdots T$}{
  

    $(z_{t,\nf+1}, \cdots, z_{t,\nf+\nf'})
  \gets$ Measurements of new features.  

    $\text{cost}_t \gets \text{cost}_t + \sum_{k= \nf+1}^{\nf+\nf'} \|z_{t,k} - h(x_t, f_k)\|_{\Sigma_{v}^{-1}}^2 $
        
    $\mu_t \gets \big(\mu_t, \ell(x_t, z_{t,\nf+1}), \cdots, \ell(x_t, z_{t,\nf+\nf'}) \big) \in \R^{\dx + (\nf+\nf') \df}$.
        
    $\mu_t 
    , \Sigma_t 
    , \text{cost}_t \gets$ 1 Gauss-Newton step on $\text{cost}_t$, about $\mu_t$ (Alg. \ref{Alg: gauss-newton}).
        

    $(z_{t,1}, \cdots, z_{t,\nf})
  \gets$ Measurements of existing features.
  
  $\text{cost}_t \gets \text{cost}_t + \sum_{k=1}^{\nf} \|z_{t,k} - h(x_t, f_k)\|_{\Sigma_{v}^{-1}}^2 $
  
  $\mu_t
  , \Sigma_t 
  , \text{cost}_t \gets$ 1 Gauss-Newton step on $\text{cost}_t$, about $\mu_t$,
  (Alg. \ref{Alg: gauss-newton}).
  
  $\hat x_t \gets \bar\mu_t \in \R^{\dx+ \nf \df}$.
    
  $\nf \gets \nf + \nf'$
  
  \If{$t < T$}{

  $\text{cost}_t \gets \text{cost}_t + \|x_{t+1} - g(x_t)\|_{\Sigma_w^{-1}}^2 $
  
  $\mu_{t+1} 
  , \Sigma_{t+1} 
  , \text{cost}_t \gets$ 1 Marginalization step on $\text{cost}_{t+1}$ with $x_M = x_t$,
  about $(\overline{\mu_t}, g(\overline{\mu_t}))$
  (Alg. \ref{Alg: marginalization}).
  
  $\text{cost}_{t+1} \gets \|x_{t+1} - \mu_{t+1}\|^2_{\Sigma_{t+1}^{-1}} $
  
  }
 }
 
 \Return{$\hat x_0, \cdots, \hat x_T$}
 \caption{EKF SLAM on Euclidean spaces, as an iterative optimization problem.}
 }
\end{algorithm}



\begin{remark}
In practice, Gauss-Newton steps for feature augmentation can be delayed and done with feature updates.
\end{remark}

\subsection{Multi-State Constrained Kalman Filter (MSCKF), on Manifolds}
\label{subsec: MSCKF}

The MSCKF algorithm maintains a full state, $\tilde{x}_t \in \X_\IMU \times (\X_p)^n$, containing the most recent IMU state,  $x_\IMU \in \X_\IMU$ and $n$ recent poses,  $(x_1, \cdots, x_n) \in (\X_p)^n$:
\begin{align*}
    \tilde{x}_t &:= (x_{t,\IMU}, x_1, \cdots, x_n) \in \X_\IMU \times (\X_p)^n,
\end{align*}
with mean $\mu_t \in \X_\IMU \times (\X_p)^n$ and covariance $\Sigma_t \in \R^{(d_\IMU + n\dx) \times (d_\IMU + n\dx)}$. 
As new poses are introduced, old poses are discarded, and features are processed and discarded to update $\tilde{x}_t$, the mean $\mu_t$, covariance $\Sigma_t$, and $n \in \N$ accordingly. 

At initialization ($t=0$), no pose has yet been recorded ($n=0$), and the full state $\tilde{x}_0$ is the initial IMU state $\tilde{x}_{0,\IMU} \in \X_\IMU$, with mean $\mu_0 \in \X_\IMU$ and covariance $\Sigma_0 \in \R^{d_\IMU \times d_\IMU}$. Thus, $\tilde{x}_0 = \mu_0$ optimizes the initial running cost $c_{MSCKF,0}: \X_\IMU \ra \R$ in our algorithm:
\begin{align*}
    c_{MSCKF,0,0}(\tilde{x}_0) = \Vert \tilde{x}_0 \boxminus \mu_0 \Vert_{\Sigma_0^{-1}}^2.
\end{align*}
Suppose that, at the current time $t$, the running cost $c_{MSCKF,t,0}: \X_\IMU \times (\X_p)^n \ra \X_\IMU \times (\X_p)^n$ is:
\begin{align*}
    c_{MSCKF,t,0}(\tilde{x}_t) = \Vert \tilde{x}_t \boxminus \mu_t \Vert_{\Sigma_t^{-1}}^2,
\end{align*}
where $\mu_t \in \X_\IMU \times (\X_p)^n$ and $\Sigma_t \in \R^{(d_\IMU + n \dx) \times (d_\IMU + n \dx)}$ denote the mean and covariance of the full state $\tilde{x}_t := (x_{t,\IMU}, x_1, \cdots, x_n) \in \X_\IMU \times (\X_p)^n$ at time $t$, consisting of the current IMU state and $n$ poses. 
    When a new image is received, the \textit{pose augmentation step} adds a new pose $x_{n+1} \in \mathcal{X}_p$ (global frame) to $\tilde{x}_t$, derived from $x_{n+1}^\IMU \in \mathcal{X}_\IMU$, the IMU position estimate in the global frame, via the map $\psi: \X_\IMU \times (\X_p)^n \times \X_\IMU \ra \X_p$, i.e.,
    \begin{align*}
        x_{n+1} &:= \psi(\tilde{x}_t,  x_{n+1}^{\IMU}) \in \X_p. 
    \end{align*}
    
    
    The \textit{feature update step} uses features measurements to update the mean and covariance of $\tilde{x}_t$. In MSCKF, (A) if a feature becomes unobserved in the current pose, it is discarded, and (B) when $n \geq N_{\max}$, a specified upper bound,  features common to $\lfloor N_{\max} \rfloor/3$ of the $n$ poses, evenly spaced in time, are processed via a feature update step, then dropped alongside the corresponding poses. Let $S_{z,1}$ and $S_{z,2}$ denote sets of pose-feature pairs $(x_i, f_j)$ from cases (A) and (B) above, respectively, and let $S_f$ denote the set of features to be processed and (Alg. \ref{Alg: MSCKF}). These constraints are then incorporated, creating a new cost $c_{MSCKF,t,2}: \X_\IMU \times (\X_p)^n \ra \R$:
    \begin{align*}
        &c_{MSCKF,t,2}(\tilde{x}_t) \\
        := \hspace{0.5mm} &\Vert \tilde{x}_t \boxminus \mu_t \Vert_{\Sigma_t^{-1}}^2  + \sum_{(x_i, f_j) \in S_{z,1} \cup S_{z,2}} \Vert z_{i,j} \boxminus h(x_i, f_j) \Vert_{\Sigma_v^{-1}}^2,
    \end{align*}
    where $z_{i,j} \in \R^{\dz}$ denotes the feature measurement of feature $j$ observed from pose $x_i \in \X_p$. By using Gauss-Newton linearization, we leverage constraints posed by the measurement residuals to construct an updated mean for $\tilde{x}_t$, denoted $\overline{\mu_t} \in \X_\IMU \times (\X_p)^n$, and an updated covariance for $\tilde{x}_t$, denoted $\overline{\Sigma}_t \in \R^{(d_\IMU + n \dx) \times (d_\IMU + n \dx)}$. As a result, our cost will be updated to $c_{MSCKF,t,3}: \X_\IMU \times (\X_p)^n \ra \R$:
    \begin{align*}
        c_{MSCKF,t,3}(\tilde{x}_t) 
        &:= \Vert \tilde{x}_t \boxminus \overline{\mu_t} \Vert_{\overline{\Sigma}_t^{-1}}^2,
    \end{align*}
    which assumes the form of $c_{MSCKF,t,0}$.
    
    
    The \textit{state propagation} step propagates the full state by assimilating dynamics residuals, creating a new cost $c_{MSCKF,t,4}: (\X_{\IMU})^2 \times (\X_p)^n \ra \R$:
    \begin{align*}
        &c_{MSCKF,t,4}(\tilde{x}_t, x_{t+1,\IMU}) \\
        := \hspace{0.5mm} &\Vert \tilde{x}_t \boxminus \overline{\mu_t} \Vert_{\overline{\Sigma}_t^{-1}}^2 + \Vert x_{t+1,\IMU} \boxminus g_\IMU(x_{t,\IMU}) \Vert_{\Sigma_t^{-1}}^2.
    \end{align*}
    In effect, $c_{MSCKF,t,4}$ appends the new IMU variable $x_{t+1,\IMU} \in \X_\IMU$ to the current full state $\tilde{x}_t \in \X_\IMU \times (\X_p)^n$, and constrains this new full state via the dynamics residuals. A marginalization step, with $\tilde{x}_{t,K} := (x_{t+1,\IMU}, x_1, \cdots, x_n) \in \X_\IMU \times (\X_p)^n$ and $\tilde{x}_{t,M} := x_{t,\IMU} \in \X_\IMU$, then removes the previous IMU state, $x_{t,\IMU}$, from the running cost. This produces a mean $\mu_{t+1} \in \X_\IMU \times (\X_p)^n$ and a covariance $\Sigma_{t+1} \in \R^{(d_\IMU + n \dx) \times (d_\IMU + n \dx)}$ for the new MSCKF full state, $\tilde{x}_{t+1} := \tilde{x}_{t,K} = (x_{t+1,\IMU}, x_1, \cdots, x_n) \in \X_\IMU \times (\X_p)^n$. The running cost 
    is updated to $c_{MSCKF,t+1,0}: \X_\IMU \times (\X_p)^n \ra \R$:
    \begin{align*}
        c_{MSCKF,t+1,0}(\tilde{x}_{t+1}) 
        &:= \Vert \tilde{x}_{t+1} \boxminus \mu_{t+1} \Vert_{\Sigma_{t+1}^{-1}}^2,
    \end{align*}
    which returns the running cost to the form of $c_{MSCKF,t,0}$.


The theorems below establish that the 
feature augmentation, 
feature update, and state propagation steps of the MSCKF, presented above in our optimization framework, correspond precisely to those presented in the standard MSCKF (Alg. \ref{Alg: MSCKF}) \cite{Mourikis2007MultiStateConstraintKalmanFilter}. (For proofs, see Appendix \ref{subsubsec: App, MSCKF}).




\begin{algorithm} \label{Alg: MSCKF, Opt}
    {
    \small
    \SetAlgoLined
    \KwData{Prior $\mathcal N(\mu_0, \Sigma_0)$ on $x_{\IMU,0} \in \X_\IMU$, noise covariances $\Sigma_w$, $\Sigma_v$, dynamics $g_\IMU
    $, measurement map $h
    $, time horizon $T$, Pose transform $\psi
    $ (IMU $\ra$ global) , $\epsilon > 0$.}
    
    \KwResult{Estimates $\hat x_t
    $ for all desired timesteps $t \in \{1, \cdots, T\}$
    .} 
    
    
    $\cost_t \gets \Vert x_0 \boxminus \mu_0 \Vert_{\Sigma_0}^2 $. (Initialize objective function).
    
    
    $S_z, S_x, S_{z,1}, S_{z,2} \gets \phi$ 
    
    $(\nx, \nf) \gets (0, 0)$ 
    
     \For{$t = 0, \cdots, T$}{
        
        \While{\emph{new pose $x_{\nx + 1} \in \X_p$ recorded, new IMU measurement not received}}
        {
            $\cost_t \gets \cost_t + \epsilon^{-1} \Vert x_{\nx + 1} \boxminus \psi(\tilde{x}_t, x_{\nx + 1}^\IMU) \Vert_2^2 $.
            
            $\mu_t  
            , \Sigma_t 
            , \cost_t \gets $ 1 Gauss-Newton 
            $\cost_t$ (Alg. \ref{Alg: gauss-newton}), about $(\mu_t, \psi(\mu_t, x_{\nx + 1}^\IMU))$ with $\epsilon \ra 0$.
            
            
            $\{z_{\nx + 1,k}
            \} \gets$ Feature measurements at $x_{\nx + 1}$
            
            
            
            $S_z \gets S_z \cup \big\{ (x_{\nx + 1}, f_{k})| f_k \text{ observed at $\nx + 1$} \big\}$ 
            
            $\nx \gets \nx + 1$
        
        \If{$\nx \geq N_{\max}-1$}{
        
            $S_x \gets \{x_i | i \text{ mod } 3 = 2, \text{ and } 1 \leq i \leq \nx. \}$
            
            $S_{z,1} \gets \big\{ (x_i, f_{k}) \in S_z \big| x_i \in S_x, \text{feature $k$ observed at each pose in } S_x \big\}$
        }
        
        $S_{z,2} \gets \big\{ (x_i, f_{k}) \in S_z | f_k \text{ not observed at } x_n
        \big\}$.
        
        
        
            $\cost_t \gets \cost_t + \sum_{(x_i, f_k) \in S_{z,1} \cup S_{z,2}} \Vert z_{i,k} \boxminus h(x_i, f_{t,k}) \Vert_{\Sigma_v^{-1}} $
            
            $\overline{\mu_t}
            $, $\overline{\Sigma_t}
            $, $\cost_t \gets$ 1 Gauss-Newton step 
            on $\cost_t$, about $\mu_t$ 
            (Alg. \ref{Alg: gauss-newton})
            
            $\hat x_t \gets \overline{\mu_t} \in \X_\IMU \times (\X_p)^\nx$.

        
        $S_z \gets S_z \backslash (S_{z,1} \cup \{(x_i, f_k) | x_i \in S_x \})$
        
        Reindex poses and features in ascending order.
        
        $(\nx, \nf) \gets (\nx-|S_x|, \nf-|S_f|)$
        
        
        
        }
        
        \If{$t < T$}{
        
            $\cost_t \gets \cost_t + \Vert x_{t+1, \IMU} \boxminus g_\IMU(x_{t,\IMU}) \Vert_{\Sigma_w^{-1}}^2 $.
        
            $\mu_{t+1} 
            , \Sigma_{t+1} 
            , \cost_t \gets$ 1 Marginalization step on $\cost_t$, about $(\overline{\mu_t}, g(\mu_{t,\IMU}))$ 
            (Alg. \ref{Alg: marginalization})
            
        
        }
     }
    
    \Return{$\hat x_0, \cdots \hat x_T \in \X_\IMU \times (\X_p)^\nx$}
     
     \caption{Multi-State Constrained Kalman Filter (MSCKF) on manifolds, as iterative optimization.}
     }
\end{algorithm}

\begin{theorem} \label{Thm: MSCKF, Pose Augmentation}
The pose augmentation step of the standard MSCKF (Alg. \ref{Alg: MSCKF, Pose Augmentation Sub-block}) is equivalent to applying a Gauss-Newton step to $c_{MSCKF,t,1}: \X_\IMU \times (\X_p)^n \times \X_\IMU \ra \R$, with:
\begin{align*}
    &c_{MSCKF,t,1}(\tilde{x}_t, x_{n+1}) \\
    = \hspace{0.5mm} &\Vert \tilde{x}_t \boxminus \mu_t \Vert_{\Sigma_t^{-1}}^2 + \epsilon^{-1} \Vert x_{n+1} \boxminus \psi(\tilde{x}_t, x_{n+1}^\IMU) \Vert_2^2,
\end{align*}
and taking $\epsilon \ra 0$ in the resulting (augmented) mean 
$\mu_t$ 
and covariance 
$\Sigma_t$.
\end{theorem}


\begin{theorem} \label{Thm: MSCKF, Feature Update}
The feature update step of the standard MSCKF (Alg. \ref{Alg: MSCKF, Feature Update Sub-block}) is equivalent to applying a Marginalization step to $c_{MSCKF,t,2}: \X_\IMU \times (\X_p)^n \times \R^{|S_f|\df} \ra \R$, with:
\begin{align} \nonumber
    &c_{MSCKF,t,2}(\tilde{x}_t, f_{S_f}) \\ \nonumber
    := &\Vert \tilde{x}_t  \boxminus \mu_t \Vert_{\Sigma_t^{-1}}^2 + \sum_{(x_i,f_j) \in S_{z,1} \cup S_{z,2}} \Vert z_{i,j} \boxminus h(x_i,f_j) \Vert_{\Sigma_v^{-1}}^2,
\end{align}
where $f_{S_f} \in \R^{|S_f|\df}$ denotes the stacked vector of all feature positions in $S_f$ (see Alg. \ref{Alg: MSCKF}).
\end{theorem}





\begin{theorem} \label{Thm: MSCKF State Propagation}
    The state propagation step of the standard MSCKF (Alg. \ref{Alg: MSCKF, State Propagation sub-block}) is equivalent to applying a Marginalization step to $c_{MSCKF,t,4}: (\X_{\IMU})^2 \times (\X_p)^n \ra \R$, with:
    \begin{align*}
        &c_{MSCKF,t,4}(\tilde{x}_t, x_{t+1,\IMU}) \\
        := \hspace{0.5mm} &\Vert \tilde{x}_t \boxminus \overline{\mu_t} \Vert_{\overline{\Sigma}_t^{-1}}^2 + \Vert x_{t+1,\IMU} \boxminus g_\IMU(x_{t,\IMU}) \Vert_{\Sigma_t^{-1}}^2.
    \end{align*}
    with $\tilde{x}_{t,K} := (x_{t+1, \IMU}, x_1, \cdots, x_n) \in \X_\IMU \times (\X_p)^n$ and $\tilde{x}_{t,M} = x_{t,\IMU} \in \X_\IMU$.
\end{theorem}



\subsection{State-of-the-Art SLAM Algorithms}
\label{subsec: State-of-the-Art SLAM Algorithms}

Our framework balances the need for computational efficiency, estimation accuracy, and map precision, tradeoffs observed in design choices for existing SLAM algorithms.

\begin{itemize}
    \item \textbf{Extended Kalman Filter (EKF)} \cite{Thrun2005ProbabilisticRobotics, Sola2014SLAMWithEKF, Zhang1997ParameterEstimationTechniques,
    }
    --The EKF iteratively updates position estimates of the current pose and all observed features; all past poses are marginalized. 
    This design favors computational speed
    over localization and mapping accuracy. A variant, the \textit{iterated Extended Kalman Filter} (iEKF), takes multiple Gauss-Newton steps before marginalization to tune the linearization point. This improves mapping and localization accuracy but increases computation time slightly.
    
    
    \item \textbf{Multi-State Constrained Kalman Filter} \cite{Mourikis2007MultiStateConstraintKalmanFilter, LiMourikis2012ImprovingTA, LiMourikis2012OptimizationBased}---The MSCKF iteratively updates a full state, with the current IMU state and $n$ past poses, while processing features observed at these poses; here $n \leq N_{\max}$, a specified upper bound that trades off accuracy and computational speed.

    \item \textbf{Sliding Window Smoother, Fixed-Lag Smoother} \cite{Maybeck1979StochasticsModels, Sibley2015SlidingWindowFilterPlanetaryLanding, DellaertKaess2006SquareRootSAM}---The fixed-lag smoother resembles the MSCKF, but performs multiple steps of Gauss-Newton descent before the marginalization step, to adjust the linearization point. 
    This improves localization and mapping accuracy, but increases the computation time.
    
    \item \textbf{Open Keyframe Visual-Inertial SLAM (OKVIS)} 
    \cite{Leutenegger2015KeyframebasedVO}---OKVIS updates a sliding window of \say{keyframes}, poses deemed most informative, 
    which may be arbitrarily spaced in time. 
    Keyframe poses leaving the sliding window, and associated landmarks, are marginalized. This design aims to improve estimation accuracy by maximizing information encoded by the stored poses, without increasing computation time.
    
    \item \textbf{GraphSLAM, BA} \cite{
    Thrun2005ProbabilisticRobotics, Grisetti2010ATutorialonGraphBasedSLAM} --These algorithms solve the full SLAM problem (no marginalization), often with high accuracy, but can be    very slow.
\end{itemize}




\section{Experiments} \label{sec: Experiments} 

This section describes the empirical performance of different marginalization schemes on pose tracking of real-world data. 
We examine the MSCKF \cite{Mourikis2007MultiStateConstraintKalmanFilter}, a standard SWF, and the keyframe-based OKVIS algorithm \cite{Leutenegger2015KeyframebasedVO}, each implemented using our unifying framework.

\subsection{Simulation Settings}

Experiments are performed on the EuRoC MAV dataset of stereo images and IMU data \cite{Burri2016EuRoC}. We standardize the front-end across all experiments, using BRISK keypoint features with brute-force matching. Outlier rejection between stereo cameras is performed via epipolar constraint tests, and outlier rejection between stereo frames at subsequent timesteps is performed via the reprojection distance test, using the latest estimate of the camera pose and feature positions. We use the GTSAM back-end in C++  to construct and update costs, compute Jacobians, and implement Gauss-Newton and marginalization steps \cite{dellaert2012gtsam, dellaert2017factor}.
To construct dynamics and measurement maps, we collect on-board IMU odometry measurements, and apply IMU pre-integration scheme as in \cite{forster2015imu} (Appendix \ref{subsec: App, simulation}) and trajectory alignment as in \cite{Umeyama1991LeastSquaresEstimationofTransformationParameters}.



\begin{figure}
    \centering
    \vspace{3mm}
    \includegraphics[width=8.7cm]{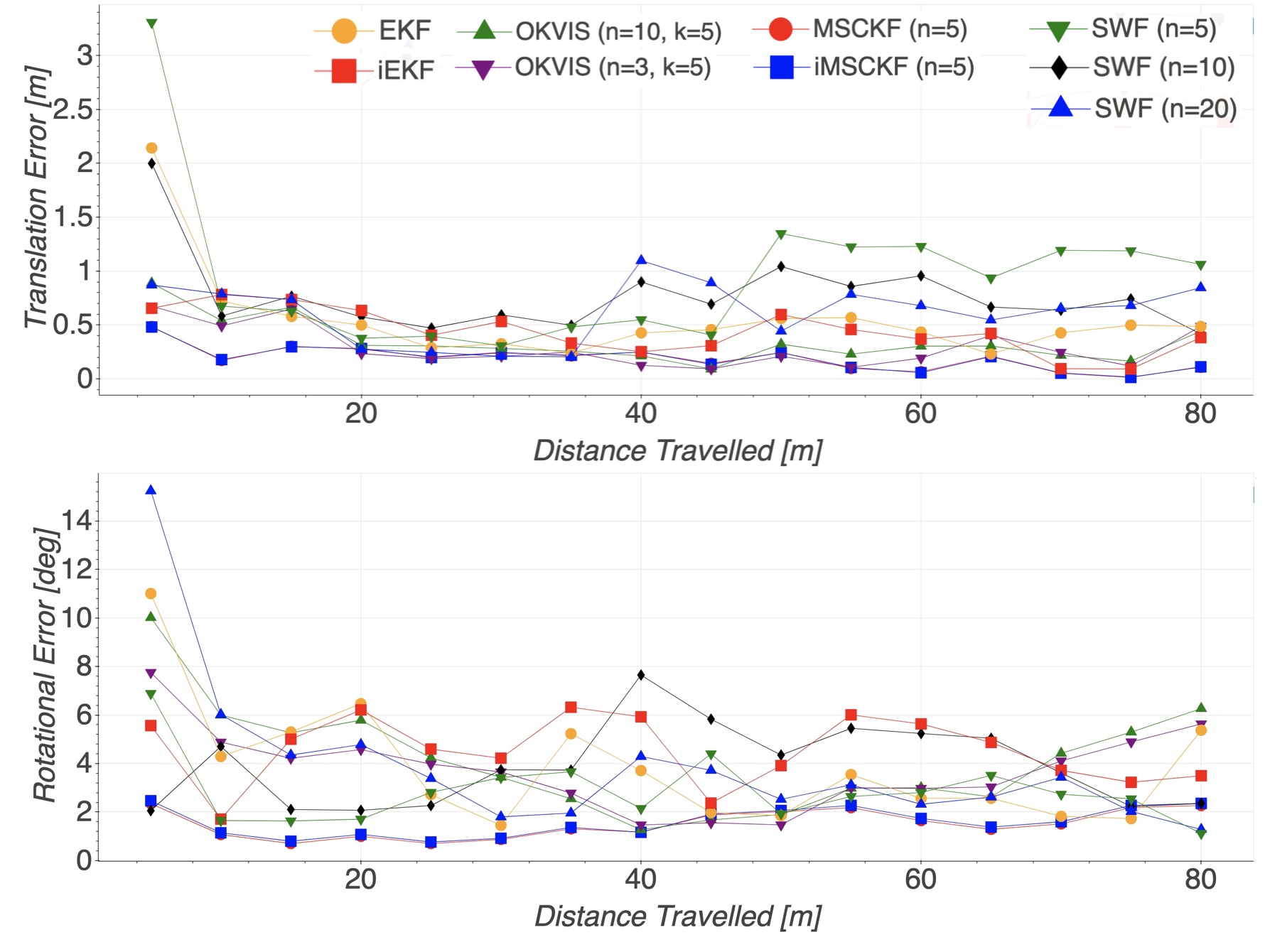}
    \caption{Localization on Vicon Room 2 (medium): Drift from, vs. distance traveled along, the ground-truth trajectory, at 5 meter intervals. 
    We apply trajectory alignment as in \cite{Umeyama1991LeastSquaresEstimationofTransformationParameters}.
    }
    \label{fig:easier}
\end{figure}

\begin{table*}[t]
    \notsotiny
    \centering
    \vspace{5mm}
    \begin{tabular}{||c c c c c c c c c c||} 
         \hline
         Data & MSCKF(n=5) & iMSCKF(n=5) & OKVIS(n=10,k=5) & OKVIS(n=3,k=5) & SWF(n=10) & SWF(n=20) & SWF(n=5) & EKF & iEKF \\ 
         \hline
         \hline
         V1\_01 & 0.09m, 2.87\degree & 0.09m, 2.87\degree & 0.20m, 3.62\degree & 0.23m, 3.68\degree & 0.36m, 3.12\degree & 0.36m, 4.72\degree & 1.00m, 8.22\degree & 1.10m, 8.79\degree & 0.04m, 59.10\degree \\ 
         \hline
         V1\_02 & 0.15m, 1.52\degree & 0.16m, 1.55\degree & - & - & 0.33m, 3.90\degree & 1.16m, 4.69\degree & 0.30m, 3.68\degree & 0.75m, 8.13\degree & 0.42m, 5.48\degree \\
         \hline
         V1\_03 & 1.00m, 4.84\degree & 1.12m, 4.90\degree & 0.42m, 6.49\degree & 0.46m, 6.99\degree & - & 6.36m, 25.85\degree & 0.79m, 6.47\degree & 14.84m, 24.58\degree & 1.83m, 8.69\degree \\
         \hline
         V2\_01 & 0.14m, 0.83\degree & 0.14m, 0.80\degree & 0.45m, 2.88\degree & 0.39m, 2.58\degree & 0.75m, 5.23\degree & 0.23m, 1.71\degree & 0.26m, 1.96\degree & 0.79m, 3.67\degree & 0.48m, 3.58\degree \\
         \hline
         V2\_02 & 0.24m, 1.67\degree & 0.24m, 1.74\degree & 0.44m, 4.69\degree & 0.38m, 4.13\degree & 0.96m, 3.71\degree & 0.75m, 4.49\degree & 1.42m, 2.99\degree & 0.83m, 4.32\degree & 0.47m, 4.14\degree \\ 
         \hline
         MH01 & 0.07m, 1.03\degree & 0.07m, 1.03\degree & 0.63m, 8.44\degree & 0.72m, 11.61\degree & 0.19m, 1.31\degree & 0.11m, 1.14\degree & 0.43m, 4.02\degree & 0.46m, 3.09\degree & 0.42m, 3.11\degree\\
         \hline
         MH02 & 0.14m, 1.05\degree & 0.19m, 1.86\degree & 0.76m, 8.79\degree & 0.93m, 9.99\degree & 0.23m, 1.89\degree & 0.23m, 1.58\degree & 0.28m, 2.80\degree & 0.34m, 2.21\degree & 0.50m, 2.68\degree \\
         \hline
         MH03 & 0.26m, 1.36\degree & 0.25m, 1.35\degree & 0.64m, 4.29\degree & 0.86m, 5.79\degree & 0.32m, 1.77\degree & 0.54m, 1.75\degree & 0.25m, 1.60\degree & 1.19m, 4.62\degree & 1.33m, 5.23\degree \\
         \hline
         MH04 & 1.11m, 1.62\degree & 1.06m, 1.52\degree & - & - & 0.79m, 1.75\degree & 0.86m, 1.79\degree & 0.59m, 1.28\degree & 4.10m, 4.58\degree & 5.21m, 6.32\degree \\
         \hline
        \end{tabular}
    \caption{Root-mean-squared error in translation and rotation on Vicon Room and Machine Hall sequences from the Euroc MAV dataset. We apply trajectory alignment as in \cite{Umeyama1991LeastSquaresEstimationofTransformationParameters}.}
    \label{tab:RMS_Error}
\end{table*}


\subsection{Results and Discussion}


Localization root-mean-squared error on Vicon Room and Machine Hall sequences from the Euroc MAV dataset are presented in Table \ref{tab:RMS_Error}. Due to space constraints, only the estimator drift on the V2\_02 sequence is plotted (Fig \ref{fig:easier}).
First, we analyze standard SWFs of window size $n=5, 10, 20$ frames. Features are marginalized when they are only visible in the oldest frame in the optimization window. EKF and iEKF are also included, and are implemented as SWFs with window size 1. For the former, only 1 Gauss-Newton step is taken, and for the latter, steps are taken until convergence.
Next, we implement MSCKF via incremental optimization (Section \ref{subsec: MSCKF}), with window size $n = 5$, and with two optimization schemes: (1) standard, with one Gauss-Newton step after marginalization, and (2) a version that takes steps until convergence (``Iterated MSCKF," or iMSCKF). Finally, we implement OKVIS with IMU window size $n = 3, 10$, keyframe window size $k=5$, and marginalization and keyframe selection schemes as in Leutenegger et al. \cite{Leutenegger2015KeyframebasedVO}.

Our experiments show that, overall, 
OKVIS
outperforms baseline SWFs, even when the latter has a larger window size. Moreover, our MSCKF implementation outperforms SWFs and OKVIS, even under challenging camera motions, despite the latter maintaining nonlinear constraints between camera poses and landmarks, and taking multiple Gauss-Newton steps per iteration. This persists even 
for SWFs
with larger window sizes. Taking multiple Gauss-Newton steps in the iMSCKF estimator did not noticeably improve performance over the standard MSCKF (Figure \ref{fig:easier}).




In contrast with SWF and OKVis implementations of comparable sizes, the MSCKF recovers better from localization errors, by employing a marginalization scheme that always maintains poses arbitrarily far in the past. This is because older poses represent higher baselines and thus supply better localization information [7]. For instance, the MSCKF maintains the first pose in the estimator for a long time, thus enforcing consistency with subsequent pose estimates and minimizing drift. In contrast, although OKVIS allows older keyframes to be maintained,
in practice keyframes are usually roughly evenly spaced and form a sliding temporal window of camera motion. Thus, earlier poses are quickly marginalized, causing estimates to drift more at the start of the trajectory. Furthermore, the MSCKF processes features in the optimization window only after they have matured. Thus, it maximally utilizes localization information with fewer updates, and ensures that each feature is always initialized through multiple-view, instead of merely stereo, triangulation. This minimizes the linearization error when features are processed and dropped.

\section{Conclusion} \label{sec: Conclusion}


This paper presents a framework for formulating and analyzing optimization and filtering-based SLAM approaches
as the iterative application of key algorithm submodules, and proves that it encompasses state-of-the-art filtering algorithms as special cases.
Experimental analysis indicate our formulation is useful for analyzing various design choices inherent in these existing SLAM algorithms, and implementing them in a modular fashion for a wide range of robotics applications, which we are eager to test on hardware. 

As future work, we wish to apply our analysis to the \textit{dynamic SLAM} problem, which concerns highly mobile features \cite{
zhang2020VDOSLAM, Cadena2016PastPresentandFuture} in practical multi-agent interactions, e.g., real-life traffic scenarios \cite{fridovich2019efficient}, by designing marginalization strategies for estimators that jointly track moving and stationary landmarks. We expect good performance on dynamic SLAM, since it enables flexible user-selected design choices.







\printbibliography


\appendix



Please use the following link to access an ArXiV version with the appendix (\url{https://arxiv.org/pdf/2112.05921.pdf}). The authors will ensure that this link stays active.

\newpage

The following supplementary material includes the appendix, which contains proofs and figures omitted in the main paper due to space limitations.

\subsection{Appendix for Section \ref{sec: Problem Formulation on General Manifolds}}
\label{subsec: App, Problem Formulation on General Manifolds}

\subsubsection{SLAM on Manifolds: Dynamics and Measurement Maps}
\label{subsubsec: App, SLAM on Manifolds: Dynamics and Measurement Maps}

To formulate SLAM on a manifold, we must alter our definitions of the state variables, features, image positions, dynamics map, and measurement map. Let $\X$ be a smooth manifold of dimension $\dx$,
on which the system state are defined. Similarly, let $\F$ be a smooth manifold of dimension $\df$, 
on which features are defined, and let $\Z$ be the smooth manifold of dimension $d_z$,
on which image measurements are defined (Often, $\mathcal{F} = \R^{\df}$ and $\mathcal{Z} \in \R^{d_z}$, e.g., with $\df = 3$ and $d_z = 2$). We then have:
\begin{align} \nonumber
    x_{t+1} &= g(x_t) \boxplus w_t, \hspace{1cm} w_t \sim \mathcal{N}(0, \Sigma_w), \\ \nonumber
    z_{t,k} &= h(x_t, f_k) \boxplus v_{t,k}, \hspace{1cm} v_{t,k} \sim \mathcal{N}(0, \Sigma_v).
\end{align}
where $x_t \in \X$ denotes the state at time $t$, $g: \X \ra \X$ denotes the discrete-time dynamics map, and $w_t \in \R^{\dx}$ denotes the dynamics noise, with covariance $\Sigma_w \in \R^{\dx \times \dx}$, $\Sigma_w \succ 0$. Moreover, $f_k \in \mathcal{F}$ denotes feature position $j$ estimated at the camera pose at time $t$, $z_{t,k} \in \Z$ denotes the image measurement of feature $j$ measured from the camera pose at time $t$, $h: \X \times \F \ra \Z$ denotes the measurement map, and $v_t \in \R^{d_z}$ denotes the measurement noise, with covariance $\Sigma_v \in \R^{d_z \times d_z}$, $\Sigma_v \succ 0$.



As before, SLAM concerns an optimization problem over a collection of poses and features, e.g., a sliding window of the most recent poses in the states $\{x_i \in \X| i = t-\nx+1, \cdots, t\}$ and features $\{f_k \in \F| j = \nf-q+1, \cdots, \nf\}$:
\begin{align*}
    &\overline{x_t} := (x_{t-\nx+1}, \cdots, x_t, f_{\nf-q+1}, \cdots, f_{\nf}) \\
    &\hspace{1cm} \in \X^\nx \times \F^q.
\end{align*}
We assume that $\overline{x_t}$ is associated with a prior distribution with mean $\mu_0 \in \X^\nx \times \F^q$ and covariance $\Sigma_0 \in \R^{(\nx \dx + q \df) \times (\nx \dx + q \df)}$.


\subsubsection{SLAM as an Optimization Problem on Manifolds}
\label{subsubsec: App, SLAM on Manifolds: the Optimization Problem}

In this subsection, we interpret the SLAM problem on manifolds as the optimization of a cost function $c: \X^\nx \times \F^q \ra \R$, constructed from residual terms of the same dimension of the minimal coordinates of $\overline{x_t}$, $x_t$ and $z_t$. In particular, we must generalize \eqref{Eqn: Cost, General} to the case where the states, dynamics and measurement maps are defined on and between manifolds. This involves replacing + and - operators with $\boxplus$ and $\boxminus$ operators, when necessary. For example, the sliding window filter window presented in Section \ref{sec: Problem Formulation on Euclidean Spaces}, would be associated with the cost $c: \X^\nx \times \F^q \ra \R$, given by:
\begin{align} \nonumber
    c(\overline{x_t}) &= \Vert \overline{x_t} \boxminus \mu_0 \Vert_{\Sigma_0^{-1}}^2 \\ \nonumber
    &\hspace{5mm} + \sum_{i=t-\nx+1}^{t-1} \Vert x_{i+1} \boxminus g(x_{i}) \Vert_{\Sigma_w^{-1}}^2 (x_{i+1} \boxminus g(x_{i})) \\ \nonumber
    &\hspace{5mm} + \sum_{j= \nf-q+1}^{\nf}\sum_{i=t-\nx+1}^{t} \Vert z_{ij} \boxminus h(x_i,f_k) \Vert_{\Sigma_v^{-1}}^2
\end{align}

Similar to Section \ref{sec: Problem Formulation on Euclidean Spaces}, we stack all residual terms into a single residual vector $C(\overline{x_t})$. For example, for the sliding window filter above, we have:
\begin{align*}
    &C(\overline{x_t}) := \Big[ \big( \Sigma_0^{-1/2}(\Tilde{x_t} \boxminus \mu_0) \big)^\top \\
    &\hspace{5mm} \big( \Sigma_w^{-1/2} (x_{t-\nx+1} \boxminus g(x_{t-\nx})) \big)^\top \cdots \big( \Sigma_w^{-1/2} (x_t \boxminus g(x_{t-1})) \big)^\top \\
    &\hspace{5mm} \big( \Sigma_v^{-1/2} ( z_{t-\nx+1, \nf-q+1} \boxminus h(x_{t-\nx+1}, f_{\nf-q+1})) \big)^\top \cdots \\
    &\hspace{5mm} \big( \Sigma_v^{-1/2} ( z_{t-\nx+1, \nf} \boxminus h(x_{t-\nx+1}, f_{\nf})) \big)^\top \cdots \\
    &\hspace{5mm} \big( \Sigma_v^{-1/2} ( z_{t, \nf-q+1} \boxminus h(x_t, f_{\nf-q+1})) \big)^\top \cdots \\
    &\hspace{5mm} \big( \Sigma_v^{-1/2} (z_{t, \nf} \boxminus h(x_t, f_{\nf})) \big)^\top \Big]^\top \\
    &\hspace{1cm} \in \R^{(2n-1)\dx + \nf \df + \nx q d_z}.
\end{align*}
As a result, $c(\overline{x_t}) = C(\overline{x_t})^\top C(\overline{x_t})$, and the SLAM problem is now reduced to the following nonlinear least squares problem:
\begin{align}
    \label{Eqn: App, Problem Formulation, Inner Product Cost, on Manifolds}
    \min_{\overline{x_t}}. c(\overline{x_t}) = \min_{\overline{x_t}}. C(\overline{x_t})^\top C(\overline{x_t})
\end{align}
Section \ref{subsec: App, Main Algorithm on manifolds} introduces the main algorithmic submodules used to find an approximate solution to \eqref{Eqn: App, Problem Formulation, Inner Product Cost, on Manifolds}.

\subsubsection{Jacobians Under Box Operators}
\label{subsubsec: App, Jacobians Under Box Operators}

In subsequent proofs, we require the following results regarding the behavior of Jacobian matrices under the $\boxplus$ and $\boxminus$ operators. In particular, we will focus on the case where the box operator acts on elements of $SO(3)$, since $SO(3)$ is the non-Euclidean-space Lie group that appears most often in this paper.

\begin{definition}[\textbf{Jacobian on $SO(3)$}]
We say that $f: SO(3) \ra SO(3)$ is differentiable at $R \in SO(3)$, with Jacobian denoted by:
\begin{align*}
    \frac{\partial}{\partial \theta} f(R) \in \R^{3 \times 3},
\end{align*}
if the following equality holds:
\begin{align*}
    \lim_{\delta \theta \ra 0} \frac{\left\Vert f(R \boxplus \delta \theta) \boxminus f(R) - \frac{\partial f}{\partial \theta} \delta \theta \right\Vert_2}{\Vert \delta \theta
    \Vert_2} = 0.
\end{align*}
\end{definition}

\begin{theorem}
Suppose $f: SO(3) \ra SO(3)$ is differentiable. Then, for any fixed $y \in SO(3)$:
\begin{align*}
    \frac{\partial}{\partial \theta} \big( y \boxminus f(R) \big) = - \frac{\partial f}{\partial \theta} + O(y \boxminus f(R)).
\end{align*}
\end{theorem}

\begin{proof}
Using the definition of the $\boxminus$ operator for SO(3), we have, for any $\delta \theta \in \R^3$:
\begin{align*}
    &\hspace{5mm} y \boxminus f(R \boxplus \delta \theta) \\
    &= y \boxminus \left( f(R) \boxplus \frac{\partial f}{\partial \theta} \delta \theta \right) + o(\delta \theta) \\
    &= y \boxminus \left( f(R) \hspace{0.5mm} \Exp \left( \frac{\partial f}{\partial \theta} \delta \theta \right) \right) + o(\delta \theta) \\
    &= \Log \left( \Exp \left(- \frac{\partial f}{\partial \theta} \delta \theta \right) f(R)^\top y \right) + o(\delta \theta) \\
    &= \Log \left( \Exp \left(- \frac{\partial f}{\partial \theta} \delta \theta \right) \Exp \big(y \boxminus f(R) \big) \right) + o(\delta \theta) \\
    &= y \boxminus f(R) + J_\ell^{-1} \big( y \boxminus f(R) \big) \cdot \left( - \frac{\partial f}{\partial \theta} \delta \theta \right) + o(\delta \theta).
\end{align*}
Here, we have defined $J_\ell^{-1}: \R^3 \backslash \{0\} \ra SO(3)$ by:
\begin{align*}
    J_\ell^{-1} (\theta) &= I - \frac{1}{2} \theta^\wedge + \left( \frac{1}{\Vert \theta \Vert^2} - \frac{1 + \cos \Vert \theta \Vert}{2 \Vert \theta \Vert \cdot \sin \Vert \theta \Vert} \right) (\theta^\wedge)^2,
\end{align*}
in accordance with \cite{Sola2014QuaternionKinematics}. Here, $\wedge: \R^3 \ra \R^{3 \times 3}$ denotes the wedge operator. Thus, we have:
\begin{align*}
    \frac{\partial}{\partial \theta} \big( y \boxminus f(R) \big) = - J_\ell^{-1} \big( y \boxminus f(R) \big) \cdot \frac{\partial f}{\partial \theta}.
\end{align*}
The theorem statement now follows.
\end{proof}

\begin{theorem}
Suppose $f: SO(3) \ra SO(3)$ is differentiable. Then for any fixed $y \in SO(3)$:
\begin{align*}
    \frac{\partial}{\partial \theta} \big( f(R) \boxminus y \big) = \frac{\partial f}{\partial \theta} + O(f(R) \boxminus y).
\end{align*}
\end{theorem}

\begin{proof}
Using the definition of the $\boxminus$ operator for SO(3), we have, for any $\delta \theta \in \R^3$:
\begin{align*}
    &\hspace{5mm} f(R \boxplus \delta \theta) \boxminus y \\
    &= \left( f(R) \boxplus \frac{\partial f}{\partial \theta} \delta \theta \right) \boxminus y + o(\delta \theta)\\
    &= \left( f(R) \hspace{0.5mm} \Exp \left( \frac{\partial f}{\partial \theta} \delta \theta \right) \right) \boxminus y + o(\delta \theta)\\
    &= \Log \left( y^\top f(R) \Exp \left(\frac{\partial f}{\partial \theta} \delta \theta \right) \right) + o(\delta \theta)\\
    &= \Log \left( \Exp \big(f(R) \boxminus y \big) \Exp \left(\frac{\partial f}{\partial \theta} \delta \theta \right) \right) + o(\delta \theta)\\
    &= f(R) \boxminus y + J_r^{-1} \big(f(R) \boxminus y \big) \cdot \left( \frac{\partial f}{\partial \theta} \delta \theta \right) + o(\delta \theta),
\end{align*}
which implies:
\begin{align*}
    \frac{\partial}{\partial \theta} \big( f(R) \boxminus y \big) = J_r^{-1} \big( f(R) \boxminus y \big) \cdot \frac{\partial f}{\partial \theta}.
\end{align*}
Here, we have defined $J_r^{-1}: \R^3 \backslash \{0\} \ra SO(3)$ by:
\begin{align*}
    J_r^{-1} (\theta) &= I + \frac{1}{2} \theta^\wedge + \left( \frac{1}{\Vert \theta \Vert^2} - \frac{1 + \cos \Vert \theta \Vert}{2 \Vert \theta \Vert \cdot \sin \Vert \theta \Vert} \right) (\theta^\wedge)^2,
\end{align*}
in accordance with \cite{Sola2014QuaternionKinematics}. The theorem statement now follows.
\end{proof}

In subsequent discussions, when we consider dynamics and measurement models of the form $y = f(R) \boxplus n$, where $n \in \R^3$ denotes small-magnitude, zero-sum noise, terms of order $O(y \boxminus f(R))$ and $O(f(R) \boxminus y)$ are often ignored.

\subsection{Appendix for Section \ref{sec: Main Algorithm}}
\label{subsec: App, Main Algorithm}

\subsubsection{Algorithm Sub-blocks}
\label{subsubsec: App, Algorithm Sub-blocks}

\begin{algorithm}
    {
    \small
    \label{Alg: gauss-newton}
    \SetAlgoLined
    \KwData{Objective $C^\top C$, linearization point $\overline{x_t^\star}$.}
    \KwResult{Mean $\mu$, covariance $\Sigma$ after a Gauss-Newton step.}
    
     
     $J \gets \frac{\partial C}{\partial x_t} \big|_{\overline{x_t^\star}}$
     
    $\Sigma_t \gets J^\top J$
        
    $\mu_t \gets \overline{x_t^\star} - (J^\top J)^{-1} J^\top C(\overline{x_t^\star})$
    
    \Return{$\mu_t, \Sigma_t$}
     
    \caption{Gauss-Newton Step.}
    }
\end{algorithm}

\begin{algorithm}
\label{Alg: marginalization}
{
    \small
\SetAlgoLined
\KwData{Objective $f = C^\top C$, vector of variables to marginalize $x_{t,M}$, linearization point $x_t^\star$.}
\KwResult{Mean $\mu_{t,K}$ and covariance $\Sigma_{t,K}$ of non-marginalized variables $\overline{x_{t,K}^\star}$.}



$C \gets$ subvector of $C$ containing entries dependent on $x_M$.

$J := \begin{bmatrix}
J_K & J_M
\end{bmatrix} \gets \begin{bmatrix}
\frac{\partial C}{\partial \overline{x_{t,K}}} \big|_{\overline{x_t^\star}} & \frac{\partial C}{\partial \overline{x_{t,M}}} \big|_{\overline{x_t^\star}}
\end{bmatrix}$.

$\Sigma_{t,K} \gets \big( J_K^\top \big[I - J_M(J_M^\top J_M)^{-1} J_M^\top \big] J_K \big)^{-1}$ 
        
$\mu_{t,K} \gets \overline{x_{t,K}^\star} - \Sigma_{t,K} J_K^\top \big[ I - J_M(J_M^\top J_M)^{-1} J_M^\top \big] C(x^\star)$

\vspace{2mm}
\Return{$\mu_{t,K}, \Sigma_{t,K}$}
 \caption{Marginalization}
 }
\end{algorithm}

\subsubsection{Proof of Theorem \ref{Thm: Gauss-Newton Eqns} (Gauss-Newton Steps)}
\label{subsubsec: App, Gauss-Newton Steps}

Here, we present the proof of Theorem \ref{Thm: Gauss-Newton Eqns}, reproduced below.

\begin{theorem}(\textbf{Gauss-Newton Step})
Let $\overline{x_t}^\star \in \R^d$ denote a given linearization point, and suppose $J := \frac{\partial C}{\partial \overline{x_t}} \in \R^{d_C \times d}$ has full column rank. Then applying a Gauss-Newton step to the cost $c(\overline{x_t})$, about $\overline{x_t}^\star \in \R^d$ yields the new cost:
\begin{align*}
    c(\overline{x_t}) = \Vert \overline{x_t}-\mu_t \Vert_{\Sigma_t^{-1}}^2 + o(\overline{x_t} - \overline{x_t}^\star),
\end{align*}
where $\mu_t \in \R^{d}$ and $\Sigma_t \in \R^{d \times d}$ are given by:
\begin{align*}
    \Sigma_t &\gets (J^\top J)^{-1}, \\
    \mu_t &\gets \overline{x_t}^{\star}  - (J^\top J)^{-1} J^\top  C(\overline{x_t}^{\star}).
\end{align*} 
\end{theorem}

\begin{proof}
We have:
\begin{align*}
    c(\overline{x_t}) &= C(\overline{x_t})^\top C(\overline{x_t}) \\
    &= \big[C(\overline{x_t}^\star) + J (\overline{x_t} - \overline{x_t}^\star) \big]^\top \big[C(\overline{x_t}^\star) + J (\overline{x_t} - \overline{x_t}^\star) \big] \\
    &\hspace{5mm} + o(\overline{x_t} - \overline{x_t}^\star) \\
    &= (\overline{x_t} - \mu_t)^\top \Sigma_t^{-1} (\overline{x_t} - \mu_t) + c_0(\overline{x_t}^\star) + o(\overline{x_t} - \overline{x_t}^\star),
\end{align*}
where $c_0(\overline{x_t}^\star) \in \R$ denotes a scalar-valued function of $\overline{x_t}^\star$ that is independent of the variable $\overline{x_t}$. This concludes the proof.
\end{proof}

\subsubsection{Proof of Theorem \ref{Thm: Marginalization Eqns} (Maringalization Steps)}
\label{subsubsec: App, Marginalization of States}

\begin{theorem}[\textbf{Marginalization Step}] 
Let $\overline{x_t}^\star \in \R^d$ denote a given linearization point, and suppose $J := \frac{\partial C}{\partial \overline{x_t}} \in \R^{d_C \times d}$ has full column rank. Define $J_K := \frac{\partial C}{\partial \overline{x_{t,K}}} \in \R^{d_C \times d_K}$ and $J_M := \frac{\partial C}{\partial \overline{x_{t,M}}} \in \R^{d_C \times d_M}$. If $C_2(\overline{x_{t, M}}, \overline{x_{t, K}})$ were a linear function of $\overline{x_t} = (\overline{x_{t, M}}, \overline{x_{t, K}})$, then applying a Marginalization step to the cost $c(\overline{x_t})$, about the linearization point $\overline{x_t}^\star = (\overline{x_{t,K}^\star}, \overline{x_{t,M}^\star}) \in \R^d$ yields:
\begin{align} \nonumber
    \min_{\overline{x_{t,M}}} c_2(\overline{x_{t,K}}, \overline{x_{t,M}}) 
    &= \Vert \overline{x_{t, K}} - \mu_{t,K} \Vert_{\Sigma_{t,K}^{-1}}^2,
\end{align}
where $\Sigma_{t,K} \in \R^{d_K \times d_K}$ and $\mu_{t,K} \in \R^{d_K}$ are given by:
\begin{align} \nonumber
    \Sigma_{t,K} &:= \big( J_K^\top \big[ I - J_M(J_M^\top J_M)^{-1} J_M^\top \big] J_K \big)^{-1} 
    , \\ \nonumber
    \mu_{t,K} &:= \overline{x_{t,K}^\star} - \Sigma_{t,K} J_K^\top \big[ I - J_M(J_M^\top J_M)^{-1} J_M^\top \big] C(\overline{x_t^\star}).
\end{align}
\end{theorem}

\begin{proof}
It suffices to show that:
\begin{align*}
    &\min_{\overline{x_{t, M}}} c_2(\overline{x_{t, K}}, \overline{x_{t, M}}) \\
    = \hspace{0.5mm} &(\overline{x_{t, K}} - \mu_K)^\top \Sigma_K^{-1} (\overline{x_{t, K}} - \mu_K)+ c'(\overline{x_t^\star}).
\end{align*}
To do so, we first note that since $C_2(\overline{x_t})$ is linear in $\overline{x_t}$:
\begin{align*}
    c_2(\overline{x_t}) &= \Vert C_2(\overline{x_t}) \Vert_2^2 = \Vert C_2(\overline{x_t^\star}) + J_2 \Delta \overline{x_t} \Vert_2^2 \\
    &= \Vert C_2(\overline{x_t^\star}) + J_K \Delta \overline{x_{t, K}} + J_M \Delta \overline{x_{t, M}} \Vert_2^2.
\end{align*}
By the method of least-squares, the optimal $\Delta \overline{x_{t, M}}$ is given by the normal equation:
\begin{align*}
    \Delta \overline{x_{t, M}} = - (J_M^\top J_M)^{-1} J_M^\top \big( C_2(\overline{x_t^\star}) + J_K \Delta \overline{x_{t, K}} \big)
\end{align*}
Substituting back into our expression for $c(\overline{x_t})$, we have:

\begin{align*}
    &\min_{\overline{x_{t, M}}}. c_2(\overline{x_t}) \\
    = \hspace{0.5mm} &\Vert \big(I - J_M (J_M^\top J_M)^{-1} J_M^\top \big) \big( C_2(\overline{x_t^\star}) + J_K \Delta \overline{x_{t, K}} \big) \Vert_2^2 \\
    = \hspace{0.5mm} &\big( C_2(\overline{x_t^\star}) + J_K \Delta \overline{x_{t, K}} \big)^\top \big[ I - J_M(J_M^\top J_M)^{-1} J_M^\top \big] \\
    &\hspace{5mm} \big( C_2(\overline{x_t^\star}) + J_K \Delta \overline{x_{t, K}} \big) 
    \\ \nonumber
    = \hspace{0.5mm} &(\overline{x_{t, K}} - \overline{x_{t,K}^\star})^\top \underbrace{J_K^\top \big[ I - J_M(J_M^\top J_M)^{-1} J_M^\top \big] J_K}_{:= \hspace{0.5mm} \Sigma_K^{-1}} \\
    &\hspace{5mm} (\overline{x_{t, K}} - \overline{x_{t,K}^\star}) \\ \nonumber
    &\hspace{5mm} + 2 (\overline{x_{t, K}} - \overline{x_{t,K}^\star})^\top J_K^\top \big[ I - J_M(J_M^\top J_M)^{-1} J_M^\top \big] C_2(\overline{x_t^\star}) \\ \nonumber
    &\hspace{5mm} + C_2(\overline{x_t^\star})^\top \big[ I - J_M(J_M^\top J_M)^{-1} J_M^\top \big] C_2(\overline{x_t^\star}) 
    \\
    = \hspace{0.5mm} &\big(\overline{x_{t, K}} \underbrace{- \overline{x_{t,K}^\star} + \Sigma_K J_K^\top \big[ I - J_M(J_M^\top J_M)^{-1} J_M^\top C_2(\overline{x_t^\star})}_{:= \hspace{0.5mm} -\mu_K} \big]  \big)^\top \Sigma_K^{-1} \\
    &\hspace{5mm} \big(\overline{x_{t, K}} \underbrace{- \overline{x_{t,K}^\star} + \Sigma_K J_K^\top \big[ I - J_M(J_M^\top J_M)^{-1} J_M^\top \big] C_2(\overline{x_t^\star})}_{:= \hspace{0.5mm} - \mu_K} \big) \\
    &\hspace{5mm} + C_2(\overline{x_t^\star}) \big(I - J_M(J_M^\top J_M)^{-1} J_M^\top \big)\\ 
    \nonumber
    &\hspace{5mm} + c'(\overline{x_t^\star}  \\
    = \hspace{0.5mm} &(\overline{x_{t, K}} - \mu_K)^\top \Sigma_K^{-1} (\overline{x_{t, K}} - \mu_K)+c'(\overline{x_t^\star}).
\end{align*}
with $\Sigma_K$ and $\mu_K$ as defined in the theorem statement, and $c'(\overline{x_t^\star} \in \R$ as a term independent of $\overline{x_t}$.

\end{proof}

\subsubsection{Main Algorithm on Manifolds}
\label{subsec: App, Main Algorithm on manifolds}

Recall the states, poses, features, dynamics and measurement maps, and costs defined in Appendix \ref{subsubsec: App, SLAM on Manifolds: the Optimization Problem}. In particular, the cost $c(\overline{x})$ is given by:
\begin{align}
    c(\overline{x}) = C(\overline{x})^\top C(\overline{x})
\end{align}
Now, let $\bar x^*$ be a chosxen linearization point. Let $\widehat C_{\bar x^*} := C \circ \pi_{\bar x^\star}^{-1}$ 
be the coordinate representation of the function $C$ near $\bar x^*$. Recall that $\widehat C_{\bar x^*}$ is simply a function from one Euclidean space to another. We can now Taylor expand to write:
\begin{align} \nonumber
 C(\bar x) &= (C \circ \pi_{\overline{x}^\star}^{-1}) \big( \pi_{\overline{x}^\star}(\overline{x}) \big) = \widehat C_{\bar x^\star}(\Delta\overline{\chi}) \\
 &= \widehat C_{\bar x^\star}(0) + J \Delta\overline{\chi} + o(\Delta \overline{\chi}), 
\end{align}
where $\Delta\overline{\chi} = \overline{x} \boxminus \overline{x}^\star$ and $J$ is the Jacobian of $\widehat C_{\bar x^*}$ with respect to $\Delta \overline{\chi}$ evaluated at zero. We then apply a modified version of the algorithms from Section \ref{subsec: Algorithm Overview}:
\begin{enumerate}
\item \emph{Gauss-Newton Descent}:
Used to update the current linearization point $\bar{x}^{\{k\}}$ to a new linearization point $\bar{x}^{\{k+1\}}$.
\begin{align} 
    \overline{x}^{(k+1)} \gets \overline{x}^{(k)} \boxplus \big( - (J^T J)^{-1} J^T C(\overline{x}^{(k)}) \big)
\end{align}
After Gauss-Newton steps have been taken, the linearization point $\overline{x}^\star$ is fixed, and all or part of the original optimization problem, reproduced below:
\begin{align*}
\min_{\overline{x}} \ c(\overline{x}) = \min_{\overline{x}} \ C(\overline{x})^\top C(\overline{x})
\end{align*}
is replaced with the following linear least squares optimization problem:
\begin{align} 
\min_{\overline{x}}. (\overline{x} \boxminus \mu)^\top \Sigma^{-1}(\overline{x} \boxminus \mu)
\end{align}
where the algorithm assigns:
\begin{align*}
    \mu &\gets \overline{x}^{\star} \boxplus (J^\top J)^{-1} J^\top  C(\overline{x}^{\star}) \\
    \Sigma &\gets (J^\top J)^{-1}.
\end{align*} 

\item \emph{Marginalization}:
Used to remove variables $\overline{x_M}$ from the optimization problem by applying linear approximation to $C$---in particular, the optimization problem:
\begin{align*}
    \min_{\overline{x}}. c(\overline{x})
\end{align*}
is approximated by:
\begin{align*}
    &\min_{\overline{x_K}}. (\overline{x_K} \boxminus \mu_K)^\top \Sigma_K^{-1} (\overline{x_K} \boxminus \mu_K),
\end{align*}
where the algorithm assigns:
\begin{align*}
    \mu_K &\gets \overline{x_K^\star}  \boxplus \\
    &\hspace{5mm} \big(-\Sigma_K J_K^\top \big[ I - J_M(J_M^\top J_M)^{-1} J_M^\top \big] C_2(\overline{x^\star})\big) \\
    \Sigma_K &\gets \big(J_K^\top \big[ I - J_M(J_M^\top J_M)^{-1} J_M^\top \big] J_K\big)^{-1},
\end{align*}

\end{enumerate} 


\subsection{Appendix for Section \ref{sec: Equivalence of Filtering and Optimization-Based Approaches}}
\label{subsec: App, Equivalence of Filtering and Optimization-Based Approaches}

\subsubsection{Continuous to Discrete Time Formulation}
\label{subsubsec: App, Continuous to Discrete Time Formulation}

The robotic systems analyzed by most mainstream SLAM algorithms are described using continuous-time dynamics models. The implementation of these algorithms thus involves discrete-time propagation. We sketch this process below, in a manner which is both mathematically aligned with the existing literature and convenient for illustrating our optimization algorithm, which operates using a discrete-time dynamics model.

To begin, let $\mathbb H$ and $\mathbb H_u$ denote the space of quaternions and the space of unit quaternions, respectively. Recall that the time derivative of a smooth curve on $\mathbb H_u$, e.g., $q \in \mathbb H_u$, can be expressed as an element of $\mathbb H$; in particular, there exists some $\omega \in \R^3$ such that:
\begin{align} \label{Eqn: Time Derivative of Unit Quaternions}
    \dot{q} = q \star \begin{bmatrix}
        0 \\ \omega
    \end{bmatrix} \in \mathbb H.
\end{align}
Let $\X$ be the $\dx$-dimensional Lie group inhabited by the robot state $x \in \X$, given by the finite Cartesian product of Euclidean spaces and unit quaternions. In other words, $\X$ can be written as:
\begin{align*}
    \X := \X_1 \times \cdots \X_N,
\end{align*}
where $\X_k$ is either $\mathbb H_u$ or an Euclidean space, for each $k \in \{1, \cdots, N\}$. Inspired by \eqref{Eqn: Time Derivative of Unit Quaternions}, we define:
\begin{align*}
    \X' := \X_1' \times \cdots \X_N',
\end{align*}
where, for each $k \in \{1, \cdots, N \}$, we have $\X_k' = \mathbb H$ if $\X_k = \mathbb H_u$, and $\X_k' = \X_k$ otherwise. 

The fundamental question underlying the time discretization process is as follows. Suppose the continuous-time dynamics model for the robot state is given by:
\begin{align} \label{Eqn: Continuous-time dynamics model}
    \dot{x}(s) = g_{ct}\big(x(s), w_c(s) \big), \hspace{1cm} \forall \hspace{0.5mm} s \in \R,
\end{align}
with $g_{ct}: \X \times \R^{\dx} \ra \X'$ smooth, and with $w_c(s) \in \R^{\dx}$ as zero-mean white noise with autocorrelation $\E[w_c(s) w_c(s')] = K \cdot \delta(s - s')$ for each $s \in \R$, where $K \in \R^{\dx \times \dx}$, $K \succ 0$, and $\delta(\cdot)$ denotes the Dirac delta function. Suppose $w_c(s)$ and deviations in $x(s)$ are treated as first-order perturbations. Can we straightforwardly construct an approximate discrete-time additive noise model, of the form:
\begin{align} \label{Eqn: Discrete-time, additive-noise dynamics model}
    x_{t+1} = g(x_t) \boxplus w(t),
\end{align}
where $x_t := x(t_0) \in \X$, $g: \X \ra \X$ smooth, and $x_{t+1} := x(t_1) \in \X$ and $w(t) \in \R^{\dx}$, whose evolution agrees with that of \eqref{Eqn: Continuous-time dynamics model} up to first-order perturbations?

We answer the above question in the affirmative, by detailing below the time discretization process used throughout the remainder of this section. With a slight abuse of notation, we write the discrete-time states as $x_t := x(t_0) \in \X$ and $x_{t+1} := x(t_1) \in \X$ for some $t_0, t_1 \in \R$, with $t_0 < t_1$. The main idea is to first perform non-linear integration on a set of nominal dynamics, which ignores error in the initial state and the dynamics noise. We then correct for these perturbations by assuming that they evolve as a linear dynamical system.

\begin{itemize}
    \item Nominal state dynamics:
    
    $\hspace{5mm}$ We define the \textit{nominal state dynamics} as:
    \begin{align*}
        \dot{\hat{x}}(s) &= g_{ct}(\hat{x}(s), 0) \hspace{1cm} \forall \hspace{0.5mm} s \in \R.
    \end{align*}
    This is the set of dynamics that would be obeyed in the absence of the white noise $w_c(s)$. If the solution $\hat{x}(s)$ uniquely exists, we can construct a smooth function $\hat{g}: \X \ra \X$ that maps $\hat{x}(t_0)$ to $\hat{x}(t_1)$, i.e.:
    \begin{align} \label{Eqn: Nominal State, Discrete-Time Propagation}
        \hat{x}(t_1) = \hat{g}\big( \hat{x}(t_0) \big),
    \end{align}
    which describes the discrete-time propagation of the nominal state $x(t)$.
    
    \item Error state dynamics:
    
    $\hspace{5mm}$ 
    To characterize the drift of the true state $x(s)$ away from the nominal state $x(s)$, we define the error state $\delta x(s) := x(s) \boxminus \delta \hat{x}(s) \in \R^{\dx}$, and characterize the corresponding \textit{error-state dynamics} as:
    \begin{align} \nonumber
        \dot{\delta x}(s) &= g_{ct}(x(s), w_c(s)) \boxminus g_{ct}(\delta x(s), 0) \\ \nonumber
        &= \frac{\partial f}{\partial x}(\hat{x}(s), 0) \cdot \delta x(s) \\
        &\hspace{5mm} + \frac{\partial f}{\partial w_c} (\hat{x}(s), 0) \cdot w_c(s) + o\big(\delta x(s) \cdot w_c(s) \big), 
    \end{align}
    By approximating the dynamics of the error state $\delta x(s) \in \R^{\dx}$ as a linear, time-varying system, we can write, for each $s \in (t_0, t_1)$:
    \begin{align} \label{Eqn: Error State, Discrete-Time Propagation}
        \delta x(s) &= \Phi(s, t_0) \delta x(t_0) \\
        &\hspace{5mm} + \int_{t_0}^s \Phi(s,\tau) \cdot \frac{\partial f}{\partial \omega_c}(\hat{x}(\tau), 0) \cdot w_c(\tau) \hspace{0.5mm} d\tau,
    \end{align}
    where $\Phi(s, t_0) \in \R^{\dx \times \dx}$ is the state transition matrix satisfying:
    \begin{align*}
        \frac{d}{ds} \Phi(s, t_0) = \frac{\partial f}{\partial x}(\hat{x}(s), 0) \cdot \Phi(s, t_0).
    \end{align*}
    
    \item True State dynamics---Discrete-time propagation of mean and covariance:
    
    $\hspace{5mm}$ From the discrete-time propagation of the nominal and error dynamics, i.e., \eqref{Eqn: Nominal State, Discrete-Time Propagation} and \eqref{Eqn: Error State, Discrete-Time Propagation}, we have:
    \begin{align*}
        x(t_1) &= \hat{x}(t_1) \boxplus \delta x(t_1) \\
        &= \hat{g}(\hat{x}(t_0)) \boxplus \Bigg( \Phi(t_1, t_0) \delta x(t_0) \\
        &\hspace{5mm} + \int_{t_0}^{t_1} \Phi(t_1, \tau) \cdot \frac{\partial f}{\partial \omega_c}(\hat{x}(\tau), 0) \cdot w_c(\tau) \hspace{0.5mm} d\tau \Bigg) \\
        &= \Big( \hat{g}(\hat{x}(t_0)) \boxplus \Phi(t_1, t_0) \delta x(t_0) \Big) \\
        &\hspace{5mm} \boxplus \left( \int_{t_0}^{t_1} \Phi(t_1, \tau) \cdot \frac{\partial f}{\partial \omega_c}(\hat{x}(\tau), 0) \cdot w_c(\tau) \hspace{0.5mm} d\tau \right) \\
        &\hspace{1.5cm} + o\big( \delta x(t_0), w_c(\cdot) \big),
    \end{align*}
    where we have used the fact that the $\boxplus$ operator, which operates through:
    \begin{align*}
        \Exp(\delta \theta + \delta \phi) = \Exp(\delta \theta) \cdot \Exp(\delta \phi) + o(\delta \theta \cdot \delta \phi)
    \end{align*}
    for any $x \in \mathbb H$ and $\delta \theta, \delta \phi \in \R^3$. 
    
    $\hspace{1cm}$ Finally, by identifying:
    \begin{align*}
        x_{t+1} &\gets x(t_1) \in \X, \\
        g(x_t) &\gets \hat{x}(t_1) \boxplus \Phi(t_1, t_0) \delta x(t_0) \in \X, \\
        w_t &\gets \int_{t_0}^{t_1} \Phi(t_1, \tau) \cdot \frac{\partial f}{\partial \omega_c}(\hat{x}(\tau), 0) \cdot w_c(\tau) \hspace{0.5mm} d\tau \in \R^{\dx},
    \end{align*}
    we obtain our discrete-time dynamics model:
    \begin{align*}
        x_{t+1} = g(x_t) \boxplus w_t.
    \end{align*}
    Notice that $w_t \in \R^{\dx \times \dx}$ is a zero-mean random variable, with covariance $\Sigma_w \in \R^{\dx \times \dx}$ given by:
    \begin{align*}
        &\Sigma_w := \E[w_t w_t^\top] \\
        = \hspace{0.5mm} &\E\Bigg[ \int_{t_0}^{t_1} \int_{t_0}^{t_1} \Phi(t_1, \tau) \frac{\partial f}{\partial \omega_c}(\hat{x}(\tau), 0) \cdot w_c(\tau) w_c(\tau')^\top \\
        &\hspace{5mm} \cdot \left( \frac{\partial f}{\partial \omega_c}(\hat{x}(\tau'), 0) \right)^\top \Phi(t_1, \tau')^\top \hspace{0.5mm} d\tau d\tau' \Bigg] \\
        = \hspace{0.5mm} &\int_{t_0}^{t_1} \Phi(t_1, \tau) \frac{\partial f}{\partial \omega_c}(\hat{x}(\tau), 0) \cdot K \cdot \left( \frac{\partial f}{\partial \omega_c}(\hat{x}(\tau), 0) \right)^\top \\
        &\hspace{5mm} \Phi(t_1, \tau)^\top \hspace{0.5mm} d\tau
    \end{align*}
    where we have used the fact that $\E[w_c(\tau) w_c(\tau')^\top] = K \cdot 
    \delta (\tau - \tau')$, with $\delta(\cdot)$ denoting the Dirac delta function. 
    
\end{itemize}


\subsubsection{EKF, Setup}
\label{subsubsec: App, EKF, Setup}

The Extended Kalman Filter (EKF), whose standard formulation is presented in Algorithm \ref{Alg: EKF}, is an iterative algorithm for updating estimates of the current pose $x_t$ (i.e. $n = 1$) and positions of all observed features at the current time, $f_t := (f_1, \cdots, f_{\nf}) \in \R^{\nf \df}$. This corresponds to the sliding window filter in our formulation, with $n = 1$ and $q = \nf$. Below, as an application of our optimization-based SLAM framework, we present the dynamics and measurement maps of the EKF algorithm in $\R^2$, as well as the associated cost functions. Dimension-wise, in its standard formulation, the 2D EKF is an instantiation of Algorithm \ref{Alg: EKF} with $\dx = 3$, $\df = 2$, and $\dz = 2$. To unify our notation, we will suppose that $\dx, \df, \dz$ assume these values throughout the rest of this section.

Let $x_t := (x_t^1, x_t^2, \theta_t) \in \R^{\dx}$ denote the \textit{robot pose}, comprising its position and angle in $\R^{\df}$, let $f_k := (f_{ k}^1, f_{ k}^2) \in \R^{\df}$ denote the position of each \textit{feature} $f_k \in \{f_1, \cdots, f_p\}$ visible at time $t$, and let $z_{t,k} := (z_{t, k}^1, z_{t, k}^2) \in \R^{\dz}$ denote the measurement of feature $f_k$ at time $t$. The dynamics map $g: \R^{\dx} \ra \R^{\dx}$, with $\dot{x}_t = g(x_t)$ is obtained by performing numerical integration on the continuous-time dynamics:
\begin{align*}
    \dot{x}_t^1 &= v \cos\theta + w_t^1, \\
    \dot{x}_t^2 &= v \sin\theta + w_t^2, \\
    \dot{\theta}_t &= \omega + w_t^3,
\end{align*}
where $w_t := (w_t^1, w_t^2, w_t^3) \in \R^{\dx}$ denotes additive zero-mean Gaussian noise on the $(x, y, \theta)$ coordinates of the state variable, respectively, with joint covariance $w_t \sim \mathcal{N}(0, \Sigma_w)$ for some covariance matrix $\Sigma_w \in \R^{\dx \times \dx}$, $\Sigma_w \succ 0$. For more details regarding the numerical integration process, see Section \ref{subsubsec: App, Continuous to Discrete Time Formulation} in the Appendix.

The measurement map $h: \R^{\dx} \times \R^{\df} \ra \R^{\dz}$ is given by:
\begin{align*}
    z_{t, k}^1 &= f_{ k}^1 - x_t^1 + v_t^1, \\
    z_{t, k}^2 &= f_{ k}^2 - x_t^2 + v_t^2,
\end{align*}
where $v_t := (v_t^1, v_t^2) \in \R^{\dz}$ denotes additive zero-mean Gaussian noise on the measurements $z_{t, j}^1$, $z_{t, j}^2 \in \R$, respectively, with joint covariance $v_t \sim \mathcal{N}(0, \Sigma_v)$ for some covariance matrix $\Sigma_v \in \R^{\dz \times \dz}$, $\Sigma_v \succ 0$. The measurement vector $z_t \in \R^{\nf \df}$ is then given by concatenating each of the $q$ residual measurements obtained at time $t$, i.e. $z_t := (z_{t, 1}, \cdots, z_{t, p}) \in \R^{\nf \dz}$.

\begin{algorithm}
    \label{Alg: EKF}
    {
    \small
    \SetAlgoLined
    \KwData{Prior distribution on $x_0 \in \R^{\dx}$: $\mathcal N(\mu_0, \Sigma_0)$, dynamics and measurement noise covariances $\Sigma_w \in \R^{\dx \times \dx}, \Sigma_v \in \R^{d_z \times d_z}$, (discrete-time) dynamics map $g: \R^{\dx} \ra \R^{\dx}$, measurement map $h: \R^{\dx} \times \R^{\nf \df} \ra \R^{d_z}$, time horizon $T \in \N$.}
    \KwResult{Estimates $\hat x_t$ for all desired timesteps $t \leq T$.}
    
    \vspace{2mm}
     \For{$t = 0, \cdots, T$}{
     
    \If{\textbf{\emph{detect new feature measurements}} $z_{t,\nf + 1:\nf + \nf'} := (z_{t,\nf + 1}, \cdots, z_{t,\nf + \nf'}) \in \R^{\nf'd_z}$}{
            $\mu_t, \Sigma_t, \nf \gets$ Alg. \ref{Alg: EKF, Feature Augmentation}, EKF feature augmentation $\big( \mu_t, \Sigma_t, \nf, z_{t,\nf + 1:\nf + \nf'}, h(\cdot) \big)$
            
        }
        
        $z_{t,1:\nf} := (z_{t,1}, \cdots, z_{t, \nf}) \in \R^{\nf d_z} \gets$ New measurements of existing features.
        
        $\overline{\mu_t}, \overline{\Sigma_t} \gets$ Alg. \ref{Alg: EKF, Feature Update}, EKF feature update $\big( \overline{\mu_t}, \overline{\Sigma_t}, z_{t,1:\nf}, h(\cdot) \big)$.
        
        
        \vspace{2mm}
        \If{$t < T$}{
            $\mu_{t+1}, \Sigma_{t+1} \gets$ Alg. \ref{Alg: EKF, State Propagation}, EKF state propagation $\big( \mu_t, \Sigma_t, g(\cdot) \big)$
        
        }
     }
    
    \Return{$\hat x_0, \cdots \hat x_T \in \R^{\dx}$.}
     
     \caption{Extended Kalman Filter SLAM, Standard Formulation.}
     }
\end{algorithm}

\begin{algorithm} \label{Alg: EKF, Feature Augmentation}
    {
    \normalsize
    \SetAlgoLined
    \KwData{Current EKF state $\tilde x_t \in \R^{\dx + \nf \df}$, with mean $\mu_t \in \R^{\dx + \nf \df}$ and covariance $\Sigma_t \in \R^{(\dx + \nf \df) \times (\dx + \nf \df)}$, current number of features $\nf$, observations of new features at current pose $z_{t,\nf+1:\nf+\nf'} := (z_{t, \nf+1}, \cdots, z_{t,\nf+\nf'}) \in \R^{\nf' \dz}$, measurement map $h: \R^{\dx} \times \R^{\df} \ra \R^{\dz}$, inverse measurement map $\ell: \R^{\dx} \times \R^{\dz} \ra \R^{\df}$.}
    
    \KwResult{Updated number of features $\nf$, updated EKF state mean $\mu_t \in \R^{\dx + \nf \df}$, covariance $\Sigma_t \in \R^{(\dx + \nf \df) \times (\dx + \nf \df)}$ (with $\nf$ already updated)}
    
        $(\mu_{t,x}, \mu_{t, f, 1:\nf}) \gets \mu_t \in \R^{\dx + \nf \df}$, with $\mu_{t,x} \in \R^{\dx}$, $\mu_{t, f, 1:\nf} \in \R^{\nf \df}$.
     
        $\ell: \R^{\dx} \times \R^{\dz} \ra \R^{\df} \gets$ Inverse measurement map, satisfying $z_{t,k} = h \big( x_t, \ell(x_t, z_{t,k}) \big)$ for each $x_t \in \R^{\dx}$, $z_{t,k} \in \R^{\dz}$, $\forall \hspace{0.5mm} k = \nf+1, \cdots, \nf+\nf'$.
        
        $\tilde\ell(\mu_{t,x}, z_{t,\nf+1}, \cdots, z_{t,\nf+\nf'}) \gets \big( \ell(\mu_{t,x}, z_{t,\nf+1}), \cdots, \ell(\mu_{t,x}, z_{t,\nf + \nf'}) \big) \in \R^{\nf' \df \times (\dx + \nf' \dz)}$
        
        $\mu_t \gets \big(\mu_t,  \tilde\ell(\mu_{t,x}, z_{t,\nf+1}, \cdots, z_{t,\nf+\nf'}) \big) \in \R^{\dx + (\nf+\nf')\df}$
        
        $\begin{bmatrix}
        \Sigma_{t,xx} & \Sigma_{t,xf} \\
        \Sigma_{t,fx} & \Sigma_{t,ff} 
        \end{bmatrix} \gets \Sigma_t \in \R^{(\dx + \nf \df) \times (\dx + \nf \df)}$, with $\Sigma_{t,xx} \in \R^{\dx \times \dx}$, $\Sigma_{t,xf} = \Sigma_{t,fx}^\top \in \R^{\dx \times \nf \df}$, $\Sigma_{t,ff} \in \R^{\nf \df \times \nf \df}$.
        
        $L_x \gets \frac{\partial \tilde \ell}{\partial x} \big|_{(\mu_t, z_t')} \in \R^{\nf'\df \times \dx}$
        
        $L_z \gets \frac{\partial \tilde \ell}{\partial z} \big|_{(\mu_t, z_t')} \in \R^{\nf'\df \times \nf'\dz}$
        
        $\tilde{\Sigma}_v \gets \text{diag}\{\Sigma_v, \cdots, \Sigma_v\} \in \R^{\nf'\dz \times \nf'\dz}$
        
        $\Sigma_t \gets \begin{bmatrix}
        \Sigma_{t,xx} & \Sigma_{t,xf} & \Sigma_{t,xx} L_x^\top \\
        \Sigma_{t,fx} & \Sigma_{t,ff} & \Sigma_{t,fx} L_x^\top \\
        L_x \Sigma_{t,xx} & L_x \Sigma_{t,xf} & L_x \Sigma_{t,xx} L_x^\top + L_z \tilde{\Sigma}_v L_z^\top
        \end{bmatrix} \in \R^{(\dx + (\nf+\nf')\df) \times (\dx + (\nf+\nf')\df)} $
        
        $\nf \gets \nf + \nf'$
    
    \Return{$\mu_t \in \R^{\dx + \nf \df}, \Sigma_t \in \R^{(\dx + \nf \df) \times (\dx + \nf \df)}, \nf \geq 0$}
     
     \caption{Extended Kalman Filter, Feature Augmentation Sub-block.}
     }
\end{algorithm}

\begin{algorithm} \label{Alg: EKF, Feature Update}
    {
    \normalsize
    \SetAlgoLined
    \KwData{Current EKF state $\tilde x_t \in \R^{\dx + \nf \df}$, with mean $\mu_t \in \R^{\dx + \nf \df}$ and covariance $\Sigma_t \in \R^{(\dx + \nf \df) \times (\dx + \nf \df)}$, new measurements of existing features $z_{t,1:\nf} := (z_{t,1}, \cdots, z_{t,\nf}) \in \R^{\nf \dz}$, measurement map $h: \R^{\dx} \times \R^{\df} \ra \R^{\dz}$}
    
    \KwResult{Updated EKF state mean $\mu_t \in \R^{\dx + \nf \df}$ and covariance $\Sigma_t \in \R^{(\dx + \nf \df) \times (\dx + \nf \df)}$}
    
    $f_{1:\nf} \gets (f_1, \cdots, f_{\nf}) \in \R^{\nf \df}$.
    
    $\Tilde{h}(x_t, f_{1:\nf}) \gets \big( h(x_t, f_1), \cdots, h(x_t, f_{\nf}) \big) \in \R^{\nf \dz}$
        
        $H_t \gets \frac{\partial \tilde{h}}{\partial (x_t, f_{1:\nf})} \Big|_{\mu_t}$  Jacobian of $\Tilde{h}: \R^{\dx} \times \R^{\nf \df} \ra \R^{\nf \dz}$ evaluated at $\mu_t \in \R^{\dx + \nf \df}$.
        
        $\Tilde{\Sigma}_v \gets \text{diag}\{\Sigma_v, \cdots, \Sigma_v \} \in \R^{\nf \dz \times \nf \dz}.$
        
        \label{Eqn: EKF, Mean Update}
        $\overline{\mu_t} \gets \mu_t + \Sigma_t H_t^T (H_t \Sigma_t H_t^T + \Tilde{\Sigma}_v)^{-1} \big(z_{t,1:\nf} - \tilde{h}(\mu_t) \big) \in \R^{\dx + \nf \df}$.
        
        \label{Eqn: EKF, Cov Update}
        $\overline{\Sigma}_t \gets \Sigma_t - \Sigma_t H_t^T (H_t \Sigma_t H_t^T + \Tilde{\Sigma}_v)^{-1} H_t \Sigma_t \in \R^{(\dx + \nf \df) \times (\dx + \nf \df)}$.
        
    
    \Return{$\overline{\mu}_t \in \R^{\dx + \nf \df}, \overline{\Sigma}_t \in \R^{(\dx + \nf \df) \times (\dx + \nf \df)}$.}
     
     \caption{Extended Kalman Filter, Feature Update Sub-block.}
     }
\end{algorithm}

\begin{algorithm} \label{Alg: EKF, State Propagation}
    {
    \normalsize
    \SetAlgoLined
    \KwData{Current EKF state $\tilde x_t \in \R^{\dx + \nf \df}$, with mean $\overline{\mu_t} \in \R^{\dx + \nf \df}$ and covariance $\overline{\Sigma_t} \in \R^{(\dx + \nf \df) \times (\dx + \nf \df)}$, (discrete-time) dynamics map $g: \R^{\dx} \ra \R^{\dx}$}
    
    \KwResult{Propagated EKF state mean $\mu_{t+1} \in \R^{\dx + \nf \df}$ and covariance $\Sigma_{t+1} \in \R^{(\dx + \nf \df) \times (\dx + \nf \df)}$}
    
    $(\overline{\mu}_{t,x}, \overline{\mu}_{t,f, 1:\nf}) \gets \overline{\mu_t}$, with $\overline{\mu}_{t,x} \in \R^{\dx}$, $\overline{\mu}_{t,f,1:\nf} \in \R^{\nf \df}$.
            
    $\begin{bmatrix}
        \overline{\Sigma}_{t,xx} & \overline{\Sigma}_{t,xf} \\
        \overline{\Sigma}_{t,fx} &
        \overline{\Sigma}_{t,ff}
    \end{bmatrix} \gets \overline{\Sigma}_t \in \R^{\dx \times \dx}$, with $\overline{\Sigma}_{t,xx} \in \R^{\dx \times \dx}, \overline{\Sigma}_{t,xf} = \overline{\Sigma}_{t,fx}^\top \in \R^{\dx \times \nf \df}$, $\overline{\Sigma}_{t,ff} \in \R^{\nf \df \times \nf \df}$.
    

    $G_t \gets \frac{\partial g}{\partial x} \Big|_{\overline{\mu}_{t,x}}$.
    
    $\mu_{t+1} \gets \big( g(\overline{\mu_{t,x}}), \overline{\mu}_{t,f,1:\nf} \big) \in \R^{\dx + \nf \df}.$   \label{Eqn: EKF, Mean Propagation}
    
    $\Sigma_{t+1} \gets \begin{bmatrix}
        G_t \overline{\Sigma}_{t,xx} G_t^\top + \Sigma_w & G_t \overline{\Sigma}_{t,xf} \\
        \overline{\Sigma}_{t,fx} G_t^\top & \overline{\Sigma}_{t,ff}
    \end{bmatrix} \in \R^{(\dx + \nf \df) \times (\dx + \nf \df)}.$ \label{Eqn: EKF, Cov Propagation}
    
    \Return{$\mu_{t+1} \in \R^{\dx + \nf \df}, \Sigma_{t+1} \in \R^{(\dx + \nf \df) \times (\dx + \nf \df)}$.}
     
     \caption{Extended Kalman Filter, State Propagation Sub-block.}
     }
\end{algorithm}

\subsubsection{Proofs from Section \ref{subsec: EKF}}
\label{subsubsec: App, EKF}

\begin{theorem} 
The feature augmentation step of the standard EKF SLAM algorithm (Alg. \ref{Alg: EKF, Feature Augmentation}) is equivalent to applying a Gauss-Newton step to $c_{EKF,t,1}: \R^{\dx + \nf \df} \ra \R$, given by:
\begin{align*}
    &c_{EKF,t,1}(\tilde{x}_t, f_{\nf+1}, \cdots, f_{\nf+\nf'}) \\
    = \hspace{0.5mm} &\Vert \tilde{x}_t - \mu_t \Vert_{\Sigma_t^{-1}}^2 + \sum_{k= \nf+1}^{\nf+\nf'} \Vert  z_{t,k} - h(x_t, f_k) \Vert_{\tilde{\Sigma}_v^{-1}}^2.
\end{align*}
\end{theorem}

\begin{proof}

To simplify the analysis below, we assume all degrees of freedom of new features are observed.
    More specifically, we assume the existence of an \textit{inverse observation map} $\ell: \R^{\dx} \times \R^{\dz} \ra \R^{\df}$, satisfying $h(x_t, \ell(x_t, z_t)) = z_t$ for each $x_t \in \R^{\dx}, z_t \in \R^{\dz}$, which directly generates position estimates of new features from their feature measurements and the current pose, by effectively \say{inverting} the measurement map $h: \R^{\dx} \times \R^{\df} \ra \R^{\dz}$ \cite{Sola2014SLAMWithEKF}. When full observations are unattainable, the missing degrees of freedom are introduced as a prior to the system \cite{Sola2014SLAMWithEKF}; in this case, similar results follow.

First, to simplify notation, define:
\begin{align*}
    z_{t,\nf+1:\nf+\nf'} &= (z_{t,\nf+1}, \cdots, z_{t,\nf+\nf'}) \in \R^{\nf' \dz}, \\
    f_{\nf+1:\nf+\nf'} &= (f_{\nf+1}, \cdots, f_{\nf+\nf'}) \in \R^{\nf' \df}, \\
    \tilde{h}(x_t, f_{\nf+1:\nf+\nf'}) &:= \big(h(x_t, f_{\nf+1}), \cdots, h(x_t, f_{\nf+\nf'}) \big) \\
    &\hspace{1cm} \in \R^{\nf' \dz}, \\
    \tilde{\Sigma}_v &= \text{diag}\{\Sigma_v, \cdots, \Sigma_v \} \in \R^{\nf'\dz \times \nf'\dz}.
\end{align*}
We can now rewrite the cost $c_{EKF,t,1}$ as:
\begin{align*}
    &c_{EKF,t,1}(\tilde{x}_t, f_{\nf+1:\nf+\nf'}) \\
    = \hspace{0.5mm} &\Vert \tilde{x}_t - \mu_t \Vert_{\Sigma_t^{-1}}^2 + \Vert z_{t,\nf+1:\nf+\nf'} - \tilde{h}(x_t, f_{\nf+1:\nf+\nf'}) \Vert_{\tilde{\Sigma}_v^{-1}}^2.
\end{align*}

To apply a Gauss-Newton step, our first task is to find a vector $C_1(\tilde{x}_t, f_{\nf+1:\nf+\nf'})$ of an appropriate dimension such that $c_{EKF,t,1}(\tilde{x}_t, f_{\nf+1:\nf+\nf'}) = C_1(\tilde{x}_t, f_{\nf+1:\nf+\nf'})^\top C_1(\tilde{x}_t, f_{\nf+1:\nf+\nf'})$. A natural choice is furnished by $C_1(\tilde{x}_t, f_{\nf+1:\nf+\nf'}) \in \R^{\dx + \nf \df + \nf' \dz}$, as defined below:
\begin{align*}
    &C_1(\tilde{x}_t, f_{\nf+1:\nf+\nf'}) \\
    := \hspace{0.5mm} &\begin{bmatrix}
        \Sigma_t^{-1/2} (\tilde{x}_t - \mu_t) \\
        \Sigma_v^{-1/2} \big( z_{t,\nf+1:\nf+\nf'} - \tilde{h}(x_t, f_{\nf+1:\nf+\nf'}) \big)
    \end{bmatrix}.
\end{align*}
Thus, our parameters for the Gauss-Newton algorithm submodule are:
\begin{align*}
    \tilde{x}_t^\star &:= (x_t^\star, f_{1:\nf}^\star, f_{\nf+1:\nf+\nf'}^\star) \\
    &= \big( \overline{\mu_t}, \ell(x_t^\star, z_{t,\nf+1}), \cdots, \ell(x_t^\star, z_{t,\nf+\nf'}) \big) \in \R^{\dx + (\nf+\nf') \df}, \\
    &\hspace{5mm} \text{where } x_t^\star \in \R^{\dx}, f_{1:\nf}^\star \in \R^{\nf \df}, f_{\nf+1:\nf+\nf'}^\star \in \R^{\nf' \df}, \\
    C_1(\tilde{x}_t^\star) &= \begin{bmatrix}
        \Sigma_t^{-1/2} (\tilde{x}_t^\star - \mu_t) \\
        \tilde{\Sigma_v}^{-1/2} \big( z_{t,\nf+1:\nf+\nf'} - \tilde{h}(x_t^\star, f_{\nf+1:\nf+\nf'}^\star) \big)
    \end{bmatrix} \\
    &= \begin{bmatrix}
    0 \\ 0
    \end{bmatrix} \in \R^{\dx + \nf \df + \nf' \dz}, \\
    J &= \begin{bmatrix}
        \Sigma_t^{-1/2} & O \\ - \tilde{\Sigma}_v^{-1/2} \tilde{H}_{t,x} \begin{bmatrix}
        I & O
        \end{bmatrix} & - \tilde{\Sigma}_v^{-1/2} \tilde{H}_{t,f}
    \end{bmatrix} \\
    &\in \R^{(\dx + \nf \df + \nf' \dz) \times (\dx + (\nf+\nf')\df)},
\end{align*}
where $\tilde{H}_t := \begin{bmatrix}
\tilde{H}_{t,x} & \tilde{H}_{t,f}
\end{bmatrix} \in \R^{\nf'\dz \times (\dx + \nf' \df)}$ is defined as the Jacobian of $\tilde{h}: \R^{\dx} \times \R^{\nf'\df} \ra \R^{\nf' \dz}$ at $(x_t^\star, f_{\nf+1:\nf+\nf'}^\star) \in \R^{\dx + \nf' \df}$, with $\tilde{H}_{t,x} \in \R^{\nf'\dz \times \dx}$ and $\tilde{H}_{t,f} \in \R^{\nf'\dz \times \nf \df}$. By Algorithm \ref{Alg: gauss-newton}, the Gauss-Newton update is thus given by:
{\small \begin{align} \nonumber
        &\hspace{5mm} \Sigma_t \\
        &\leftarrow (J^\top J)^{\dagger} \\ \nonumber
        &= \Bigg(\begin{bmatrix}
            \Sigma_t^{-1/2} & - \begin{bmatrix}
            I \\ O
            \end{bmatrix} \tilde{H}_{t,x}^\top \tilde{\Sigma}_v^{-1/2}  \\ O & - \tilde{\Sigma}_v^{-1/2} \tilde{H}_{t,f}
        \end{bmatrix} \\
        &\hspace{5mm}
        \begin{bmatrix}
            \Sigma_t^{-1/2} & O \\ - \tilde{\Sigma}_v^{-1/2} \tilde{H}_{t,x} \begin{bmatrix}
            I & O
            \end{bmatrix} & - \tilde{\Sigma}_v^{-1/2} \tilde{H}_{t,f}
        \end{bmatrix} \Bigg)^{\dagger} \\ \nonumber
        &= \begin{bmatrix}
            \Sigma_t^{-1} + \begin{bmatrix}
                I \\ O
            \end{bmatrix} \tilde{H}_{t,x}^\top \tilde{\Sigma}_v^{-1} \tilde{H}_{t,x} \begin{bmatrix}
                I & O
            \end{bmatrix} & \begin{bmatrix}
                I \\ O
            \end{bmatrix} \tilde{H}_{t,x}^\top \tilde{\Sigma}_v^{-1/2} \tilde{H}_{t,f} \\
            \tilde{H}_{t,f}^\top \tilde{\Sigma}_v^{-1} \tilde{H}_{t,x} \begin{bmatrix}
                I & O
            \end{bmatrix} & \tilde{H}_{t,f}^\top \tilde{\Sigma}_v^{-1} \tilde{H}_{t,f}
        \end{bmatrix}^{\dagger} \\ 
        \label{Eqn: EKF SLAM Proof, Feature Aug, J T J inverse}
        &= \begin{bmatrix}
            \Omega_{t,xx} + \tilde{H}_{t,x}^\top \tilde{\Sigma}_v^{-1} \tilde{H}_{t,x} & \Omega_{t,xf} & \tilde{H}_{t,x}^\top \tilde{\Sigma}_v^{-1} \tilde{H}_{t,f} \\
            \Omega_{t, fx} & \Omega_{t, ff} & O \\
            \tilde{H}_{t,f}^\top \tilde{\Sigma}_v^{-1} \tilde{H}_{t,x} & O & \tilde{H}_{t,f}^\top \tilde{\Sigma}_v^{-1} \tilde{H}_{t,f}
        \end{bmatrix}^{\dagger}, \\ \nonumber
        \overline{\mu_t} &\gets \tilde{x}_t^\star - (J^\top J)^{\dagger} J^\top C_1(\tilde{x}_t^\star) \\ \nonumber
        &= \big( \overline{\mu_t}, \ell(x_t^\star, z_{t,\nf+1}), \cdots, \ell(x_t^\star, z_{t,\nf+\nf'}) \big),
    \end{align}}
    where $\dagger$ denotes the Moore-Penrose pseudoinverse.
    
    Here, we have defined $\Omega_{t,xx} \in \R^{\dx \times \dx}, \Omega_{t,xf} = \Omega_{t,fx}^\top \in \R^{\dx \times \nf \df}$ and $\Omega_{t,ff} \in \R^{\nf \df \times \nf \df}$ by:
    \begin{align} 
    \label{Eqn: EKF SLAM proof, Omega def}
         \begin{bmatrix}
             \Omega_{t,xx}  & \Omega_{t,xf} \\
            \Omega_{t, fx} & \Omega_{t, ff}
         \end{bmatrix} := \begin{bmatrix}
             \Sigma_{t,xx}  & \Sigma_{t,xf} \\
            \Sigma_{t, fx} & \Sigma_{t, ff}
         \end{bmatrix}^{-1}
    \end{align}
    
    To conclude the proof, we must show that \eqref{Eqn: EKF SLAM Proof, Feature Aug, J T J inverse} is identical to the update equations for covariance matrix in the standard formulation of the Extended Kalman Filter algorithm, i.e., we must show that:
    \begin{align*}
        &\begin{bmatrix}
        \Sigma_{t,xx} & \Sigma_{t,xf} & \Sigma_{t,xx} L_x^\top \\
        \Sigma_{t,fx} & \Sigma_{t,ff} & \Sigma_{t,fx} L_x^\top \\
        L_x \Sigma_{t,xx} & L_x \Sigma_{t,xf} & L_x \Sigma_{t,xx} L_x^\top + L_z \Sigma_v L_z^\top
        \end{bmatrix} \\
        &\hspace{1cm} = \begin{bmatrix}
            \Omega_{t,xx} + \tilde{H}_{t,x}^\top \tilde{\Sigma}_v^{-1} \tilde{H}_{t,x} & \Omega_{t,xf} & \tilde{H}_{t,x}^\top \tilde{\Sigma}_v^{-1} \tilde{H}_{t,f} \\
            \Omega_{t, fx} & \Omega_{t, ff} & O \\
            \tilde{H}_{t,f}^\top \tilde{\Sigma}_v^{-1} \tilde{H}_{t,x} & O & \tilde{H}_{t,f}^\top \tilde{\Sigma}_v^{-1} \tilde{H}_{t,f}
        \end{bmatrix}^{\dagger}
    \end{align*}
    This follows by applying \eqref{Eqn: EKF SLAM proof, Omega def}, as well as the matrix equalities resulting from taking the derivative of the equation $z_t := h \big(x_t, \ell(x_t, z_t) \big)$ with respect to $x_t \in \R^{\dx}$ and $z_t \in \R^{\dz}$, respectively:
    \begin{align*}
        I &= \tilde{H}_{t,f} L_z, \\
        O &= \tilde{H}_{t,x} + \tilde H_{t,f} L_x. 
    \end{align*}
\end{proof} 

\begin{theorem} 
The feature update step of the standard EKF SLAM algorithm (Alg. \ref{Alg: EKF, Feature Update}) is equivalent to applying a Gauss-Newton step on $c_{EKF,t,2}: \R^{\dx + \nf \df} \ra \R$, given by:
    \begin{align*}
        &c_{EKF,t,2}(\tilde{x}_t) \\
        := \hspace{0.5mm} &\Vert \tilde{x}_t-\mu_t \Vert_{\Sigma_t^{-1}}^2 + \sum_{k=1}^{\nf} \Vert z_{t,k}-h(x_t, f_k) \Vert_{\Sigma_v^{-1}}^2.
    \end{align*}
\end{theorem}

\begin{proof}
First, to simplify notation, define:
\begin{align*}
    z_{t,1:\nf} &:= (z_{t,1}, \cdots, z_{t,\nf}) \in \R^{\nf \dz}, \\
    f_{1:\nf} &:= (f_1, \cdots, f_{\nf}) \in \R^{\nf \df}, \\
    \Tilde{h}(x_t, f_{1:\nf}) &:= \big( h(x_t, f_1), \cdots, h(x_t, f_{\nf}) \big) \in \R^{\nf \dz}, \\
    \Tilde{\Sigma}_v &:= \text{diag}\{\Sigma_v, \cdots, \Sigma_v \} \in \R^{\nf \dz \times \nf \dz}.
\end{align*}
We can then rewrite the cost as:
\begin{align*}
    c_{EKF,t,2}(\tilde{x}_t) = 
    \Vert \tilde{x}_t^\star - \mu_t \Vert_{\Sigma_t^{-1}}^2 + \Vert z_{t,1:\nf} - \tilde{h}(\tilde{x}_t^\star)  \Vert_{\tilde{\Sigma}_v^{-1}}^2.
\end{align*}

To apply a Gauss-Newton step, our first task is to find a vector $C_2(\tilde{x}_t)$ of an appropriate dimension such that $c_{EKF,t,2}(\tilde{x}_t) = C_2(\tilde{x}_t)^\top C_2(\tilde{x}_t)$. A natural choice is furnished by $C_2(\tilde{x}_t) \in \R^{\dx + \nf \df + \nf \dz}$, as defined below:
\begin{align*}
    C_2(\tilde{x}_t) := \begin{bmatrix} 
    \Sigma_t^{-1/2}(\tilde{x}_t - \mu_t) \\ \tilde{\Sigma}_v^{-1/2}(z_{t,1:\nf} - \Tilde{h}(\tilde{x}_t)) \end{bmatrix}.
\end{align*}
Thus, our parameters for the Gauss-Newton algorithm submodule are:
\begin{align*}
    \tilde{x}_t^\star &= \mu_t \in \R^{\dx + \nf \df}, \\
    C_2(\tilde{x}_t^\star) &= 
    \begin{bmatrix} 
    \Sigma_
    t^{-1/2}(\tilde{x}_t^\star - \mu_t) \\ \tilde{\Sigma}_v^{-1/2}(z_{t,1:\nf} - \Tilde{h}(\tilde{x}_t^\star)) \end{bmatrix} = \begin{bmatrix} 
    0 \\ \tilde{\Sigma}_v^{-1/2}(z_{t,1:\nf} - \tilde{h}(\mu_t) \end{bmatrix} \\
    &\in \R^{\dx + \nf \df + \nf \dz}, \\
    J &= \begin{bmatrix}
        \Sigma_t^{-1/2} \\
        -\tilde{\Sigma}_v^{-1/2} H_t 
    \end{bmatrix} \in \R^{(\dx + \nf \df + \nf \dz) \times (\dx + \nf \df)},
\end{align*}
where $\tilde{H}_t \in \R^{\nf \dz} \times \R^{\dx + \nf \df}$ is defined as the Jacobian of $\Tilde{h}: \R^{\dx} \times \R^{\nf \df} \ra \R^{\nf \dz}$ at $\tilde{x}_t^\star \in \R^{\dx + \nf \df}$. By Algorithm \ref{Alg: gauss-newton}, the Gauss-Newton update is thus given by:
    \begin{align*}
        \overline{\Sigma}_t &\leftarrow (J^\top J)^{-1} \\
        &= (\Sigma_t^{-1} + H_t^\top \tilde \Sigma_v^{-1} H_t)^{-1} \\
        &= \Sigma_t - \Sigma_t H_t^\top (\tilde \Sigma_v + H_t \Sigma_t H^\top)^{-1} H_t \Sigma_t, \\
        \overline{\mu_t} &\gets \mu_t - (J^\top J)^{-1} J^\top C_2(\tilde{x}_t^\star) \\
        &= \mu_t - (\Sigma^{-1}_t + H_t^\top \tilde \Sigma_v^{-1} H_t)^{-1} \begin{bmatrix}
        \Sigma_t^{-1/2} & - H_t^\top \tilde \Sigma_v^{-1/2}
        \end{bmatrix} \\
        &\hspace{2cm}
        \begin{bmatrix}
        0 \\ \Sigma_v^{-1/2}(z_{t,1:\nf} - \tilde{h}(\mu_t))
        \end{bmatrix} \\
        &= \mu_t + (\Sigma^{-1}_t + H_t^\top \tilde \Sigma_v^{-1} H_t)^{-1} H_t^\top \tilde \Sigma_v^{-1} \big( z_{t,1:\nf} - \tilde{h}(\mu_t) \big),\\
        &= \mu_t + \tilde \Sigma_v^{-1} H_t^\top (H_t \Sigma_t H_t^\top + \tilde \Sigma_v)^{-1} \big( z_{t,1:\nf} - \tilde{h}(\mu_t) \big),
    \end{align*}
    which are identical to the feature update equations for the mean and covariance matrix in the Extended Kalman Filter algorithm, i.e. \eqref{Eqn: EKF, Mean Update} and \eqref{Eqn: EKF, Cov Update} respectively. Note that, in the final step, we have used a variant of the Woodbury Matrix Identity.
\end{proof}

\begin{theorem} 
The state propagation step of the standard EKF SLAM algorithm (Alg. \ref{Alg: EKF, State Propagation}) is equivalent to applying a Marginalization step to $c_{EKF,t,4}: \R^{2 \dx + \nf \df} \ra \R$, given by:
    \begin{align*}
        &c_{EKF,t,4}(\tilde{x}_t, x_{t+1}) \\
        := \hspace{0.5mm} &\Vert \tilde{x}_t - \overline{\mu}_t \Vert_{\overline \Sigma_t^{-1}}^2 + \Vert x_{t+1}-g(x_t) \Vert_{\Sigma_w^{-1}}^2.
    \end{align*}
\end{theorem}

\begin{proof}
Intuitively, the state propagation step marginalizes out $x_t \in \R^{\dx}$ and retain $x_{t+1} \in \R^{\dx}$. In other words, in the notation of our Marginalization algorithm submodule, we have:
\begin{align*}
    \tilde{x}_{t,K} &= \tilde x_{t+1} \in \R^{\dx + \nf \df}, \\
    \tilde{x}_{t,M} &= x_t \in \R^{\dx + \nf \df}.
\end{align*}
To apply a marginalization step, our first task is to find vectors $C_{K}(x_K) = C_{K}(\tilde{x}_t)$ and $C_{M}(x_K, x_M) = C_{M}(\tilde{x}_t, x_{t+1})$ of appropriate dimensions such that $c_{EKF,t,4}(\tilde{x}_t, x_{t+1}) = C_{K}(x_{t+1})^\top C_{K}(x_{t+1}) + C_{M}(\tilde{x}_t, x_{t+1})^\top C_{M}(\tilde{x}_t, x_{t+1})$. A natural choice is furnished by $C_{K}(x_{t+1}) \in \R$ and $C_{M}(\tilde{x}_t, x_{t+1}) \in \R^{\dx}$, as defined below:
\begin{align*}
    c_{K}(x_{t+1}) &= 0 \\
    c_{M}(\tilde{x}_t, x_{t+1}) &= \Vert \tilde{x}_t - \overline{\mu_t} \Vert_{\overline{\Sigma}_t^{-1}}^2 + \Vert x_{t+1}-g(x_t) \Vert_{\Sigma_w^{-1}}^2.
\end{align*}
    where we have identified the following parameters, in the language of a Marginalization step (Section \ref{sec: Problem Formulation on Euclidean Spaces}):
    \begin{align*}
        C_{K}(\tilde{x}_{t,K})&=0 \in \R \\
        C_{M}(\tilde{x}_{t,K},\tilde{x}_{t,M}) &= \begin{bmatrix} \bar\Sigma_t^{-1/2}(\tilde{x}_t - \overline{\mu_t}) \\ \Sigma_w^{-1/2} \big(x_{t+1} - g(x_t) \big) \end{bmatrix} \in \R^{2\dx + \nf \df}.
    \end{align*}
    For convenience, we will define the pose and feature track components of the mean $\mu_t \in \R^{\dx + \nf \df}$ by $\mu_t := (\mu_{t,x}, \mu_{t,f}) \in \R^{\dx + \nf \df}$, with $\mu_{t,x} \in \R^{\dx}$ and $\mu_{t,f} \in \R^{\nf \df}$, respectively. This mirrors our definition of $x_t \in \R^{\dx}$ and $f_{1:\nf} \in \R^{\nf \df}$ as the components of the full state $\tilde{x}_t := (x_t, f_{1:\nf}) \in \R^{\dx + \nf \df}$. In addition, we will define the components of $\bar\Sigma_t^{-1/2} \in \R^{(\dx + \nf \df) \times (\dx + \nf \df)}$ and $\bar\Sigma_t^{-1} \in \R^{(\dx + \nf \df) \times (\dx + \nf \df)}$ by:
    \begin{align*}
       \begin{bmatrix}
            \Omega_{t,xx} & \Omega_{t,xf} \\
            \Omega_{t,fx} & \Omega_{t,ff}
        \end{bmatrix} &:=  \bar\Sigma_t^{-1} \in \R^{(\dx + \nf \df) \times (\dx + \nf \df)}, \\
        \begin{bmatrix}
            \Lambda_{t,xx} & \Lambda_{t,xf} \\
            \Lambda_{t,fx} & \Lambda_{t,ff}
        \end{bmatrix} &:= \bar\Sigma_t^{-1/2} \in \R^{(\dx + \nf \df) \times (\dx + \nf \df)},
    \end{align*}
    where $\Sigma_{t,xx}, \Lambda_{t,xx} \in \R^{\dx \times \dx}$, $\Sigma_{t,xf}, \Lambda_{t,xf} \in \R^{\dx \times \nf \df}$, $\Sigma_{t,fx}, \Lambda_{t,fx} \in \R^{\nf \df \times \dx}$, and $\Sigma_{t,ff}, \Lambda_{t,ff} \in \R^{\nf \df \times \nf \df}$. Using the above definitions, we can reorder the residuals in $C_K \in \R$ and $C_M \in \R^{2\dx+\nf \df}$, and thus redefine them by:
    \begin{align*}
        &C_{K}(\tilde{x}_{t,K}) =0 \in \R \\
        &C_{M}(\tilde{x}_{t,K},\tilde{x}_{t,M}) \\
        &= \begin{bmatrix} 
            \Lambda_{t,xx}(x_t - \mu_{t,x}) + \Lambda_{t,xf}(f_{1:\nf} - \mu_{t,f}) \\
            \Sigma_w^{-1/2}(x_{t+1} - g(x_t)) \\
            \Lambda_{t,fx}(x_t - \mu_{t,x}) + \Lambda_{t,ff}(f_{1:\nf} - \mu_{t,f})
        \end{bmatrix} \in \R^{2\dx + \nf \df}.
    \end{align*}
    Our state variables and cost functions for the Gauss-Newton algorithm submodule are:
    \begin{align*}
        \overline{x_M^\star} &= \tilde{x}_t^\star = \overline{\mu_t} \in \R^{\dx + \nf \df}, \\
        \overline{x_K^\star} &= g(\tilde{x}_t^\star) = g(\overline{\mu_t}) \in \R^{\dx + \nf \df}, \\
        C_{K}(\tilde{x}_{t,K}^\star) &= 0 \in \R, \\
        C_{M}(\tilde{x}_{t,K}^\star,\tilde{x}_{t,M}^\star) &= 
        \begin{bmatrix}
        0 \\ 0
        \end{bmatrix} \in \R^{2 \dx + \nf \df}, \\
        J_K &= \begin{bmatrix} 
            O & \Lambda_{xf} \\ \Sigma_w^{-1/2} & O \\
            O & \Lambda_{ff}
        \end{bmatrix} \in \R^{(2 \dx + \nf \df) \times (\dx + \nf \df)} \\
        J_M &= \begin{bmatrix}
            \Lambda_{xx} \\ -\Sigma_w^{-1/2} G_t \\
            \Lambda_{xf}
        \end{bmatrix} \in \R^{(2 \dx + \nf \df) \times \dx},
    \end{align*}
    where we have defined $G_t$ to be the Jacobian of $g: \R^{\dx} \ra \R^{\dx}$ at $\overline{\mu_{t,x}} \in \R^{\dx}$, i.e.:
    \begin{align*}
        G_t := \frac{\partial g}{\partial x_t} \Bigg|_{x_t = \overline{\mu_{t,x}}}
    \end{align*}
    Applying the Marginalization equations, we thus have:
    \begin{align*}
    \mu_{t+1} &\gets \tilde{x}_{t,K} - \Sigma_{t+1} J_K^\top \big[ I - J_M(J_M^\top J_M)^{-1} J_M^\top \big] \\
    &\hspace{1cm} C_{M}(\overline{x_K^\star}, \overline{x_M^\star}) \\
    &= g(\overline{\mu_t}), \\
    \Sigma_{t+1} &\gets \big(J_K^\top \big[ I - J_M(J_M^\top J_M)^{-1} J_M^\top \big] J_K\big)^{-1}, \\ 
    &= \big(J_K^\top J_K - J_K^\top J_M(J_M^\top J_M)^{-1} J_M^\top J_K\big)^{-1}, \\ 
    &= \Bigg(\begin{bmatrix}
    \Sigma_w^{-1} & O \\
    O & \Lambda_{fx} \Lambda_{xf} + \Lambda_{ff}^2
    \end{bmatrix} 
    - \begin{bmatrix}
    - \Sigma_w^{-1} G_t \\
    \Lambda_{fx} \Lambda_{xx} + \Lambda_{ff} \Lambda_{fx}
    \end{bmatrix} \\
    &\hspace{1cm} (\Lambda_{xx}^2 + \Lambda_{xf} \Lambda_{fx} + G_t^\top \Sigma_w^{-1} G_t)^{-1} \\
    &\hspace{1cm} \cdot \begin{bmatrix}
    -G_t^\top \Sigma_w^{-1} & \Lambda_{xx} \Lambda_{xf} + \Lambda_{fx} \Lambda_{ff}
    \end{bmatrix} \Bigg)^{-1} \\
    &= \Bigg(\begin{bmatrix}
            \Sigma_w^{-1} & O \\
            O & \Omega_{ff}
            \end{bmatrix} 
            - \begin{bmatrix}
            - \Sigma_w^{-1} G_t \\
            \Omega_{fx}
            \end{bmatrix} \\ 
            &\hspace{1cm}(\Omega_{xx} + G_t^\top \Sigma_w^{-1} G_t)^{-1} \begin{bmatrix}
            -G_t^\top \Sigma_w^{-1} & \Omega_{xf}
        \end{bmatrix} \Bigg)^{-1}
    \end{align*}
    To show that this is indeed identical to the propagation equation for the covariance matrix in the Extended Kalman Filter algorithm, i.e. Algorithm \ref{Alg: EKF}, Line \ref{Eqn: EKF, Cov Propagation}, we must show that:
    \begin{align*}
        &\Bigg(\begin{bmatrix}
            \Sigma_w^{-1} & O \\
            O & \Omega_{ff}
            \end{bmatrix} 
            - \begin{bmatrix}
            - \Sigma_w^{-1} G_t \\
            \Omega_{fx}
            \end{bmatrix} (\Omega_{xx} + G_t^\top \Sigma_w^{-1} G_t)^{-1} \\
            &\hspace{1cm}
            \begin{bmatrix}
            -G_t^\top \Sigma_w^{-1} & \Omega_{xf}
        \end{bmatrix} \Bigg)^{-1} \\
        = \hspace{0.5mm} & \begin{bmatrix}
                G_t \overline{\Sigma}_{t,xx} G_t^\top + \Sigma_w & G_t \overline{\Sigma}_{t,xf} \\
                \overline{\Sigma}_{t,xf} G_t^\top & \overline{\Sigma}_{t,ff}
            \end{bmatrix}
    \end{align*}
    This follows by brute-force expanding the above block matrix components, and applying Woodbury's Matrix Identity, along with the definitions of $\Sigma_{t,xx}, \Lambda_{t,xx}$, $\Sigma_{t,xf}, \Lambda_{t,xf}$, $\Sigma_{t,fx}, \Lambda_{t,fx}$, $\Sigma_{t,ff}$, and $\Lambda_{t,ff}$.
\end{proof}

\subsubsection{MSCKF, Setup}
\label{subsubsec: App, MSCKF, Setup}

The Multi-State Constrained Kalman Filter (MSCKF) algorithm iteratively refines the mean and covariance of a MSCKF full state, consisting of the most recent IMU state and a sliding window of $n$ poses. 
(Algorithm \ref{Alg: MSCKF})
In particular, when a set of new IMU measurements is obtained, the MSCKF full state is propagated forward in time. When a new image measurement arrives, the current pose is appended to the MSCKF full state vector. Features not observed in the current pose are marginalized. If the number of poses maintained in the MSCKF full state, denoted $n$, exceeds a pre-specified upper bound $N_{\max}$, then features common to every third currently maintained pose, starting from the second oldest pose, are marginalized. Below, we discuss the key components of the MSCKF algorithm---the IMU state, poses, MSCKF full states,  features, image measurements, dynamics, pose augmentation and measurement maps---in more detail.

\begin{algorithm} \label{Alg: MSCKF}
    {
    \small
    \SetAlgoLined
    \KwData{Prior distribution on $x_{\IMU,0} \in \X_p$: $\mathcal N(\mu_0, \Sigma_0)$, dynamics and measurement noise covariances $\Sigma_w \in \R^{\dx \times \dx}$, $\Sigma_v \in \R^{\dz \times d_z}$, discrete-time dynamics map $g_\IMU: \R^{d_\IMU} \times \R^{d_\IMU}$, measurement map $h: \X_p \times \R^{\df} \ra \R^{d_z}$, time horizon $T$, pose transformation $\psi: \X_\IMU \times (\X_p)^{\nx} \times \X_p \ra \X_p$ (IMU $\ra$ global).}
    
    \KwResult{Estimates $\hat x_t \in \X_\IMU \times (\X_p)^{\nx}$ for all desired timesteps $t \leq T$, where $\nx:=$ number of poses in $\hat{x}_t$ at time $t$.} 
    
    
    $S_z, S_x, S_{z,1}, S_{z,2} \gets \phi$ 
    
    $(\nx, \nf) \gets (0, 0)$ 
    
     \For{$t = 0, \cdots, T$}{
        
        \While{\emph{new image $\mathcal{I}$ with new pose $x_{n+1} \in \X_p$ recorded, next IMU measurement not yet received}}
        {
            $\mu_t \in \X_\IMU \times (\X_p)^{\nx}, \Sigma_t \in \R^{(d_\IMU + \nx \dx) \times (d_\IMU + \nx \dx)} \gets $ Alg. \ref{Alg: MSCKF, Pose Augmentation Sub-block} ($\tilde{x}_t$, $\mu_t$, $\Sigma_t$, $x_{n+1}$, $x_{n+1}^\IMU$, $\psi(\cdot)$)
            
            
            $\{z_{n+1,j}| \text{ feature } j \text{ is observed at } x_{n+1} \} \gets$ Feature measurements at $x_{n+1}$
            
            $\{f_j | \text{ Feature } j \text{ is observed at } x_{n+1}\} \gets$ Feature position estimates at $x_{n+1}$.
            
            Record new estimates of existing features and first estimate of new features at $x_{n+1} \in \X_p$.
            
            $S_z \gets S_z \cup \big\{ (x_{n+1}, f_{j})| \text{ Feature } j \text{ observed at $n+1$} \big\}$ 
            
            $n \gets n+1$
        
        \If{$n \geq N_{\max}-1$}{
        
            $S_x \gets \{x_i | i \text{ mod } 3 = 2, \text{ and } 1 \leq i \leq n. \}$
            
            $S_{z,1} \gets \big\{ (x_i, f_{j}) \in S_z \big| x_i \in S_x, \text{feature $j$ observed at each pose in } S_x \big\}$
        }
        
        $S_{z,2} \gets \big\{ (x_i, f_{j}) \in S_z | x_i \in x_{1:\nx}, \text{feature } j \text{ observed at } x_i \text{ but not at } x_n \big\}$.
        
        $S_f \gets \big\{ f_{j} \big| \exists \hspace{0.5mm} x_i \in x_{1:\nx} \text{ s.t. } (x_i, f_{j}) \in S_{z,1} \cup S_{z,2} \big\}$
        
        \If{$S_f \ne \phi$}{
            
            $\overline{\mu_t} \in \X_\IMU \times (\X_p)^{\nx}, \overline{\Sigma_t} \in \R^{(d_\IMU + \nx \dx) \times (d_\IMU + \nx \dx)} \gets $ Alg. \ref{Alg: MSCKF, Feature Update Sub-block} ($\tilde{x}_t$, $\mu_t$, $\Sigma_t$, $x_{n+1}$, $S_{z,1} \cup S_{z,2}$, $S_f$, $h(\cdot)$)
            
            
            $\hat x_t \gets \overline{\mu_t} \in \X_\IMU \times (\X_p)^{\nx}$.

        }
        
        $S_z \gets S_z \backslash (S_{z,1} \cup \{(x_i, f_j) | x_i \in S_x \})$
        
        Reindex poses and features, in ascending order of index, i.e., $\{x_1, \cdots, x_{\nx-|S_x|}\}$ and $\{f_1, \cdots, f_{\nf-|S_f|}\}$.
        
        $(\nf,\nx) \gets (\nf-|S_f|, \nx-|S_x|)$
        
        
        
        }
        
        \If{$t < T$}{
        
            $\mu_{t+1} \in \X_\IMU \times (\X_p)^{\nx}, \Sigma_{t+1} \in \R^{(d_\IMU + \nx \dx) \times (d_\IMU + \nx \dx)} \gets$ Alg. \ref{Alg: MSCKF, State Propagation sub-block}, MSCKF State Propagation ($\tilde{x}_t$, $\overline{\mu_t}$, $\overline{\Sigma_t}$)
            
        
        }
     }
    
    \Return{$\hat x_0, \cdots \hat x_T \in \X_\IMU \times (\X_p)^{\nx}$}
     
     \caption{Multi-State Constrained Kalman Filter, Standard Formulation.}
     }
\end{algorithm}

    


    
    The IMU state $x_{t,\IMU}$ takes the form:
    \begin{align}
        x_{t,\IMU} := (q_{WS}, v_S, b_g, b_a, r_{WS})(t) \in \R^{3} \times \mathbb H_u \times \R^9,
    \end{align}
    where we use $S$ and $W$ to represent the sensor frame and world frame, respectively. Here, $r_{WS} \in \R^3$ denotes the position of the IMU sensor frame represented in the world frame, $q_{WS} \in \mathbb H_u$ denotes the unit quaternion of axis rotation from world frame to IMU sensor frame, and $R(q_{WS}) \in SO(3)$ denotes the rotation matrix associated with $q_{WS}$. Moreover, $v_S \in \R^3$ denotes the linear velocity of the IMU sensor frame relative to the world frame, as represented in the world frame, while $b_g \in \R^3$ and $b_a \in \R^3$ denote the sensor biases of the gyroscope and accelerometer, respectively. Finally, $\tilde \omega_S \in \R^3$ and $\tilde a_S \in \R^3$ denote gyroscope and accelerometer measurements, respectively.
    
    
    For convenience, define $\X_\IMU := \R^3 \times \mathbb H_u \times \R^9$ and $\X_\IMU' := \R^3 \times \mathbb H \times \R^9$.
    The continuous-time IMU dynamics map $g_{\IMU, ct}: \X_\IMU \ra \X_\IMU'$ is given by:
    \begin{align*}
        \dot{q}_{WS} &= q_{WS} \star \frac{1}{2} \begin{bmatrix}
        0 \\ \Tilde{\omega}_S - b_g - w_g
        \end{bmatrix}, \\
        \dot{b}_g &= w_{b_g}, \\ 
        \dot{v}_S &= R(q_{WS})^\top \big( \Tilde{a}_S - b_a + w_a \big) + g_W, \\
        \dot{b}_a &= 
        w_{b_a}, \\
        \dot{r}_{WS} &= v_S.
    \end{align*}
    where $\star$ denotes quaternion multiplication, and $w_g, w_a, w_{b_g}, w_{b_a} \in \R^3$ denote zero-mean standard Gaussian noise.

    
    Each pose $x_k \in \{x_1, \cdots, x_n\}$ currently maintained in the sliding window of poses takes the form $x_k := (q_{WC_k}, r_{WC_k}) \in \mathbb H_u \times \R^3$, where $r_{WC_k} \in \R^3$ denotes the position of the camera at pose $x_k$ in the world frame, while $q_{WC_k} \in \mathbb H_u$ denotes the quaternion associated with the axis rotation from the world frame to the camera frame at pose $x_k \in \mathbb H \times \R^3$.
    For convenience, we define $\X_p := \mathbb H_u \times \R^3$.


    The MSCKF full state maintained throughout the operation of the MSCKF algorithm contains the IMU state at the current time, as well as a collection of $n$ poses, where $n$ is constrained to remain below a pre-specified, fixed upper bound $N_{\max}$:
    \begin{align} \label{Eqn: MSCKF, MSCKF full state def}
        \tilde{x}_t := (x_{t,\IMU}, x_1, \cdots, x_n) \in \X_\IMU \times (\X_p)^n.
    \end{align}
    The state space is thus $\X := \X_\IMU \times (\X_p)^n$.
    
    When a new image measurement arrives, the estimate of the current camera pose in the IMU frame, denoted $x_{\nx + 1}^\IMU \in \X_\IMU$, is transformed to the global frame, and appended to the MSCKF full state $\tilde{x}_t$. This coordinate transformation is realized by the map $\psi: \X_\IMU \times (\X_p)^n \ra \X_p$, defined by:
    \begin{align*}
        &\psi \big( q_{WS}, v_S, b_g, b_a, r_{WS}, q_{WC_1}, r_{WC_1}, \cdots, \\
        &\hspace{1cm} q_{WC_n}, r_{WC_n}, q_{WI_{\nx + 1}}, r_{WI_{\nx + 1}} \big) \\
        = \hspace{0.5mm} &\big( q_{IC} \star q_{WI_{\nx + 1}}, r_{WI_{\nx + 1}} + C(q_{WI_{\nx + 1}}) r_{IC} \big) \\
        := \hspace{0,5mm} &(q_{WC_{\nx + 1}}, r_{WC_{\nx + 1}}),
    \end{align*}
    where $q_{IC}$ denotes the quaternion encoding the (fixed) transformation from the IMU frame to the camera frame. In summary, the MSCKF algorithm defines the new pose $x_{\nx + 1}$ and updates the MSCKF full state $\tilde{x}_t$ as follows:
    \begin{align*}
        x_{\nx + 1} &\gets \psi(\tilde{x}_t,  x_{\nx + 1}^{\IMU}) \in \X_p, \\
        \tilde{x}_t &\gets (\tilde{x}_t, x_{\nx + 1}) = \big(\tilde{x}_t, \psi(\tilde{x}_t,  x_{\nx + 1}^{\IMU}) \big) \in \X_\IMU \times (\X_p)^{\nx + 1},
    \end{align*}
    with the map $\psi$ as defined above.
    
    When new feature position estimates are detected from a new image measurement, the new camera pose corresponding to this image measurement is appended to $\tilde{x}_t$, and $n$ is incremented by 1. If $n = N_{\max}$ the upper limit $N_{\max}$, a third of all old poses in $\tilde{x}_t$ is discarded, starting from the second oldest pose. Then, feature measurements, corresponding to features unobserved at the current pose, are marginalized and used to update the mean and covariance of the new MSCKF full state $\tilde{x}_t$.
    

    As is the case with the EKF algorithm, we assume that the image measurement space and feature space are given by $\R^{\dz}$ and $\R^{\df}$, respectively, with $\dz = 2$ and $\df = 3$. Throughout the duration of the MSCKF algorithm, poses and features are added into, dropped from, and marginalized from the MSCKF full state. Suppose at a given time, the MSCKF maintains $n$ poses in the \textit{MSCKF full state} $\tilde{x}_t$, and retains measurements of $\nf$ features. For each pose $i \in \{1, \cdots, n\}$ and feature $j \in \{1, \cdots, p\}$ currently maintained in the SLAM algorithm, if feature $j$ were detected at pose $i$, let $z_{i, j} \in \R^{\dz}$ denote the associated feature measurement. For the MSCKF, the measurement map $h: \X_\IMU \times \R^{\df} \ra \R^{\dz}$ is given by:
    \begin{align*}
        z_{i, j} &= h(x_i, f_{j}) \\
        &:= \frac{1}{(R(q_{WC_k}) f_j - r_{WC_k})_z} \begin{bmatrix}
            (R(q_{WC_k}) f_j - r_{WC_k})_x \\ (R(q_{WC_k}) f_j - r_{WC_k})_y
        \end{bmatrix}(t) \\
        &\hspace{1cm} + v_{i, j}.
    \end{align*}
    where $R(q_{WC_k}) \in SO(3)$ denotes the rotation matrix associated with the quaternion $q_{WC_k}$, $f_j \in \R^3$ denotes the position of feature $j$ in the world frame, while the subscript indices \say{$x, y, z$} refer to the respective coordinates of the vector $R(q_{WS}) f_j - r_{WS} \in \X_p$. Meanwhile, $v_{i,j} \in \R^{\dz}$ denotes zero-mean standard Gaussian noise in the measurement at time $t$, with covariance matrix $\Sigma_v \in \R^{\dz \times \dz}$, $\Sigma_v \succ 0$. 
    
    When a new image measurement is received, the MSCKF algorithm performs marginalization, described in Section \ref{subsec: Marginalization of States}, using two sets of feature measurements---the set of all feature measurements common to old poses $x_i$ to be dropped, denoted $S_{z,1}$, as well as the set of all feature measurements of features $f_j$ not seen in the current pose, denoted $S_{z,2}$. These are more precisely defined in Section \ref{subsec: MSCKF}. The measurement vector used for marginalization, denoted $\tilde{z} \in \R^{|S_{z,1} \cup S_{z,2}|\dz}$, is then given by concatenating the $q$ residual measurements obtained at times $t-\nx + 1, \cdots, t$, i.e.:
    \begin{align*}
        \tilde{z} := \{z_{i,j} | (x_i, f_j) \in S_{z,1} \cup S_{z,2} \} \in \R^{|S_{z,1} \cup S_{z,2}|\dz}.
    \end{align*}


\begin{algorithm} \label{Alg: MSCKF, Pose Augmentation Sub-block}
    {
    \normalsize
    \SetAlgoLined
    \KwData{MSCKF state $\tilde x_t \in \X_\IMU \times (\X_p)^n$, with mean $\mu_t \in \X_\IMU \times (\X_p)^n$ and covariance $\Sigma_t \in \R^{(d_\IMU + \nx \dx) \times (d_\IMU + \nx \dx)}$, New pose $x_{\nx + 1} \in \X_p$, measurement of new pose in IMU frame $x_{\nx + 1}^\IMU \in \X_p$, Transformation of poses from IMU frame to global frame $\psi: \R^{(d_\IMU + \nx \dx)} \times \X_p \ra \X_p$}
    
    \KwResult{Updated MSCKF state mean $\mu_t \in \X_\IMU \times (\X_p)^n$ and covariance $\Sigma_t \in \R^{(d_\IMU + \nx \dx) \times (d_\IMU + \nx \dx)}$, updated number of poses $n$.}

            
    $\tilde{x}_t \gets (\tilde{x}_t, x_{\nx + 1}) \in \R^{d_\IMU + (\nx + 1) \dx}$, where $x_{\nx + 1} \in \X_p$ is the new pose vector.
            
    $\{z_{\nx + 1,j} | \text{ Feature } j \text{ is observed at pose } \nx + 1 \} \gets$ Feature measurements at pose $x_{\nx + 1}$
            
    $\{f_j^\star | \text{ Feature } j \text{ is observed at pose } x_{\nx + 1}\} \gets$ Feature position estimates at pose $x_{\nx + 1}$.
            
    $\mu_t \gets (\mu_t, \psi(\mu_t, x_{\nx + 1}^\IMU)) \in \R^{d_\IMU + (\nx + 1) \dx}$, where $\mu_{t,\IMU} \in \R^{d_\IMU} :=$ IMU component of $\mu_t$, $x_{\nx + 1}^\IMU \in \X_p :=$ pose estimate of $x_{\nx + 1}$ from the IMU frame.
            
    $\Sigma_t \gets
        \begin{bmatrix}
            I_{d_\IMU + (\nx + 1)\dx} \\
            \frac{\partial \psi}{\partial (\tilde{x}_t, x_{\nx + 1}^\IMU)} 
        \end{bmatrix}
        \Sigma_t
        \begin{bmatrix}
            I_{d_\IMU + (\nx + 1)\dx} \\
            \frac{\partial \psi}{\partial (\tilde{x}_t, x_{\nx + 1}^\IMU)} 
        \end{bmatrix}^\top $
    
    \Return{$\mu_t \in \X_\IMU \times (\X_p)^n$, $\Sigma_t \in \R^{(d_\IMU + \nx \dx) \times (d_\IMU + \nx \dx)}$, $\nx \geq 0$}
     
     \caption{Multi-State Constrained Kalman Filter, Pose Augmentation Sub-block.}
     }
\end{algorithm}

\begin{algorithm} \label{Alg: MSCKF, Feature Update Sub-block}
    {
    \normalsize
    \SetAlgoLined
    \KwData{MSCKF state $\tilde x_t \in \X_\IMU \times (\X_p)^n$, with mean $\mu_t \in \X_\IMU \times (\X_p)^n$ and covariance $\Sigma_t \in \R^{(d_\IMU + \nx \dx) \times (d_\IMU + \nx \dx)}$, Set of image measurements for marginalization $S_{z,1} \cup S_{z,2}$, Set of features to marginalize $S_f$, measurement map $h: \X_p \times \R^{\df} \ra \R^{\dz}$.}
    
    \KwResult{Updated MSCKF state mean $\overline{\mu_t} \in \X_\IMU \times (\X_p)^n$ and covariance $\overline{\Sigma_t} \in \R^{(d_\IMU + \nx \dx) \times (d_\IMU + \nx \dx)}$.}
    
    $f_{S_f} \in \R^{|S_f| \df} \gets$ Concatenation of all features in $S_f$
            
            $f_{S_f}^\star \in \R^{|S_f| \df} \gets$ Concatenation of position estimate of all features in $S_f$
            
            $\Tilde{h}(\tilde{x}_t, f_{S_f}) \in \R^{|S_{z,1} \cup S_{z,2}| \dz} \gets $ Concatenation of measurement map outputs $\big\{ h(x_i, f_{j}) | (x_i, f_{j}) \in S_{z,1} \cup S_{z,2} \big\}$.
             
            $\tilde{z} \in \R^{|S_{z,1} \cup S_{z,2}| \dz} \gets$ Concatenation of feature measurements $\big\{ z_{ij} | (x_i, f_{j}) \in S_{z,1} \cup S_{z,2} \big\}$.
            
            $\tilde{H}_{t,x} \gets \frac{\partial \Tilde{h}}{\partial \tilde{x}_t} (\mu_t, f_{S_f}^\star) \in \R^{|S_{z,1} \cup S_{z,2}| \dz \times (d_\IMU + \nx \dx)}.$ 
            
            $\tilde{H}_{t,f} \gets \frac{\partial \Tilde{h}}{\partial f_{S_f}} (\mu_t, f_{S_f}^\star) \in \R^{|S_{z,1} \cup S_{z,2}| \dz \times |S_f| \df}.$
            
            $\{a_1, \cdots, a_{|S_{z,1} \cup S_{z,2}| \dz - |S_f| \df} \} \subset \R^{|S_{z,1} \cup S_{z,2}| \dz} \gets \text{Orthonormal basis for } N(\tilde{H}_{t,f}^\top).$
            
            $A \gets \begin{bmatrix}
            a_1 & \cdots & a_{|S_{z,1} \cup S_{z,2}| \dz - |S_f| \df}
            \end{bmatrix} \in \R^{|S_{z,1} \cup S_{z,2}| \dz \times (|S_{z,1} \cup S_{z,2}| \dz - |S_f| \df)}$.
            
            \label{Eqn: MSCKF, QR decomposition of A transpose Hx}
            $QT \gets$ QR Decomposition of $A^\top \tilde{H}_{t,x}$, with $Q \in \R^{(|S_{z,1} \cup S_{z,2}| \dz - |S_f| \df) \times (|S_{z,1} \cup S_{z,2}| \dz - |S_f| \df)}$, $T \in \R^{(|S_{z,1} \cup S_{z,2}| \dz - |S_f| \df) \times (d_\IMU + \nx \dx)}$.
            
            
            \label{Eqn: MSCKF Alg, Covariance Update}
            $\overline{\Sigma}_t^{-1} \gets \Sigma_t^{-1} + T^\top (Q^\top A^\top RAQ)^{-1} T \in \R^{(d_\IMU + \nx \dx) \times (d_\IMU + \nx \dx)}$.
            
            \label{Eqn: MSCKF Alg, Mean Update}
            $\overline{\mu_t} \gets \mu_t \boxplus \big( \Sigma_t^{-1} + T^\top (Q^\top A^\top RAQ)^{-1} T \big)^{-1} T^\top (Q^\top A^\top R AQ)^{-1} \big(\tilde{z} \boxminus \Tilde{h}(\tilde{x}_t) \big) \in \X_\IMU \times (\X_p)^n$.
            
            $\hat x_t \gets \overline{\mu_t} \in \X_\IMU \times (\X_p)^n$.
    
    \Return{$\overline{\mu_t} \in \X_\IMU \times (\X_p)^n$, $\overline{\Sigma_t} \in \R^{(d_\IMU + \nx \dx) \times (d_\IMU + \nx \dx)}$}
     
     \caption{Multi-State Constrained Kalman Filter, Feature Update Sub-block.}
     }
\end{algorithm}

\begin{algorithm} \label{Alg: MSCKF, State Propagation sub-block}
    {
    \normalsize
    \SetAlgoLined
    \KwData{MSCKF state $\tilde x_t \in \X_\IMU \times (\X_p)^n$, with mean $\mu_t \in \X_\IMU \times (\X_p)^n$ and covariance $\Sigma_t \in \R^{(d_\IMU + \nx \dx) \times (d_\IMU + \nx \dx)}$, (discrete-time) dynamics map $g: \R^{d_\IMU} \ra \R^{d_\IMU}$.}
    
    \KwResult{Updated MSCKF state mean $\mu_{t+1} \in \X_\IMU \times (\X_p)^n$ and covariance $\Sigma_{t+1} \in \R^{(d_\IMU + \nx \dx) \times (d_\IMU + \nx \dx)}$.}

    $(\overline{\mu}_{t,\IMU}, \overline{\mu}_{t,x, 1:n}) \gets \overline{\mu_t}$, with $\overline{\mu}_{t,\IMU} \in \R^{d_\IMU}$, $\overline{\mu}_{t,x,1:n} \in \R^{\nx \dx}$.
            
            $G_t \gets $ Jacobian of $g_\IMU: \R^{d_\IMU} \ra \R^{d_\IMU}$ evaluated at $\overline{\mu}_{t,\IMU} \in \R^{d_\IMU}$.
            
            $\mu_{t+1} \gets \big( g_\IMU(\overline{\mu}_{t,\IMU}), \overline{\mu}_{t,x,1:n} \big) \in \X_\IMU \times (\X_p)^n.$ \label{Eqn: MSCKF, Mean Propagation}
            
            $\Sigma_{t+1} \gets
            \begin{bmatrix}
                G_t & O \\
                O & I_{\nx \dx}
            \end{bmatrix}
            \overline{\Sigma_t}
            \begin{bmatrix}
                G_t^\top & O \\
                O & I_{\nx \dx}
            \end{bmatrix} + \begin{bmatrix}
                \Sigma_w & O \\
                O & O
            \end{bmatrix} \in  \R^{(d_\IMU + \nx \dx) \times (d_\IMU + \nx \dx)}$. \label{Eqn: MSCKF, Cov Propagation}
         
    \Return{$\mu_{t+1} \in \X_\IMU \times (\X_p)^n$, $\Sigma_{t+1} \in \R^{(d_\IMU + \nx \dx) \times (d_\IMU + \nx \dx)}$.}
     
     \caption{Multi-State Constrained Kalman Filter, State Propagation Sub-block.}
     }
\end{algorithm}

\subsubsection{Proofs from Section \ref{subsec: MSCKF}}
\label{subsubsec: App, MSCKF}

\begin{theorem}
The pose augmentation step of the standard MSCKF SLAM algorithm (Alg. \ref{Alg: MSCKF, Pose Augmentation Sub-block}) is equivalent to applying a Gauss-Newton step to $c_{MSCKF,t,1}: \X_\IMU \times (\X_p)^n \ra \R$, given by:
\begin{align*}
    &c_{MSCKF,t,1}(\tilde{x}_t, x_{\nx + 1}) \\
    = \hspace{0.5mm} &\Vert \tilde{x}_t \boxminus \mu_t \Vert_{\Sigma_t^{-1}}^2 + \epsilon^{-1} \Vert x_{\nx + 1} \boxminus \psi(\tilde{x}_t, x_{\nx + 1}^\IMU) \Vert_2^2,
\end{align*}
and taking $\epsilon \ra 0$ in the resulting (augmented) mean 
$\mu_t$ 
and covariance 
$\Sigma_t$.
\end{theorem}

\begin{proof}
We claim that from an optimization perspective, the state augmentation step is equivalent to applying one Gauss-Newton step to the cost function $c_{MSCKF,t,1}(\tilde{x}_t, x_{\nx + 1})$, specified above, and then taking the limit $\epsilon \ra 0$ in the resulting augmented mean $\mu_t(\epsilon) \in \X_\IMU \times (\X_p)^{(\nx + 1)}$ and augmented covariance $\mu_t(\epsilon) \in \R^{(d_\IMU + (\nx + 1)\dx) \times (d_\IMU + (\nx + 1)\dx)}$. 

To apply a Gauss-Newton step, our first task is to find a vector $C(\tilde{x}_t, x_{\nx + 1})$ of an appropriate dimension such that $c_{MSCKF,t,1}(\tilde{x}_t, x_{\nx + 1}) = C_1(\tilde{x}_t, x_{\nx + 1})^\top C_1(\tilde{x}_t, x_{\nx + 1})$. A natural choice is furnished by $C_1(\tilde{x}_t, x_{\nx + 1}) \in \R^{d_\IMU + (\nx + 1) \dx}$, as defined below:
\begin{align*}
    C_1(\tilde{x}_t, x_{\nx + 1}) &:= \begin{bmatrix}
        \Sigma_t^{-1/2} (\tilde{x}_t \boxminus \mu_t) \\
        \epsilon^{-1/2} \big( x_{\nx + 1} \boxminus \psi(\tilde{x}_t, x_{\nx + 1}^\IMU) \big)
    \end{bmatrix}.
\end{align*}
Thus, our parameters for the Gauss-Newton algorithm submodule are:
\begin{align*}
    (\tilde{x}_t^\star, x_{\nx + 1}^\star) &:= (\mu_t, \psi(\mu_t, x_{\nx + 1}^\IMU)) \in \X_\IMU \times (\X_p)^n, \\
    C_1(\tilde{x}_t^\star, x_{\nx + 1}^\star) &= \begin{bmatrix}
        \Sigma_t^{-1/2} (\tilde{x}_t^\star - \mu_t) \\
        \epsilon^{-1/2} \big( x_{\nx + 1}^\star - \psi(\tilde{x}_t^\star, x_{\nx + 1}^\IMU) \big)
    \end{bmatrix} = \begin{bmatrix}
    0 \\ 0
    \end{bmatrix} \\
    &\hspace{1cm} \in \R^{d_\IMU + (\nx + 1) \dx}, \\
    J &= \begin{bmatrix}
        \Sigma_t^{-1/2} & O \\ -\epsilon^{-1/2} \Psi & \epsilon^{-1/2} I_{\dx}
    \end{bmatrix} \\
    &\hspace{1cm} \in \R^{(d_\IMU + (\nx + 1)\dx) \times (d_\IMU + (\nx + 1)\dx)},
\end{align*}
where $\Psi \in \R^{\dx \times (d_\IMU + \nx \dx)}$ is defined as the Jacobian of $\psi: \X_\IMU \times (\X_p)^n \ra \X_p$ with respect to $\tilde{x}_t$ at $(\tilde{x}_t^\star, x_{\nx + 1}^\IMU) \in \R^{d_\IMU + (\nx + 1) \dx}$. By Algorithm \ref{Alg: gauss-newton}, the Gauss-Newton update is thus given by:
    \begin{align} \nonumber
        \Sigma_t(\epsilon) \leftarrow (J^\top J)^{-1} &= \begin{bmatrix}
        \Sigma_t^{1/2} & O \\
        \Psi \Sigma_t^{1/2} & \epsilon^{1/2} I_{\dx}
        \end{bmatrix}
        \begin{bmatrix}
        \Sigma_t^{1/2} & \Sigma_t^{1/2} \Psi^\top \\
        O & \epsilon^{1/2} I_{\dx}
        \end{bmatrix} \\ \nonumber
        &= \begin{bmatrix}
            \Sigma_t & \Sigma_t \Phi^\top \\
            \Psi \Sigma_t & \Psi \Sigma_t \Psi^\top + \epsilon I_{\dx}
        \end{bmatrix}, \\ \nonumber
        \mu_t(\epsilon) \gets \tilde{x}_t^\star - (J^\top &J)^{-1} J^\top C_1(\tilde{x}_t^\star, x_{\nx + 1}^\star) \\ \nonumber
        = 0. \hspace{1cm} &
    \end{align}
    Taking $\epsilon \ra 0$ concludes the proof.
\end{proof} 

\begin{theorem}
The feature update step of the standard MSCKF algorithm (Alg. \ref{Alg: MSCKF, Feature Update Sub-block}) is equivalent to applying a marginalization step to $c_{MSCKF,t,2}: \X_\IMU \times (\X_p)^n \times \R^{|S_f|\df} \ra \R$, given by:
\begin{align} \nonumber
    &c_{MSCKF,t,2}(\tilde{x}_t, f_{S_f}) \\ \nonumber
    := &\Vert \tilde{x}_t  \boxminus \mu_t \Vert_{\Sigma_t^{-1}}^2 + \sum_{(x_i,f_j) \in S_{z,1} \cup S_{z,2}} \Vert z_{i,j} \boxminus h(x_i,f_j) \Vert_{\Sigma_v^{-1}}^2,
\end{align}
where $f_{S_f} \in \R^{|S_f|\df}$ denotes the stacked vector of all feature positions in $S_f$ (see Algorithm \ref{Alg: MSCKF}).
\end{theorem}

\begin{proof}
First, we rewrite $c_{MSCKF,t,2}$ as:
\begin{align} \nonumber
    &c_{MSCKF,t,2}(\tilde{x}_t, f_{S_f}) \\  \nonumber
    := \hspace{0.5mm} &\Vert \tilde{x}_t  \boxminus \mu_t \Vert_{\Sigma_t^{-1}}^2 + \Vert \tilde{z} \boxminus \tilde{h}(\tilde{x}_t, f_{S_f}) \Vert_{\tilde{\Sigma}_v^{-1}}^2,
\end{align}
where $\tilde{z} \in \R^{|S_{z,1} \cup S_{z,2}| \dz}$, $\tilde{h}: \X_\IMU \times (\X_p)^n \times \R^{|S_f|\df} \ra \R^{|S_{z,1} \cup S_{z,2}| \dz}$: are defined as follows---$\tilde{z}$ denotes the stacked measurement vectors in $\{z_{i,j} | (x_i, f_j) \in S_{z,1} \cup S_{z,2} \} \in \R^{|S_{z,1} \cup S_{z,2}| \dz}$, $\tilde{h}(\tilde{x}_t, f_{S_f})$ denotes the stacked outputs of the measurement map in $\{h(x_i, f_j) | (x_i, f_j) \in S_{z,1} \cup S_{z,2} \} \in \R^{|S_{z,1} \cup S_{z,2}| \dz}$, and $\tilde{\Sigma}_v := \diag\{\Sigma_v, \cdots, \Sigma_v\} \in \R^{|S_z|\dz \times |S_z|\dz}$.

Essentially, by marginalizing the feature position estimates, this step utilizes information from feature measurements to constrain our state estimates. To accomplish this, we choose our algorithm variables as follows:
\begin{align*}
    \tilde{x}_{t,K} &:= \tilde{x}_t = (x_{t,\IMU}, x_1, \cdots, x_n) \\
    &\in \R^{d_\IMU + \nx \dx + |S_f|\df}, \\
    \tilde{x}_{t,M} &:= f_{S_f} \in \R^{|S_f|\df}, \\
    \overline{x} &:= (\tilde{x}_{t,K}, \tilde{x}_{t,M}) \\ 
    &\in \R^{d_\IMU + \nx \dx + |S_f|\df}, \\
    C_M(\tilde{x}_{t,K}, \tilde{x}_{t,M}) &:= \begin{bmatrix}
        \Sigma_t^{-1/2} (\tilde{x}_t \boxminus \mu_t) \\
        \tilde{\Sigma}_v^{-1x/2} \big( \tilde{z} \boxminus
        \tilde{h}(\tilde{x}_t, f_{S_f}) \big)
    \end{bmatrix} \\
    &\in \R^{d_\IMU + \nx \dx + |S_{z,1} \cup S_{z,2}|\dz}.
\end{align*}

The Marginalization algorithm block then implies that:
\begin{align*}
    J_K &:= \frac{\partial C_M}{\partial \tilde{x}_t} (\overline{\mu_t}, f_{S_f}^\star) = \begin{bmatrix}
    \Sigma_t^{-1/2} \\ - \tilde{\Sigma}_v^{-1/2} \tilde{H}_{t,x}
    \end{bmatrix} \\
    &\in \R^{(d_\IMU + \nx \dx + |S_{z,1} \cup S_{z,2}|\dz) \times (d_\IMU + \nx \dx)}, \\
    J_M &:= \frac{\partial C_M}{\partial f_{S_f}}(\overline{\mu_t}, f_{S_f}^\star) = \begin{bmatrix}
    O \\ -\tilde{\Sigma}_v^{-1/2} \tilde{H}_{t,f}
    \end{bmatrix} \\
    &\in \R^{(d_\IMU + \nx \dx + |S_{z,1} \cup S_{z,2}|\dz) \times |S_f|\df},
\end{align*}
where we have defined:
\begin{align*}
    f_{S_f}^\star \in \R^{|S_f| \df} &\gets \text{Stacked position estimates of features in } S_f, \\    
    \tilde{H}_{t,x} &:=  \frac{\partial \tilde{h}}{\partial \tilde{x}_t} \tilde{h}(\overline{\mu_t}, f_{S_f}^\star) \in \R^{|S_{z,1} \cup S_{z,2}|\dz \times (d_\IMU + \nx \dx)}, \\
    \tilde{H}_{t,f} &:= \frac{\partial \tilde{h}}{\partial f_{S_f}}(\overline{\mu_t}, f_{S_f}^\star) \in \R^{|S_{z,1} \cup S_{z,2}|\dz \times |S_f|\df}.
\end{align*}

Recall that the marginalization equations \eqref{Eqn: Main Alg, Marginalization, mu K} and \eqref{Eqn: Main Alg, Marginalization, Sigma K} in our formulation read:
\begin{align*}
    \mu_K &\leftarrow \mu_K - \Sigma_K J_K^\top \big[I - J_M(J_M^\top J_M)^{-1} J_M^\top \big] C_M\big(\tilde{x}_{t,K}, \tilde{x}_{t,M} \big), \\
    \Sigma_K &\leftarrow \big(J_K^\top (I - J_M(J_M^\top J_M)^{-1} J_M^\top) J_K \big)^{-1}.
\end{align*}
Substituting in the above expressions for $J_K, J_M$, and $C_M\big(\overline{\mu_t}, f_{S_f}^\star \big)$, we have:
\begin{align*}
    \overline{\Sigma}_t &\leftarrow (J_K^\top (I - J_M(J_M^\top J_M)^{-1} J_M^\top ) J_K \big)^{-1}, \\
    &= \Bigg( \begin{bmatrix}
    \Sigma_t^{-1/2} & -\tilde{H}_{t,x}^\top \tilde{\Sigma}_v^{-1/2}
    \end{bmatrix} \\
    &\hspace{1cm}
    \begin{bmatrix}
    I & O \\
    O & I - \tilde{\Sigma}_v^{-1/2} \tilde{H}_{t,f} (\tilde{H}_{t,f}^\top \tilde{\Sigma}_v^{-1} \tilde{H}_{t,f})^{-1} \tilde{H}_{t,f}^\top \tilde{\Sigma}_v^{-1/2}
    \end{bmatrix} \\
    &\hspace{1cm}
    \begin{bmatrix}
    \Sigma_t^{1/2} \\ -\tilde{\Sigma}_v^{-1/2} \tilde{H}_{t,x}
    \end{bmatrix} \Bigg)^{-1} \\
    &= \big( \Sigma_t^{-1} + \tilde{H}_{t,x}^\top \tilde{\Sigma}_v^{-1/2} \big[ I - \tilde{\Sigma}_v^{-1/2} \tilde{H}_{t,f} (\tilde{H}_{t,f}^\top \tilde{\Sigma}_v^{-1} \tilde{H}_{t,f})^{-1} \\
    &\hspace{1cm} \tilde{H}_{t,f}^\top \tilde{\Sigma}_v^{-1/2} \big] \tilde{\Sigma}_v^{-1/2} \tilde{H}_{t,x} \big)^{-1} \\
    \overline{\mu_t} &\leftarrow \mu_K - \Sigma_K J_K^\top  \big[I - J_M(J_M^\top J_M)^{-1} J_M^\top \big] C_M\big(\overline{\mu_t}, f_{S_f}^\star \big) \\ 
    &= \mu_t + \big( \Sigma_t^{-1} + \tilde{H}_{t,x}^\top \tilde{\Sigma}_v^{-1/2} \\
    &\hspace{5mm} \big[ I - \tilde{\Sigma}_v^{-1/2} \tilde{H}_{t,f} (\tilde{H}_{t,f}^\top \tilde{\Sigma}_v^{-1} \tilde{H}_{t,f})^{-1} \tilde{H}_{t,f}^\top \tilde{\Sigma}_v^{-1/2} \big] \\
    &\hspace{1cm} \cdot \tilde{\Sigma}_v^{-1/2} \tilde{H}_{t,x} \big)^{-1} \\
    &\hspace{1cm} \cdot \tilde{H}_{t,x}^\top \tilde{\Sigma}_v^{-1/2} \big[ I - \tilde{\Sigma}_v^{-1/2} \tilde{H}_{t,f} (\tilde{H}_{t,f}^\top \tilde{\Sigma}_v^{-1} \tilde{H}_{t,f})^{-1} \\
    &\hspace{1cm} \cdot \tilde{H}_{t,f}^\top \tilde{\Sigma}_v^{-1/2} \big] \tilde{\Sigma}_v^{-1/2} \big( \tilde{z} - \tilde{h}(\tilde{x}_t, f_{S_f}) \big).
\end{align*}
Comparing with the update step in the MSCKF algorithm, i.e., \eqref{Eqn: MSCKF Alg, Mean Update} and \eqref{Eqn: MSCKF Alg, Covariance Update}, reproduced below:
\begin{align*}
    \overline{\Sigma}_t^{-1} &\gets \Sigma_t^{-1} + T^\top (Q^\top A^\top \tilde{\Sigma}_v AQ)^{-1} T, \\
    \overline{\mu_t} &\gets \mu_t + \big( \Sigma_t^{-1} + T^\top (Q^\top A^\top \tilde{\Sigma}_v AQ)^{-1} T \big)^{-1} \\
    &\hspace{1cm} T^\top (Q^\top A^\top \tilde{\Sigma}_v AQ)^{-1} \big( \tilde{z} - \tilde{h}(\tilde{x}_t, f_{S_f}) \big)
\end{align*}
we find that it suffices to show:
\begin{align*}
    &T^\top (Q^\top A^\top \tilde{\Sigma}_v AQ)^{-1} \\
    = \hspace{0.5mm} &\tilde{H}_{t,x}^\top \tilde{\Sigma}_v^{-1/2} \big[ I - \tilde{\Sigma}_v^{-1/2} \tilde{H}_{t,f} (\tilde{H}_{t,f}^\top \tilde{\Sigma}_v^{-1} \tilde{H}_{t,f})^{-1} \tilde{H}_{t,f}^\top \tilde{\Sigma}_v^{-1/2} \big] \\
    &\hspace{1cm} \cdot \tilde{\Sigma}_v^{-1/2} \\
    = \hspace{0.5mm} &\tilde{H}_{t,x}^\top \tilde{\Sigma}_v^{-1} - \tilde{H}_{t,x}^\top \tilde{\Sigma}_v^{-1} \tilde{H}_{t,f} (\tilde{H}_{t,f}^\top \tilde{\Sigma}_v^{-1} \tilde{H}_{t,f})^{-1} \tilde{H}_{t,f}^\top \tilde{\Sigma}_v^{-1}.
\end{align*}
To see this, recall that $A$ is defined as a full-rank matrix whose columns span $N(\tilde{H}_{t,f}^\top)$. Thus:
\begin{align*}
    (\tilde{\Sigma}_v^{-1/2} \tilde{H}_{t,f})^\top \cdot \tilde{\Sigma}_v^{1/2} AQ = \tilde{H}_{t,f}^\top AQ = O.
\end{align*}
In other words, the columns of $\tilde{\Sigma}_v^{-1/2} \tilde{H}_{t,f}$ and of $\tilde{\Sigma}_v^{1/2} AQ$ form bases of orthogonal subspaces whose direct sum equals $\R^{\nx q\dz}$. We thus have:
\begin{align*}
    &\tilde{\Sigma}_v^{-1/2} \tilde{H}_{t,f}(\tilde{H}_{t,f}^\top \tilde{\Sigma}_v^{-1} \tilde{H}_{t,f})^{-1} \tilde{H}_{t,f}^\top \tilde{\Sigma}_v^{-1/2} \\
    &\hspace{1cm} + \tilde{\Sigma}_v^{1/2} AQ (Q^\top A^\top \tilde{\Sigma}_v AQ)^{-1} Q^\top A^\top \tilde{\Sigma}_v^{1/2} = I,
\end{align*}
which in turn implies that:
\begin{align*}
    &T^\top (Q^\top A^\top \tilde{\Sigma}_v AQ)^{-1} \\
    = \hspace{0.5mm} &\tilde{H}_{t,x}^\top AQ(Q^\top A^\top \tilde{\Sigma}_v AQ)^{-1} Q^\top A^\top \\
    = \hspace{0.5mm} &\tilde{H}_{t,x}^\top \tilde{\Sigma}_v^{-1/2} (\tilde{\Sigma}_v^{1/2} AQ) \\
    &\hspace{1cm} (Q^\top A^\top \tilde{\Sigma}_v^{1/2} \cdot \tilde{\Sigma}_v^{1/2} AQ)^{-1} (Q^\top A^\top \tilde{\Sigma}_v^{1/2}) \tilde{\Sigma}_v^{-1/2} \\
    = \hspace{0.5mm} &\tilde{H}_{t,x}^\top \tilde{\Sigma}_v^{-1/2} \big(I - \tilde{\Sigma}_v^{-1/2} \tilde{H}_{t,f}(\tilde{H}_{t,f}^\top \tilde{\Sigma}_v^{-1} \tilde{H}_{t,f})^{-1} \tilde{H}_{t,f}^\top \tilde{\Sigma}_v^{-1/2} \big)\\
    &\hspace{1cm} \tilde{\Sigma}_v^{-1/2} \\
    = \hspace{0.5mm} &\tilde{H}_{t,x}^\top \tilde{\Sigma}_v^{-1} - \tilde{H}_{t,x}^\top \tilde{\Sigma}_v^{-1} \tilde{H}_{t,f} (\tilde{H}_{t,f}^\top \tilde{\Sigma}_v^{-1} \tilde{H}_{t,f})^{-1} \tilde{H}_{t,f}^\top \tilde{\Sigma}_v^{-1},
\end{align*}
as claimed.

\end{proof}

\begin{theorem} 
The state propagation step of the standard MSCKF SLAM algorithm (Alg. \ref{Alg: MSCKF, State Propagation sub-block}) is equivalent to applying a Marginalization step once to $c_{MSCKF,t,4}: \R^{2d_\IMU + \nx \dx} \ra \R$, given by:
    \begin{align*}
        &c_{MSCKF,t,4}(\tilde{x}_t, x_{t+1,\IMU}) \\
        := \hspace{0.5mm} &\Vert \tilde{x}_t \boxminus \overline{\mu_t} \Vert_{\overline{\Sigma}_t^{-1}}^2 + \Vert x_{t+1,\IMU} \boxminus g_\IMU(x_{t,\IMU}) \Vert_{\Sigma_t^{-1}}^2.
    \end{align*}
\end{theorem}

\begin{proof}
We claim that from an optimization perspective, the update step is equivalent to applying one marginalization step to the cost function $c_{MSCKF,t,4}(\tilde{x}_t, x_{t+1,\IMU})$ specified above. In particular, we wish to marginalize out $x_{t,\IMU} \in \X_\IMU$ and retain $x_{t+1,\IMU} \in \X_\IMU$; in other words, in the notation of our Marginalization algorithm submodule, we have:
\begin{align*}
    \tilde{x}_{t,K} &:= (x_{t+1,\IMU}, x_1, \cdots, x_n) \in \X_\IMU \times (\X_p)^n, \\
    \tilde{x}_{t,M} &:= x_{t,\IMU} \in \X_\IMU.
\end{align*}
    
To apply a marginalization step, our first task is to find vectors $C_{K}(x_K) = C_{K}(\tilde{x}_t)$ and $C_{M}(x_K, x_M) = C_{M}(\tilde{x}_t, x_{t+1,\IMU})$ of appropriate dimensions such that $c_{MSCKF,t,4}(\tilde{x}_t, x_{t+1,\IMU}) = C_{K}(x_{t+1,\IMU})^\top C_{K}(x_{t+1,\IMU}) + C_{M}(\tilde{x}_t, x_{t+1,\IMU})^\top C_{M}(\tilde{x}_t, x_{t+1,\IMU})$. A natural choice is furnished by $C_{K}(x_{t+1,\IMU}) \in \R$ and $C_{M}(\tilde{x}_t, x_{t+1,\IMU}) \in \X_p$, as defined below:
    \begin{align*}
        C_{K}(\tilde{x}_{t,K})&=0 \in \R \\
        C_{M}(\tilde{x}_{t,K},\tilde{x}_{t,M}) &= \begin{bmatrix} \bar\Sigma_t^{-1/2}(\tilde{x}_t - \overline{\mu_t}) \\ \Sigma_w^{-1/2} \big(x_{t+1,\IMU} - g_\IMU(x_{t,\IMU}) \big) \end{bmatrix} \\
        &\hspace{1cm} \in \R^{2d_\IMU + \nx \dx}.
    \end{align*}
    For convenience, we will define the IMU state and pose components of the mean $\mu_t \in \X_\IMU \times (\X_p)^n$ by $\mu_t := (\mu_{t,\IMU}, \mu_{t,\IMU}) \in \X_\IMU \times (\X_p)^n$, with $\mu_{t,\IMU} \in \X_p$ and $\mu_{t,x} \in (\X_p)^n$, respectively. This mirrors our definition of $x_t \in \X_p$ and $x_{\nx + 1} \in (\X_p)^n$ as the components of the full state $\tilde{x}_t := (x_t, x_{\nx + 1}) \in \X_\IMU \times (\X_p)^n$. In addition, we will define the components of $\bar\Sigma_t^{-1/2} \in \R^{(d_\IMU + \nx \dx) \times (d_\IMU + \nx \dx)}$ and $\bar\Sigma_t^{-1} \in \R^{(d_\IMU + \nx \dx) \times (d_\IMU + \nx \dx)}$ by:
    \begin{align*}
       &\begin{bmatrix}
            \Omega_{t,\IMU,\IMU} & \Omega_{t,\IMU,x} \\
            \Omega_{t,x,\IMU} & \Omega_{t,x,x}
        \end{bmatrix} :=  \bar\Sigma_t^{-1} \\
        &\hspace{1cm} \in \R^{(d_\IMU + \nx \dx) \times (d_\IMU + \nx \dx)}, \\
        &\begin{bmatrix}
            \Lambda_{t,\IMU,\IMU} & \Lambda_{t,\IMU,x} \\
            \Lambda_{t,x,\IMU} & \Lambda_{t,x,x}
        \end{bmatrix} := \bar\Sigma_t^{-1/2} \\
        &\hspace{1cm} \in \R^{(d_\IMU + \nx \dx) \times (d_\IMU + \nx \dx)},
    \end{align*}
    with the dimensions of the above block matrices given by $\Sigma_{t,\IMU,\IMU}, \Lambda_{t,\IMU,\IMU} \in \R^{d_\IMU \times d_\IMU}$, $\Sigma_{t,\IMU,x}, \Lambda_{t,\IMU,x} \in \R^{d_\IMU \times \nx \dx}$, $\Sigma_{t,x,\IMU}, \Lambda_{t,x,\IMU} \in \R^{\nf \dx \times d_\IMU}$, and $\Sigma_{t,x,x}, \Lambda_{t,x,x} \in \R^{\nx \dx \times \nx \dx}$. Using the above definitions, we can reorder the residuals in $C_K \in \R$ and $C_M \in \R^{2d_\IMU + \nx \dx}$, and thus redefine them by:
    \begin{align*}
        &C_{K}(\tilde{x}_{t,K})=0 \in \R \\
        &C_{M}(\tilde{x}_{t,K},\tilde{x}_{t,M}) \\
        = \hspace{0.5mm} & \begin{bmatrix} 
            \Lambda_{t,\IMU,\IMU}(x_{t,\IMU} - \mu_{t,\IMU}) + \Lambda_{t,\IMU,x}(x_{1:n} - \mu_{t,x}) \\
            \Sigma_w^{-1/2}(x_{t+1,\IMU} - g_\IMU(x_{t,\IMU})) \\
            \Lambda_{t,x,\IMU}(x_{t,\IMU} - \mu_{t,\IMU}) + \Lambda_{t,x,x}(x_{1:n} - \mu_{t,x})
        \end{bmatrix} \in \R^{2d_\IMU + \nx \dx},
    \end{align*}
    where $x_{1:n} := (x_1, \cdots, x_n) \in (\X_p)^n$.

    Our state variables and cost functions for the Gauss-Newton algorithm submodule are:
    \begin{align*}
        \overline{x_M^\star} &= \tilde{x}_t^\star = \overline{\mu_t} \in \X_\IMU \times (\X_p)^n, \\
        \overline{x_K^\star} &= g(\tilde{x}_t^\star) = g(\overline{\mu_t}) \in \X_\IMU \times (\X_p)^n, \\
        C_{K}(\tilde{x}_{t,K}^\star) &= 0 \in \R, \\
        C_{M}(\tilde{x}_{t,K}^\star,\tilde{x}_{t,M}^\star) &= 
        \begin{bmatrix}
        0 \\ 0
        \end{bmatrix} \in \R^{2 d_\IMU + \nx \dx}, \\
        J_K &= \begin{bmatrix} 
            O & \Lambda_{\IMU,x} \\ \Sigma_w^{-1/2} & O \\
            O & \Lambda_{xx}
        \end{bmatrix} \\
        &\in \R^{(2 d_\IMU + \nx \dx) \times (d_\IMU + \nx \dx)} \\
        J_M &= \begin{bmatrix}
            \Lambda_{\IMU,\IMU} \\ -\Sigma_w^{-1/2} G_t \\
            \Lambda_{x,\IMU}
        \end{bmatrix} \in \R^{(2 d_\IMU + \nx \dx) \times \dx},
    \end{align*}
    where we have defined $G_t$ to be the Jacobian of $g_\IMU: \X_\IMU \ra \X_\IMU$ at $\overline{\mu_{t,\IMU}} \in \X_\IMU$, i.e.:
    \begin{align*}
        G_t := \frac{\partial g}{\partial x_{t,\IMU}} \Bigg|_{x_{t,\IMU} = \overline{\mu_{t,\IMU}}}
    \end{align*}
    Applying the Marginalization update equations, we thus have:
    \begin{align*}
    \mu_{t+1} &\gets \tilde{x}_{t,K} - \Sigma_{t+1} J_K^\top \big[ I - J_M(J_M^\top J_M)^{-1} J_M^\top \big] \\
    &\hspace{1cm} C_{M}(\overline{x_K^\star}, \overline{x_M^\star}) \\
    &= g(\overline{\mu_t}), \\
    \Sigma_{t+1} &\gets \big(J_K^\top \big[ I - J_M(J_M^\top J_M)^{-1} J_M^\top \big] J_K\big)^{-1}, \\ 
    &= \big(J_K^\top J_K - J_K^\top J_M(J_M^\top J_M)^{-1} J_M^\top J_K\big)^{-1}, \\ 
    &= \Bigg(\begin{bmatrix}
    \Sigma_w^{-1} & O \\
    O & \Lambda_{x,\IMU} \Lambda_{\IMU,x} + \Lambda_{xx}^2
    \end{bmatrix} 
    \\
    &\hspace{1cm} - \begin{bmatrix}
    - \Sigma_w^{-1} G_t \\
    \Lambda_{x,\IMU} \Lambda_{\IMU,\IMU} + \Lambda_{xx} \Lambda_{x,\IMU}
    \end{bmatrix} \cdot \\
    &\hspace{1cm} \cdot (\Lambda_{\IMU,\IMU}^2 + \Lambda_{\IMU,x} \Lambda_{x,\IMU} + G_t^\top \Sigma_w^{-1} G_t)^{-1} \\
    &\hspace{1cm} \cdot \begin{bmatrix}
    -G_t^\top \Sigma_w^{-1} & \Lambda_{\IMU,\IMU} \Lambda_{\IMU,x} + \Lambda_{x,\IMU} \Lambda_{xx}
    \end{bmatrix} \Bigg)^{-1} \\
    &= \Bigg(\begin{bmatrix}
            \Sigma_w^{-1} & O \\
            O & \Omega_{xx}
            \end{bmatrix} 
            - \begin{bmatrix}
            - \Sigma_w^{-1} G_t \\
            \Omega_{x,\IMU}
            \end{bmatrix} \\
            &\hspace{1cm} (\Omega_{\IMU,\IMU} + G_t^\top \Sigma_w^{-1} G_t)^{-1} \\
            &\hspace{1cm} \begin{bmatrix}
            -G_t^\top \Sigma_w^{-1} & \Omega_{\IMU,x}
        \end{bmatrix} \Bigg)^{-1}
    \end{align*}
    To show that this is indeed identical to the propagation equation for the covariance matrix in the Extended Kalman Filter algorithm, i.e. Algorithm \ref{Alg: EKF}, Line \ref{Eqn: EKF, Cov Propagation}, we must show that:
    \begin{align*}
        &\Bigg(\begin{bmatrix}
            \Sigma_w^{-1} & O \\
            O & \Omega_{xx}
            \end{bmatrix} 
            - \begin{bmatrix}
            - \Sigma_w^{-1} G_t \\
            \Omega_{x,\IMU}
            \end{bmatrix} \\
            &\hspace{5mm} (\Omega_{\IMU,\IMU} + G_t^\top \Sigma_w^{-1} G_t)^{-1} \begin{bmatrix}
            -G_t^\top \Sigma_w^{-1} & \Omega_{\IMU,x}
        \end{bmatrix} \Bigg)^{-1} \\
        = \hspace{0.5mm} & \begin{bmatrix}
                G_t \overline{\Sigma}_{t,\IMU,\IMU} G_t^\top + \Sigma_w & G_t \overline{\Sigma}_{t,\IMU,x} \\
                \overline{\Sigma}_{t,\IMU,x} G_t^\top & \overline{\Sigma}_{t,x,x}
            \end{bmatrix}
    \end{align*}
    This follows by brute-force expanding the above block matrix components, and applying Woodbury's Matrix Identity, along with the definitions of $\Sigma_{t,\IMU,\IMU}, \Lambda_{t,\IMU,\IMU}$, $\Sigma_{t,\IMU,x}, \Lambda_{t,\IMU,x}$, $\Sigma_{t,x,\IMU}, \Lambda_{t,x,\IMU}$, $\Sigma_{t,x,x}$, and $\Lambda_{t,x,x}$.
    
    
    
    
    
\end{proof}

\subsection{Simulation Settings}
\label{subsec: App, simulation}

\subsubsection{Dataset}

All experiments are performed on the EuRoC MAV dataset \cite{Burri2016EuRoC}, a popular public SLAM dataset of stereo image sequences and inertial measurement unit (IMU) measurements. The stereo images and IMU measurements arrive at rates of 20Hz and 200Hz, respectively. The sensor suite is mounted on board a micro aerial vehicle. Ground-truth poses, recorded using a Vicon moion capture system and IMU biases, are also available for each sequence. Our experiments are carried out on the Vicon Room 2 01 and 02 sequences, which contain about 2300 stereo images each and span about 2 minutes of real-time operation. The first of these sequences is easier to analyze via our SLAM algorithms, as it corresponds to simple and slow evolution of the camera, while the latter contains some jerky and quick motions that prove challenging to some algorithms.

\subsubsection{Front-end and Back-end}

Since the focus of this work is the SLAM back-end, we standardize the front-end across all experiments, altering only the back-end used to process the abstracted data produced by the front-end. We use keypoint features as environment landmarks, as is standard in visual SLAM. First, BRISK features are extracted from both images of the input stereo pair. Then, feature matching is carried out between the left and right image frames. We then carry out brute force matching using Hamming distance on the binary BRISK descriptors, and filter outliers via an epipolar constraint check, using the known relative pose between the two cameras in the stereo set-up. Only keypoints for which a stereo match was found are kept. Next, a four-way consistence check is carried out. i.e. a match between two stereo frames $S_1, S_2$ is accepted if and only if both observations of a given feature in $S_1$ are matched to the respective observations of the same feature in $S_2$. Finally, outlier matches are rejected by projecting the best estimate of the matched feature onto the best estimate of the current camera pose, and rejecting matches that have a high reprojection error. Any stereo matches in the current frame that were not matched with a previously seen is recorded as a newly detected landmark, and initialized using stereo triangulation from the best estimate of the current camera pose. The front-end maintains data structures allowing two-way access between features and camera poses: for each feature index, it is possible to look up all camera poses from which that feature is visible, and likewise for each camera pose it is possible to query which features are visible in that frame.

For book-keeping the cost function in the back-end, computing Jacobians, and implementing Gauss-Newton optimization, we use GTSAM in C++ \cite{dellaert2012gtsam, dellaert2017factor}.


\subsubsection{Dynamics and Image Measurement Models}

We use an on-board IMU to collect odometry measurements, i.e., body-frame angular velocity and linear acceleration, and apply the IMU pre-integration scheme detailed in \cite{forster2015imu}, as summarized below. The objective of IMU preintegration is to establish a discrete-time dynamics map $x_{t+1} = g(x_t)$ that allows us to predict the pose of the robot at time $x_{t+1}$ given the pose at time $t$ and the IMU measurement at time $t$. Here, each timestep corresponds to a new \textit{image} measurement. Since IMU measurements arrive at a faster rate than image measurements, to compute this map, we must stack several IMU measurements into a relative state measurement, which can then be concatenated with the state at time $x_t$ to get the predicted state at time $x_{t+1}$. 

The robot state $x_t$ consists of the orientation, position, and velocity of the body frame relative to the world frame, and the IMU biases, so that $x_t = (R_t, p_t, v_t, b_t)$, where $(R_t, p_t) \in SE(3)$, $v_t \in \mathbb R^3$ and $b_t = (b_t^g, b_t^a) \in \mathbb R^3\times\mathbb R^3 \simeq \mathbb R^6$ are the IMU biases in the gyroscope and accelerometer respectively. The IMU biases are slowly varying and generally unknown, so they are included in the robot state and are also jointly estimated.

The IMU measures angular velocity and accelerations. The measurements are denoted ${}_B\tilde {a}$ and ${}_B \tilde \omega_{WB}$. Here, $B$ refers to the robot's body frame and $W$ the world frame. The prefix $B$ means that the quantity is expressed in the $B$ frame, and the suffix $WB$ denotes that the quantity represents the motion of the $B$ frame relative to $W$. So the pose of the robot is $(R_{WB}, {}_Wp) \in SE(3)$. The measurements are affected by additive white noise $\eta$ and the slowly varying IMU biases:
\begin{align*}
{}_B\tilde\omega_{WB}(t) &= {}_B\omega_{WB}(t) + b^g(t) + \eta^g(t) \\
{}_B\tilde a(t) &= {R}_{WB}^\top(t) ({}_W a(t) - {}_W g) + b^g(t) + \eta^g(t)
\end{align*}
where ${}_W\omega_{WB}$ is the true angular velocity, ${}_W a$ the true acceleration, and ${}_W g$ the gravity acceleration vector in the world frame. Assume that between timestep $t$ and $t+1$, we received $m$ IMU measurements, at constant time increments $\Delta t$. \cite{forster2015imu} provides expressions for $(x_{t+1} \boxminus g(x_t))$ directly in terms of the IMU measurements. Additionally, expressions for noise propagation are also provided, which allows us to compute the covariance $\Sigma_v$ over the error $(x_{t+1} \boxminus g(x_t))$ in terms of the measurement noise $\eta$, which is what we need to compute the required cost function $\|x_{t+1} \boxminus g(x_t)\|_{\Sigma_v^{-1}}^{2}$ and Jacobians.

 
Given a camera pose $x_t = (R_t, p_t)$ and a 3D feature location $f_j = (f_j^x, f_j^y, f_j^z)$, the camera measurement model $h(x_t, f_j)$ predicts the projected pixel location of the point in both stereo images. We assume the stereo camera pair is calibrated and rectified, so that both cameras have the same pinhole camera matrix $K$, and the epilines are horizontal. As such, the two measurements of a point in the two cameras will share the same $v$-coordinate in $(u, v)$ image space. Therefore, an image measurement will be stored as a 3-vector $(u_L, u_R, v)$, where the coordinates of the measurement in the left and right image are $(u_L, v)$ and $(u_R, v)$ respectively. The measurement map $h$ predicts the image location $(u_L, u_R, v)$ by projecting the point $f_j$ onto both image frames (with the standard pinhole projection), using the known poses of the two cameras in the robot's body-frame. Due to noise, the actual measurement will not have the exact same $v$ coordinate, so the image measurement is collected by averaging the two $v$-coordinates of the two keypoints. This measurement $z_{tj}$ is then compared to the predicted coordinates by a Mahalanobis distance in $\mathbb R^3$ space to get the cost function $\|z_{tj} - h(x_t, f_j)\|_{\Sigma_w^{-1}}^2$. The covariance $\Sigma_w$ is the expected noise in image space, which is a design parameter. In our experiments we choose $\Sigma_w = \sigma I_3$ where $I_3$ is the $3\times 3$ identity matrix and $\sigma = 0.05$ pixels, which was found to work well in practice.




\end{document}